%% file: mainNeutral.tex
\documentclass[11pt]{article}              

\usepackage[margin=25mm]{geometry}
\usepackage{graphics}
\usepackage{colortbl}
\usepackage{caption}
\usepackage{subfig}
\usepackage{graphicx}
\usepackage{fancyhdr}
\usepackage{amssymb,amsfonts,amsmath,amsthm}
\usepackage{natbib}
\newcommand{\ve}[1]{\boldsymbol{#1}}

\newcommand{\cA}{\ensuremath{\mathcal{A}}}
\newcommand{\cD}{\ensuremath{\mathcal{D}}}
\newcommand{\cR}{\ensuremath{\mathcal{R}}}

\newcommand{\cT}{\ensuremath{\mathcal{T}}}
\newcommand{\cV}{\ensuremath{\mathcal{V}}}

\newcommand{\cH}{\ensuremath{\mathcal{H}}}
\newcommand{\cM}{\ensuremath{\mathcal{M}}}
\newcommand{\cm}{\ensuremath{\mathcal{M}}}
\newcommand{\cU}{\ensuremath{\mathcal{U}}}
\newcommand{\cX}{\ensuremath{\mathcal{X}}}
\newcommand{\cP}{\ensuremath{\mathcal{P}}}
\newcommand{\cY}{\ensuremath{\mathcal{Y}}}
\newcommand{\eqdef}{\stackrel{\text{def}}{=}}
\newcommand{\di}[1]{{\rm d}#1} 

\newcommand{\Esp}[1]{{\mathbb E}\left[ #1 \right]}
\newcommand{\Var}[1]{{\rm Var}\left[ #1 \right]}
\newcommand{\iu}{{\boldsymbol{\mathsf{u}}}}
\newcommand{\iv}{{\boldsymbol{\mathsf{v}}}}
\newcommand{\iw}{{\boldsymbol{\mathsf{w}}}}

\newcommand{\BParams}{\mathbf{X}}

\newcommand{\vx}{\ve{x}}
\newcommand{\vX}{\ve{X}}

\newcommand{\vu}{\ve{u}}
\newcommand{\vU}{\ve{U}}
\newcommand{\Rr}{{\mathbb R}}
\newcommand{\innerprod}[2]{\left\langle #1, #2\right\rangle}
\newcommand{\indfun}[1]{\ve{1}_{#1}}
\newcommand{\Ntot}{\ensuremath{N_{ED}}}

\newcommand{\maxDegSSE}{p_{\mathrm{max}}^{\mathrm{SSE}}}
\newcommand{\maxDegPCE}{p_{\mathrm{max}}^{\mathrm{PCE}}}
\newcommand{\egr}{{\em e.g.},\ }
\newcommand{\ies}{{\em i.e.},\ }
\newcommand{\highlightdiff}[1]{\textcolor{black}{#1}}

\DeclareMathOperator*{\argmin}{arg\,min}

\usepackage{authblk}
	
\graphicspath{{./},{./Figures/}}

\begin{document}
\title{Stochastic Spectral Embedding} 

\author[1]{S. Marelli} \author[1]{P.-R. Wagner} \author[1]{C. Lataniotis} \author[1]{B. Sudret} 

\affil[1]{Chair of Risk, Safety and Uncertainty Quantification,
  
  ETH Zurich, Stefano-Franscini-Platz 5, 8093 Zurich, Switzerland}

\date{08.04.2020}
\maketitle

\abstract{Constructing approximations that can accurately mimic the behavior of complex models at reduced computational
	costs is an important aspect of uncertainty quantification.
	Despite their flexibility and efficiency, classical surrogate models such as Kriging or 
	polynomial chaos expansions tend to struggle with highly non-linear, localized or 
	non-stationary computational models.
	
	We hereby propose a novel sequential adaptive surrogate modeling method based on recursively embedding locally spectral
	expansions. 
	It is achieved by means of disjoint recursive partitioning of the input domain, which consists in sequentially splitting the latter into smaller subdomains, and constructing a simpler local spectral expansions in each, exploiting the trade-off complexity vs. locality. 
	The resulting expansion, which we refer to as ``stochastic spectral embedding'' (SSE), is a piece-wise continuous approximation of the model response that shows promising approximation capabilities, and good scaling with both the problem dimension and the
	size of the training set.
	
	We finally show how the method compares favorably against state-of-the-art sparse polynomial chaos expansions on a set of models with different complexity and input dimension.\\[1em] 

}

\maketitle

%%%%%%%%%%%%%%%%%%%%%%%%%%%%%%%%%%%%%%%%%%%%%%%%%%%%%%%%%%%%%%%%%%%%%%%%
\input{01_Introduction}

\input{02_Methodology}

\input{03_Applications}

\input{04_Conclusions}

\section*{Acknowledgements}
The PhD thesis of the second author is supported by ETH grant \#44 17-1.

\appendix
\input{05_Appendix}
%%%%%%%%%%%%%%%%%%%%%%%%%%%%%%%%%%%%%%%%%%%%%%%%%%%%%%%%%%%%%%%%%%%%%%%%%%%%
%% References

\bibliographystyle{chicago}
\bibliography{References}
\end{document}

%% file: 01_Introduction.tex
\section{Introduction}
\label{sec:Introduction}

In the era of \textit{machine learning} (ML) and \textit{uncertainty quantification} (UQ), it is not surprising to see their boundary
getting progressively blurred.
Cross-fertilization between the two disciplines is nowadays the norm, rather than an exception, and for good reasons.
Physics-informed neural networks are reaching unprecedented approximation power in UQ applications 
(see, \textit{e.g.}, \citep{Raissi2019JCP,Guofei2019SIAMJCP}), while sparse polynomial chaos expansions are used as denoising regressors in 
\cite{Torre2019PCE4ML}, as high-dimensional regression tools in \cite{Lataniotis2020}, and on real-world experimental data in \cite{Abbiati2020MSSP}. 
UQ-born Gaussian process modeling \cite{Santner2003,Rasmussen2006} is now a 
staple tool in ML \cite{Rasmussen2006}, while support vector machines 
\citep{Vapnik2013} found their way in rare event estimation 
\citep{Bourinet2016RESS,MoustaphaJRUES2018}. 

\highlightdiff{More in general, the adoption of both surrogate models and ML 
is becoming 
mainstream in applied sciences and engineering, with applications in entirely 
different fields. A few examples from the recent literature include: 
macroeconomics \cite{Harenberg2019}, wind turbine modeling \cite{Slot2020}, 
nuclear engineering \cite{Radaideh2020}, smart grid engineering 
\cite{Wang2020}, crash test simulations \cite{MoustaphaJRUES2018}, dam 
stability assessment \cite{Guo2020}, and many more.}
This list could be extended arbitrarily, as does the rich literature on these 
topics, but this task lies outside the scope of the current paper.

A common aspect across all of these works is the use of efficient and accurate functional approximation tools.
Regardless of the specific technique, the general concept is straightforward: 
given a finite set of input realizations and their corresponding model responses, known as the \textit{training set} (ML) 
or \textit{experimental design} (UQ), a suitable parametric function is calibrated such that it accurately approximates 
the underlying (possibly unknown) input-output map.
For the sake of consistency, and a little bias towards UQ, we will refer to this process as \textit{surrogate modeling}, 
acknowledging that it is also known as \textit{emulation}, \textit{metamodeling}, \textit{reduced order-} or 
\textit{response surface- modeling}, or sometimes simply \textit{regression}. 
A variety of methods is available in the surrogate modeling literature, which we cluster here in two classes:
\begin{itemize}
    \item \textit{Localized surrogates:} this includes interpolants (\textit{e.g.} Gaussian process modeling \cite{Santner2003}, spline interpolation \cite{reinsch1967Spline}, sparse grids \citep{bungartz_griebel_2004}), but also local regression methods (\textit{e.g.} Gaussian process regression \cite{Rasmussen2006}, multivariate moving averages \cite{lowry1992multivariate} or support vector machines \citep{Vapnik2013}). 
    These techniques rely on the availability of local information, e.g. through kernels on point-wise distance measures or support vectors, to provide predictions that are more accurate closer to the points in the training set. 
    They therefore tend to perform better in interpolation, rather than extrapolation, tasks.
    
    \item \textit{Global surrogates:} they provide global approximations without capitalizing 
    on locally available information. Examples in this class include spectral methods (\textit{e.g.} polynomial chaos expansions (PCE)
    \cite{Xiu2002,BlatmanJCP2011} \highlightdiff{and
    Pointcar\'e expansions \cite{Roustant2017}),} linear regression methods 
    (\textit{e.g.} 
    compressive sensing \cite{donoho2006compressed,Luethen2020sparse}, 
    generalized linear models \cite{Nelder1972GLM})\highlightdiff{, artificial 
    neural networks (ANNs) \cite{Goodfellow2016}}. 
    These techniques tend to achieve better global accuracy (\textit{e.g.} in terms of generalization error), thus offering some degree of 
    extrapolation capabilities, but also worse local accuracy than their localized counterparts.
\end{itemize}

Each of the two classes have advantages and disadvantages, but they both tend to perform well on models that show homogeneous complexity 
throughout the input parameter space. 
Some models of practical engineering relevance, however, can show a highly localized behavior in 
different regions of the parameter space. 
Common examples include likelihood functions used in Bayesian inference \cite{NagelJCP2016}, crash test simulations
\cite{Serna2009Weimar}, snap-through models \cite{Hrinda2010SnapThrough}, and discontinuous models in general.

Different approaches with varying degree of complexity have been proposed in the UQ and ML literature 
to address this kind of behavior. 
Examples include regression trees \citep{Chipman2010,breiman2017classification}, multivariate adaptive regression splines (MARS, \citep{Friedman1991}), various combinations of Kriging and PCE (PC-Kriging, \cite{SchoebiIJUQ2015,KersaudySudret2015}), multi-resolution/multi-element polynomial chaos expansions \cite{LeMaitre2004JCP,WanKarniadakis2006,FooKarniadakis2008JCP} and deep neural networks \cite{Goodfellow2016}, among others. 
Such methods can be broadly classified in two macro-families: \emph{global approximations with local refinements} (\textit{e.g.} PC-Kriging), or \emph{domain-decomposition-based methods} (regression trees, MARS, multi-element polynomial chaos expansions).
The class of global approximations with local refinements rely on efficiently combining global surrogates (\textit{e.g.} spectral decompositions as polynomial chaos expansions, or global regression models) with local interpolation techniques (\textit{e.g.} Gaussian processes or splines), to provide surrogates with acceptable extrapolation capabilities and good local accuracy.
The class of domain-decomposition-based methods relies instead on the idea of partitioning the input parameter space into (often disjoint) subdomains, followed by the use of regression-based surrogates in each subdomain. This \textit{divide-and-conquer} approach is particularly effective in reducing the complexity of the computational model in each subdomain, hence allowing relatively simple techniques to be used as local approximants. 
A prime example of this class of methods is given by regression trees \citep{Chipman2010}, where the local surrogates are as simple as constant values.

A common trait of most surrogate models used in a UQ context is that they rely on some form of regularity of the underlying computational model (\textit{e.g.} smoothness or symmetry) to achieve an efficient representation based on an experimental design of relatively small size. 
It is therefore not surprising that they often show limited scalability with both the number of input dimensions (the well known \textit{curse of dimensionality}) and with the size of the experimental design. 
Indeed, most local surrogates and interpolants rely on either kernel or clustering methods, neither of which scales linearly with the number of dimensions. Moreover, their training requires the solution of complex optimization problems that often have at least as many parameters as input dimension \cite{Rasmussen2006,Vapnik2013}.
Global regression methods, on the other hand, require the optimization and storage of a large number of parameters or coefficients, which also rarely scales linearly in high dimension for non-trivial models.

To step further into scalability considerations, the number of available samples in the experimental design deserves some discussion. 
Historically, UQ-based surrogate modeling has taken a \textit{parsimonious} approach: focus on small but informative experimental designs ($N_{ED} \approx 10^{1-2}$), because of the high computational costs associated to engineering models, and to their smooth behavior. 
On the other hand, ML has seen its expansion in the era of \textit{big data}, focusing on large experimental designs ($N_{ED}\approx 10^{5-7}$), with often noisy data and highly non-smooth behavior. 
Albeit the gap is closing over time, a \textit{no-man's land} in between the two still exists: computational models that show a complexity that is too high for classical surrogate modeling (\textit{e.g.} extremely non-linear, or highly localized), but are expensive enough to only allow for $N_{ED}\approx 10^{3-4}$, regardless of the input dimension.

It is with this class of problems in mind that we propose a new surrogate modeling technique, namely stochastic spectral embedding (SSE), that combines global spectral representations and adaptive domain decompositions. We demonstrate that SSE can efficiently approximate models with varying degrees of complexity across the input space, while maintaining favorable scaling properties with both the input dimension and the size of the experimental design.

The paper is organized as follows: we first describe the general rationale and the details of the algorithm in Section~\ref{sec:SSE main}. 
Then, in Section~\ref{sec:SSE regression} we tackle the issue of constructing an SSE from an experimental design, in a regression context.
In Section~\ref{sec:Applications}, we choose a reference spectral decomposition technique (polynomial chaos expansions, PCE) and we apply SSE to two highly complex analytical functions to showcase its capability to adapt to models with non-homogeneous complexity, and its scalability to high dimensions and large experimental designs. 
Finally, we also tackle two models of engineering complexity that are known to be challenging for classical surrogate modeling methods.
We present concluding remarks in Section~\ref{sec:Conclusions} and discuss extensions of the algorithm that could further improve its performance.

%% file: 02_Methodology.tex
%\section{Methodology}
%\label{sec:Methodology}

%%%%%%%%%%%%%%%%%%%%%%%%%%%%%%%%%%%%%%%%%%%%%%%%%%%%%%%%%%%%%%%%%%%%%%%%%%%%

% The approach bears many similarities with 
% \emph{regression trees} and \emph{recursive partitioning} \citep{Breiman1984} 
% but extends these methods by using spectral basis functions. Additionally the 
% similarities to multi-element generalized polynomial chaos (ME-gPC, 
% \citet{PCE:Wan2006}) have to be noted here. The difference to this method being 
% the hierarchical nature of SSE, where PCEs are embedded into each other and not 
% constructed in a segmented input space.

\section{Stochastic spectral embedding: rationale and main algorithm}
\label{sec:SSE main}

As the name suggests, stochastic spectral embedding (SSE) is a combination of two main ingredients: a \textit{stochastic spectral representation}-based surrogate model and some form of \textit{embedding}, which implies the sequential construction of subdomains of the full input space.
In other words, SSE consists in iteratively refining a spectral surrogate model by means of \textit{embedding} additional surrogate models in subdomains of the parent expansion.
In a sense, SSE can be seen as an extension of regression trees \cite{breiman2017classification,Chipman2010} to a much wider class of regression models, with the addition of a strong stochastic component due to the use of spectral representations.

\subsection{Spectral expansions}
\label{sec:spectral expansions}
We will consider herein the Hilbert space $\cH$ of random variables of the form $Y = \cM(\vX)$ with finite second moments ($\Esp{Y^2}<\infty$), where $\vX$ is an $M-$dimensional random vector with joint distribution $\vX \sim f_{\vX}(\vx)$. Let the space be equipped with the inner product:
\begin{equation}
    \label{eqn:innerProd}
    \innerprod{g(\vX)}{h(\vX)}_{\cH} \eqdef  \Esp{g(\vX)h(\vX)} = \int\limits_{\cD_{\vX}} g(\vx)h(\vx) f_{\vX}(\vx)\,\di\vx,
\end{equation}
where $\cD_{\vX} \subseteq \Rr^{M}$ is the support of $\vX$. Then, every $Y \in \cH$ admits a spectral representation $\cm_S$ of the form:
\begin{equation}
    \label{eqn:spectral}
    Y = \cm_S\left(\vX\right) \eqdef \sum\limits_{j = 1}^\infty a_j \Psi_j(\vX),
\end{equation}
where the $a_j \in \Rr$ are real coefficients, and the $\Psi_j$'s form a countably infinite orthonormal basis of the space:
\begin{equation}
    \label{eqn:orho}
    \Esp{\Psi_i(\vX)\Psi_j(\vX)} = \innerprod{\Psi_i(\vX)}{\Psi_j(\vX)}_{\cH} = \delta_{ij},
\end{equation}
where $\delta_{ij}$ is the Kronecker delta. For notational simplicity, the inner product subscript $\cH$ is omitted hereinafter.

Spectral decompositions of the form of Eq.~\eqref{eqn:spectral} have a property that is particularly important for surrogate modelling, namely the fact that due to the orthogonality of the basis in Eq.~\eqref{eqn:orho}, their (finite) second moment is given by:
\begin{equation}
    \label{eqn:spectral variance}
    \Esp{\cM(\vX)^2} = \innerprod{\cM_S(\vX)}{\cM_S(\vX)} = \sum\limits_{j=1}^\infty a_j^2 < +\infty .
\end{equation}
The converging sum in Eq.~\eqref{eqn:spectral variance} implies therefore that the coefficients $a_j$ must decay at least geometrically when sorted by decreasing absolute value.
This property is sometimes referred to as \textit{compressibility}, because it essentially means that most of the information on the model variability is contained in a finite set of coefficients/basis elements. 
This allows one to truncate the spectral decomposition in Eq.~\eqref{eqn:spectral} even if in principle it has an infinite number of terms. The truncated version of Eq.~\eqref{eqn:spectral} is given by:
\begin{equation}
    \label{eqn:spectral truncated}
    \cM_S(\vX) \approx \widehat{\cM}_S(\vX) =  \sum\limits_{j \in \cA} a_j \Psi_j(\vX),
\end{equation}
where $\cA$ is a truncation set (typically related to the complexity of the basis functions, \egr maximum frequency in Fourier expansions, or maximum polynomal degree in PCE). 
The rapid decay in the coefficients of spectral expansions is the main reason why many powerful surrogate modeling techniques that belong to the so-called class of \textit{compressive sensing} \citep{donoho2006compressed,Candes2008a}, have proven to be very effective in various recent applications \citep{BlatmanJCP2011,Torre2019PCE4ML,Luethen2020sparse}. 
Compressive sensing uses sparse regression tools to identify the best truncation set $\cA$ based on the available information in the experimental design. \\ 
Because of the truncation introduced in Eq.~\eqref{eqn:spectral truncated}, the expansion is in general not exact, hence we define the \textit{residual} $\cR(\vX)$ as:
\begin{equation}
    \label{eqn:residual}
    \cR(\vX) = \cM(\vX) - \widehat{\cM}_S(\vX).
\end{equation}
Due to the convergent behavior of the truncated expansion, it follows that $\Var{\cR(\vX)}\ll \Var{\widehat{\cM}_S(\vX)}$.
By definition, spectral expansions belong to the class of global representations, i.e. the basis functions in Eq.~\eqref{eqn:spectral truncated} have support on the entire domain $\cD_{\vX}$. 
Therefore, highly localized models, or those with  inhomogeneous behavior throughout the input domain tend to require an extremely large number of terms in the truncated expansion to achieve satisfactory approximation accuracy (Gibbs phenomenon). 
As an example, the number of terms in the well-established polynomial chaos expansion \citep{Xiu2002,BlatmanJCP2011} can grow very fast when the underlying model has strongly localized behavior, because a high polynomial degree is required for an accurate representation.

\subsection{A sequential partitioning approach}
\label{sec:sequential partitioning}
To alleviate this limitation, while still capitalizing on the powerful 
convergence properties of spectral methods, SSE constructs a sequence of 
spectral expansions of manageable complexity on increasingly smaller subdomains 
of the original domain. 
\highlightdiff{Such subdomains are denoted $\cD_{\vX}^{\ell,p} \subseteq 
\cD_{\vX}$, where 
$\ell$ is the expansion level and $p$ is a subdomain index within the 
level.} The expansion is performed only on the local residual from the 
previous level.

\begin{figure}[h]
	\centering
	\subfloat[$\ell=0$]{
		\includegraphics[width=0.33\linewidth,clip=true,trim=0 0 0 
		0]{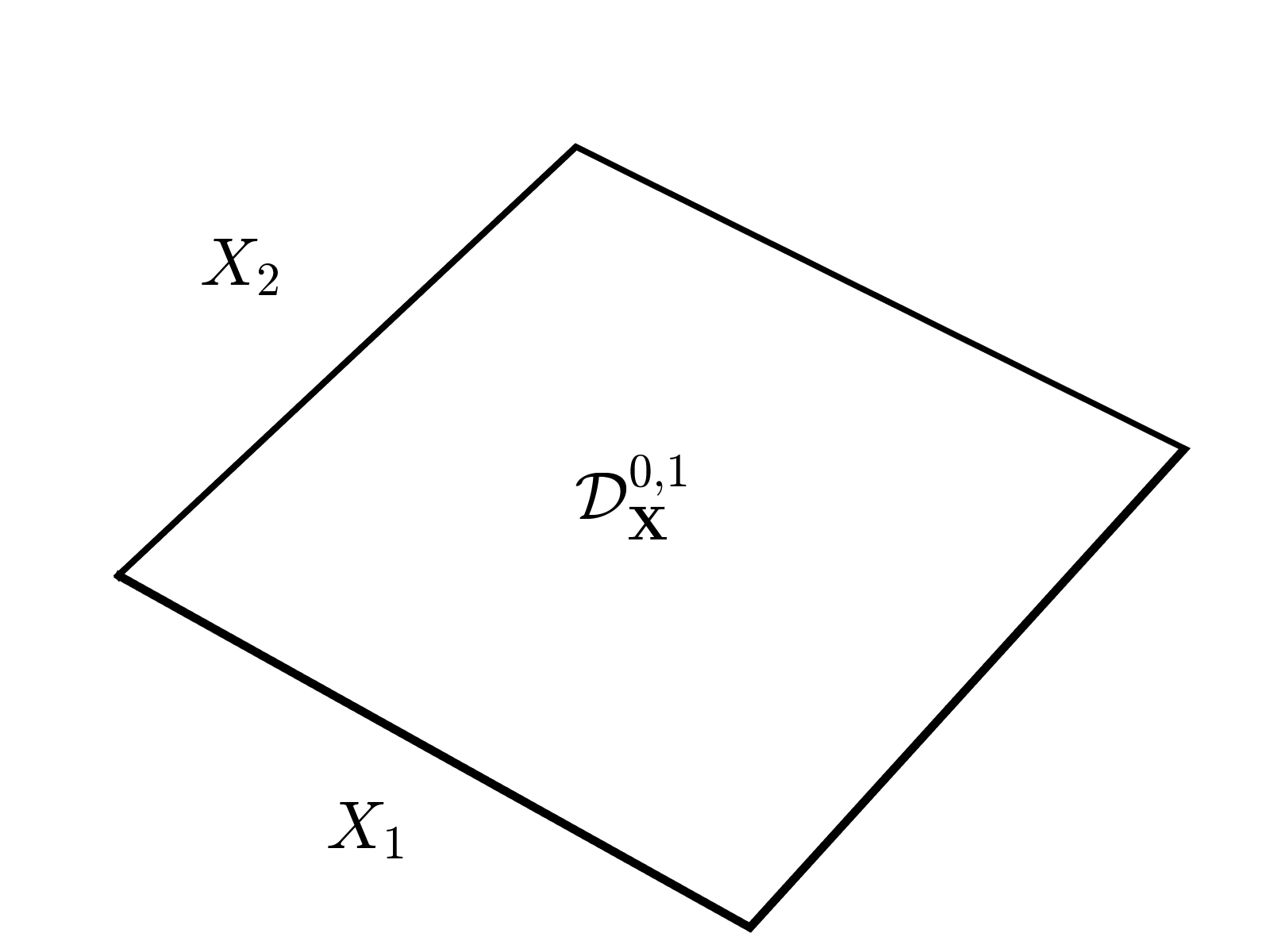}
	}%
	\subfloat[$\ell=1$]{
		\includegraphics[width=0.33\linewidth,clip=true,trim=0 0 0 
		0]{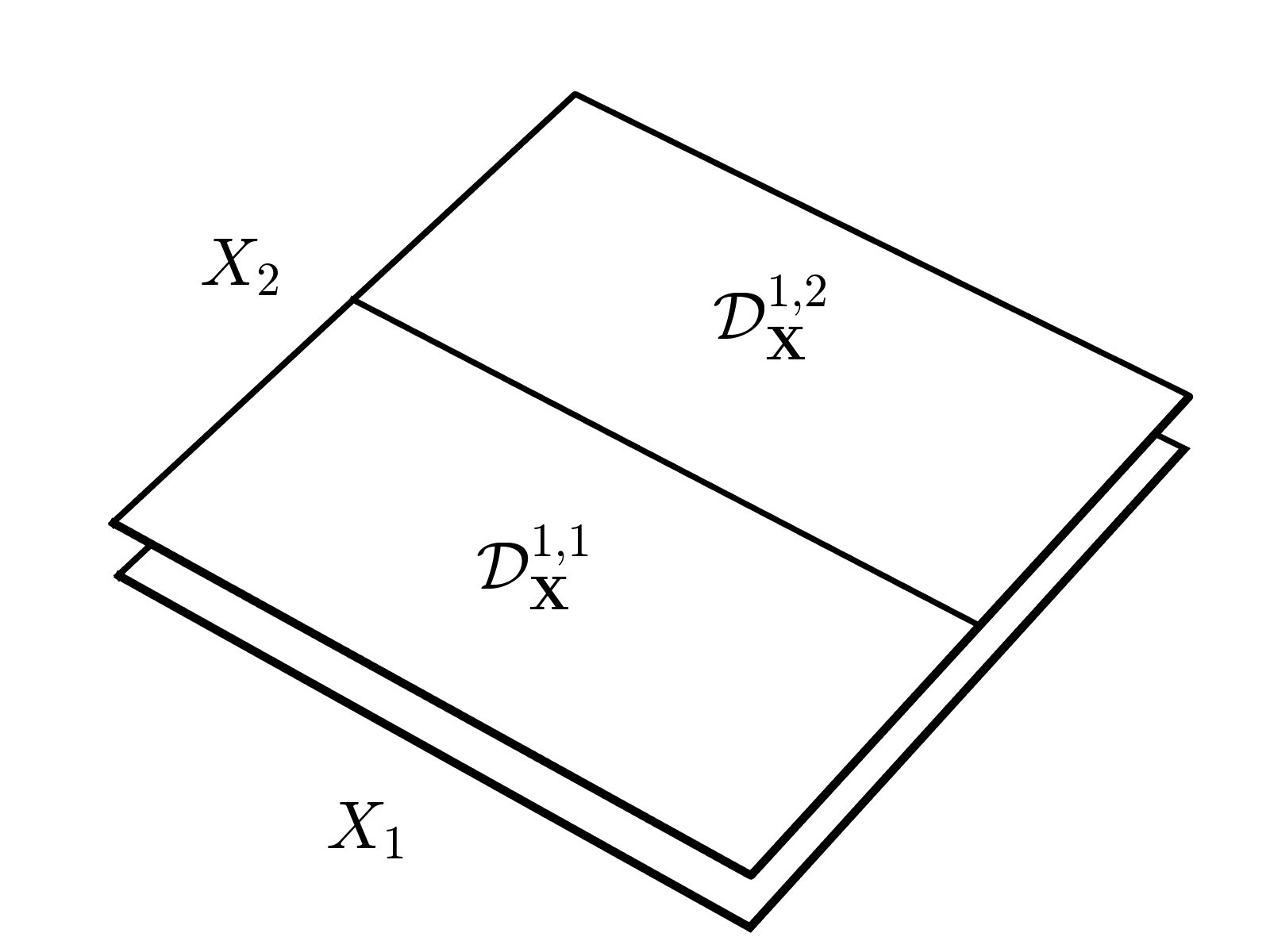}
	}%
	\subfloat[$\ell=2$]{
		\includegraphics[width=0.33\linewidth,clip=true,trim=0 0 0 
		0]{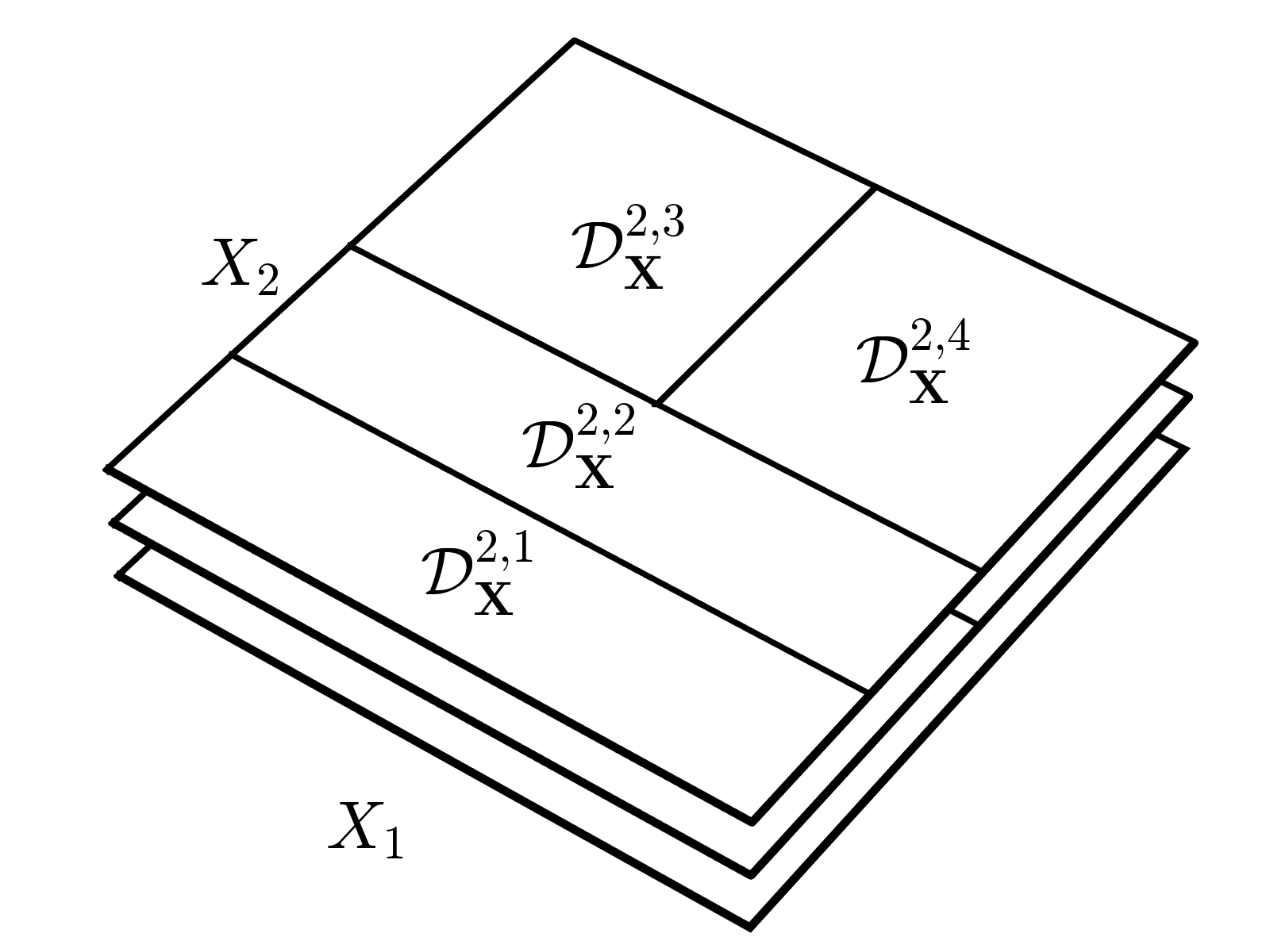}
	}%
	\caption{\highlightdiff{Example of a possible partitioning sequence for 
	an SSE with $L=2$ 
	and $P_\ell=2^{\ell}$, and bounded uniform marginal distributions. 
	In this simple example, each subdomain is split in two equal parts across a 
	random direction.}}
	\label{fig:SSEconstructionScheme}
\end{figure}

\highlightdiff{For illustration purposes, 
Figure~\ref{fig:SSEconstructionScheme} shows an 
example of sequential partitioning for a simple 2D bounded domain, obtained by 
splitting each subdomain in two equal subdomains across a random dimension.
When $\ell = 0$, there is only a single subdomain $\cD_{\vX}^{0,1} \eqdef 
\cD_{\vX}$ and the residual is $\widehat{\cR}_S^{0,1}(\vX) = 
\widehat{\cm}_S(\vX)$ from Eq.~\eqref{eqn:residual}.}

While the idea of partitioning the input space in smaller subdomains and constructing a surrogate model in each is certainly not new in UQ (see, e.g. \cite{LeMaitre2004JCP,WanKarniadakis2006}), the use of a sequence of residuals in SSE sets it apart from other divide and conquer methods.

In more formal terms, SSE in its basic form can be written as a multi-level expansion of the form:
\begin{equation}
    \label{eqn:SSE}
    \cm_\text{SSE}(\vX) = \sum\limits_{\ell = 0}^{L} \sum\limits_{p=1}^{P_\ell} \indfun{\cD_{\vX}^{\ell,p}}(\vX)\, \widehat{\cR}_S^{\ell,p}(\vX),
\end{equation}
where $L$ is the total number of expansion levels considered, $P_\ell$ is the 
number of subdomains at level $\ell \in \left\{1,\dots, L\right\}$, 
$\indfun{\cD_{\vX}^{\ell,p}}(\vX)$ is the 
indicator function of the subdomain $\cD_{\vX}^{\ell,p}$. Finally, 
$\widehat{\cR}_S^{\ell,p}(\ve{X})$ is the truncated expansion of the residual 
of the SSE up to level $\ell-1$, $\cR^{\ell}(\vX)$, %constructed in the 
%partition of the input support $\cD_{\vX}^{\ell,p}$:
\begin{equation}
    \label{eqn:SSE residual}
    \cR^{\ell}(\vX) = \cm(\vX) - \sum\limits_{k = 0}^{\ell-1}\sum\limits_{p=1}^{P_k} \indfun{\cD_{\vX}^{k,p}}(\vX) \widehat{\cR}_S^{k,p}(\vX).
\end{equation}

In the above equations, each residual term is expanded onto a local orthonormal basis as follows:
\begin{equation}
    \label{eqn:SSE residual expansion}
    \widehat{\cR}^{k,p}_S(\vX) = \sum\limits_{j \in \cA^{k,p}} a_j^{k,p} \Psi_j^{k,p}(\vX).
\end{equation}

A local inner product is defined in the domain $\cD_{\vX}^{k,p}$:
\begin{equation}
    \label{eqn:inner prod k,p}
    \innerprod{\Psi_i^{k,p}(\vX)}{\Psi_j^{k,p}(\vX)}_{k,p} = \int\limits_{\cD_{\vX}^{k,p}}\Psi_i^{k,p}(\vx)\Psi_j^{k,p}(\vx) f_{\vX}^{k,p}(\vx)\,\di\vx,
\end{equation}
where:
\begin{equation}
\label{eqn:partition pdf}
    f_{\vX}^{k,p}(\vx) = \indfun{\cD_{\vX}^{k,p}}(\vx)\frac{f_{\vX}(\vx)}{\cV^{k,p}}
\end{equation}
is the joint PDF of the input parameters restricted to the subdomain $\cD_{\vX}^{k,p}$ and rescaled by its probability mass $\cV^{k,p}$:
\begin{equation}
\label{eqn:pMass partition}
    \cV^{k,p} = \int\limits_{\cD_{\vX}^{k,p}} f_{\vX}(\vx)\,\di\vx.
\end{equation}

A crucial aspect of SSE is that by partitioning the entire input domain $\cD_{\vX}$ into smaller subdomains $\cD_{\vX}^{\ell,p}$, it trades the complexity of the single, often global expansion in Eq.~\eqref{eqn:spectral truncated} for a (possibly large) number of local expansions with much smaller truncation sets. 
In cases where the spectral basis is continuous in Eq.~\eqref{eqn:spectral truncated}, SSE results in a final piecewise continuous approximation, but no continuity is ensured on the boundaries of the subdomains.
Mean-square convergence of the procedure is guaranteed by the spectral convergence in each level, which implies that the residual local variance in each subdomain is in expectation decreasing rapidly. 
In other words, for each increasing level $\ell$ in Eq.~\eqref{eqn:SSE}, new discontinuity bounds are generated during the partitioning step, but the variance of the overall residual is reduced, thus resulting, in expectation, in lower amplitude discontinuities. 
This behavior is analogous to that of regression trees \cite{Friedman1991,breiman2017classification}.

\subsection{The SSE algorithm}
\label{sec:SSE algo}
Algorithmically, SSE consists of a local refinement sequence of a global spectral expansion into sequentially smaller subdomains $\cD_{\vX}^{\ell,p}$.
For notational simplicity, we introduce here a set of local random vectors distributed according to the local PDF in Eq.~\eqref{eqn:partition pdf}: $\vX^{\ell,p} \sim f_{\vX}^{\ell,p}(\vx)$. We further choose a certain \emph{partitioning strategy} that is discussed in Section~\ref{sec:SSE partitioning strategy}.

Then, the SSE algorithm can be written as:
\begin{enumerate}\itemsep0pt
    \item \textbf{Initialization:}%\vspace{-.4cm}
        \begin{enumerate}\itemsep0pt\small
            \item $\ell = 0$, $p = 1$
			\item $\cD_{\vX}^{\ell,p} = \cD_{\vX}$
			\item ${\cR}^{\ell}(\vX) = \cm(\vX)$
        \end{enumerate}
    \item \textbf{For each subdomain $\cD_{\vX}^{\ell,p}, p = 1,\cdots,P_\ell$:}%\vspace{-.4cm}
        \begin{enumerate}\itemsep0pt\small
            \item \label{algo:2a} Calculate the truncated expansion $\widehat{\cR}_S^{\ell,p}(\vX^{\ell,p})$ of the residual ${\cR}^{\ell}(\vX^{\ell,p})$ in the current subdomain
            \item \label{algo:2b} Update the residual in the current subdomain ${\cR}^{\ell+1}(\vX^{\ell,p}) = {\cR}^{\ell}(\vX^{\ell,p}) - \widehat{\cR}_S^{\ell,p}(\vX^{\ell,p})$
			%            \item Select a splitting direction $d^{\ell,p} \in \left\{1,\cdots, M\right\}$
			%            \item Split the current partition $\cD_{\vX}^{\ell,p}$ along $d^{\ell,p}$ in $N_S$ subdomains  $\cD_{\vX}^{\ell+1,\{1,\cdots,N_S\}}$ 
			\item \label{algo:2c} Split the current subdomain $\cD_{\vX}^{\ell,p}$ in $N_S$ subdomains  $\cD_{\vX}^{\ell+1,\{s_1,\cdots,s_{N_S}\}}$ based on a partitioning strategy
            \item If $\ell < L$, $\ell \leftarrow \ell + 1$, go back to  \ref{algo:2a}, otherwise terminate the algorithm
        \end{enumerate}%\vspace{-.4cm}
    \item \textbf{Termination}%\vspace{-.4cm}
    \begin{enumerate}\itemsep0pt\small
        \item Return the full sequence of $\cD_{\vX}^{\ell,p}$ and  $\widehat{\cR}_S^{\ell,p}(\vX^{\ell,p})$ needed to compute Eq.~\eqref{eqn:SSE}.
    \end{enumerate}
\end{enumerate}

Note that in steps \ref{algo:2a} and \ref{algo:2b} of the previous algorithm the residual ${\cR}^{\ell}(\vX^{\ell,p})$ is only indexed by $\ell$, but not by the subdomain index $p$. 
This is because the residual is fully defined with respect to the previous level $\ell$, which is independent on the particular subdomain under consideration (see Figure~\ref{fig:SSEconstructionScheme}).

% From a construction perspective, the SSE algorithm requires only a few ingredients: a spectral expansion $\widehat{\cR}_S^{\ell,p}(\vX^{\ell,p})$ of the residual in each partition (which includes a spectral basis and its coefficients as given in Eq.~\eqref{eqn:SSE residual expansion}), a splitting strategy for each partition at each level, and the maximum number of levels $L$.

%%
\section{Building a stochastic spectral embedding from data}
\label{sec:SSE regression}

For it be useful in practical applications, SSE needs to be  ``trainable'' from a finite-size experimental design.
Hereinafter, we consider an experimental design $\cX = \left\{ \vx^{(1)},\cdots,\vx^{(N)}\right\}$ and its corresponding model evaluations $\cY = \left\{y^{(1)},\cdots,y^{(N)}\right\}$ as the only data available for training.

Upon closer inspection of the algorithm in Section~\ref{sec:SSE algo}, the training phase of the SSE representation consists in estimating the following quantities from the available experimental design:
\begin{enumerate}\itemsep0pt
	\item The expansion coefficients of the local residual in each level and subdomain $\ve{a}^{\ell,p}$ (Eq.~\eqref{eqn:SSE residual expansion}).
	\item A partitioning strategy at each level.
	\item The total number of splitting levels, $L$.
\end{enumerate}
In the following sections we introduce a comprehensive adaptive strategy based on sparse linear regression to perform each of these steps from a given experimental design.

\subsection{Calculating the residual expansion coefficients}
\label{sec:SSE residual expansion}

For a specific subdomain $\cD^{\ell,p}_{\vX}$, a local spectral expansion of the residual $\cR_S^{\ell}$ needs to be constructed from the available experimental design.
We therefore define a so-called \textit{local experimental design} $\cX^{\ell,p} \subseteq \cX$, the subset of the original experimental design lying within the subdomain $\cD^{\ell,p}_{\vX}$:
\begin{equation}
\label{eqn:local exp design}
\cX^{\ell,p} \eqdef \left\{\vx^{(j)}, j=1,\cdots,N^{\ell,p},~ \text{such that}~ \vx^{(j)} \in \left( \cX \cap \cD_{\vX}^{\ell,p} \right)\right\}.
\end{equation}
A similar notation is used to identify the corresponding model responses, $\cY^{\ell,p}$.
Using the auxiliary local random vector $\vX^{\ell,p}$ introduced in the previous section, the residual expansion in Eq.~\eqref{eqn:SSE residual expansion} reads:
\begin{equation}
\label{eqn:local residual expansion}
    \widehat{\cR}^{\ell,p}_S(\vX^{\ell,p}) = \sum\limits_{j \in \cA^{\ell,p}} a_j^{\ell,p} \Psi_j^{\ell,p}(\vX^{\ell,p}).
\end{equation}
Given the local experimental design $\cX^{\ell,p}$ and a truncated local spectral basis $\Psi_j^{\ell,p},\, j\in\cA^{\ell,p}$, the task of identifying the coefficients $\ve{a}^{\ell,p} \eqdef \left\{a_j^{\ell,p}, j\in \cA^{\ell,p}\right\}$ can then be cast as a linear regression problem (see, \egr \cite{Berveiller2006}):
\begin{equation}
\label{eqn:local regression}
	\ve{a}^{\ell,p}\approx\widehat{\ve{a}}^{\ell,p} = \argmin_{\ve{a}} \sum\limits_{\vx^{(i)}\in \cX^{\ell,p} } \left(\cR^\ell(\vx^{(i)}) -  \sum\limits_{j \in \cA^{\ell,p}} a_j^{\ell,p} \Psi_j^{\ell,p}(\vx^{(i)}) \right)^2 .
\end{equation}
While in principle the regression problem in Eq.~\eqref{eqn:local regression} can be solved through ordinary least squares, recent literature on the topic of compressive sensing has amply demonstrated that sparse regression approaches can provide great benefits in terms of accuracy, especially for relatively small experimental designs \cite{donoho2006compressed,BlatmanJCP2011,Luethen2020sparse}. A review of the available techniques for this purpose lies outside the scope of this paper and is extensively explored for one popular class of spectral representations (polynomial chaos expansions) in \cite{Luethen2020sparse}.

\subsection{Partitioning strategy}
\label{sec:SSE partitioning strategy}

A second step necessary to construct SSE from data is to identify a proper partitioning strategy between levels. 
Any strategy for the partitioning of the input domain $\cD_{\vX}$ can be employed for Eq.~\eqref{eqn:SSE}, under the sole condition that at each level $\ell$:
\begin{equation}
    \bigcup\limits_{p=1}^{P_\ell} \cD_{\vX}^{\ell,p} = \cD_{\vX}.
\end{equation}

While a comprehensive study on different partitioning strategies would be interesting, for the sake of simplicity we adopt hereinafter a rather simple approach, similar in spirit to regression trees \cite{breiman2017classification}.
In other words, we split every subdomain in two parts of equal probability mass along one of the input directions, $d^{\ell,p}\in\{1,\cdots,M\}$.
Note that the direction in itself can be different for each subdomain, even on the same level.

\begin{figure}
	\centering
	\subfloat[$\ell=0$]{
	    \begin{minipage}{0.33\linewidth}
		\includegraphics[width=\linewidth,clip=true,trim=0 0 0 0]{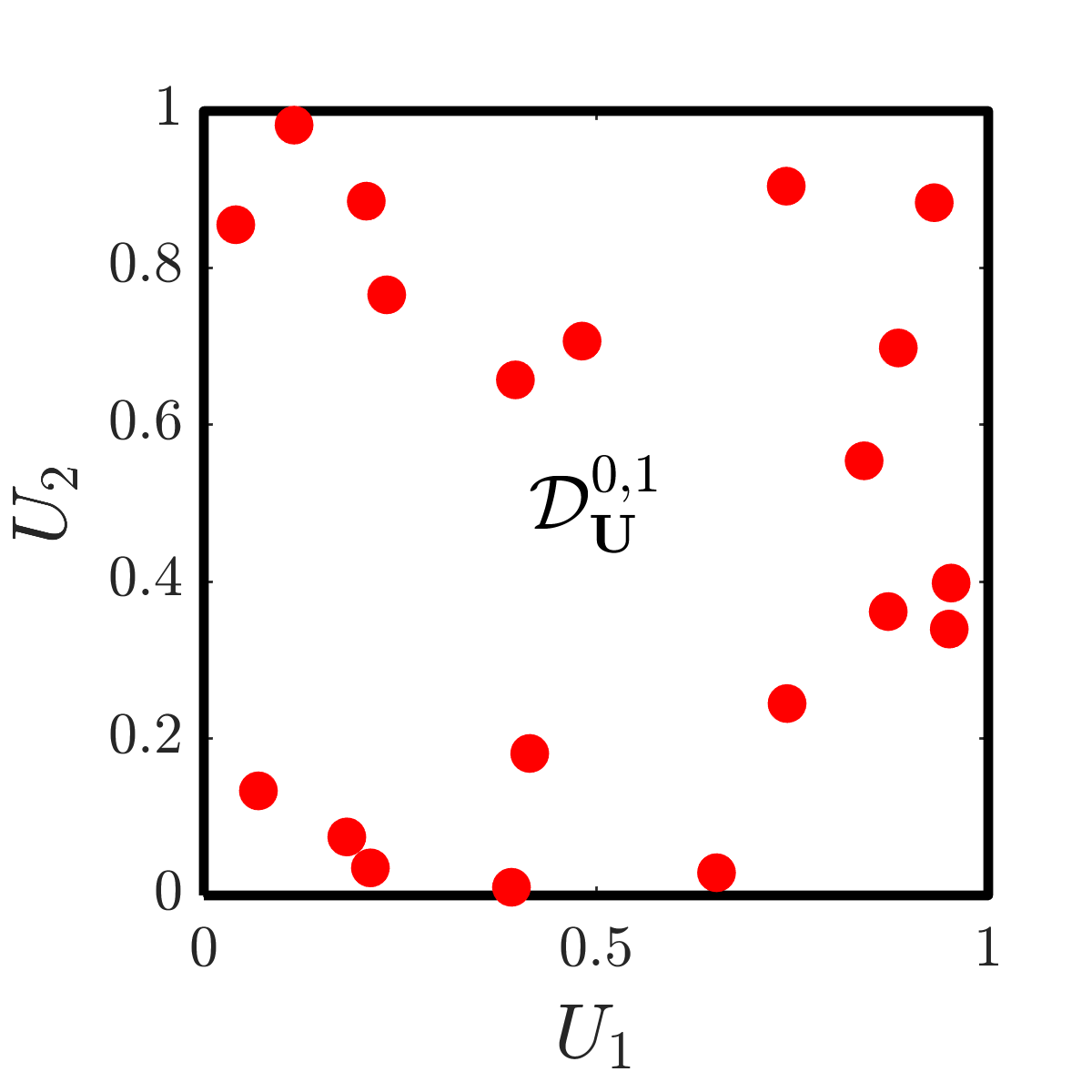}
		
		\includegraphics[width=\linewidth,clip=true,trim=0 0 0 0]{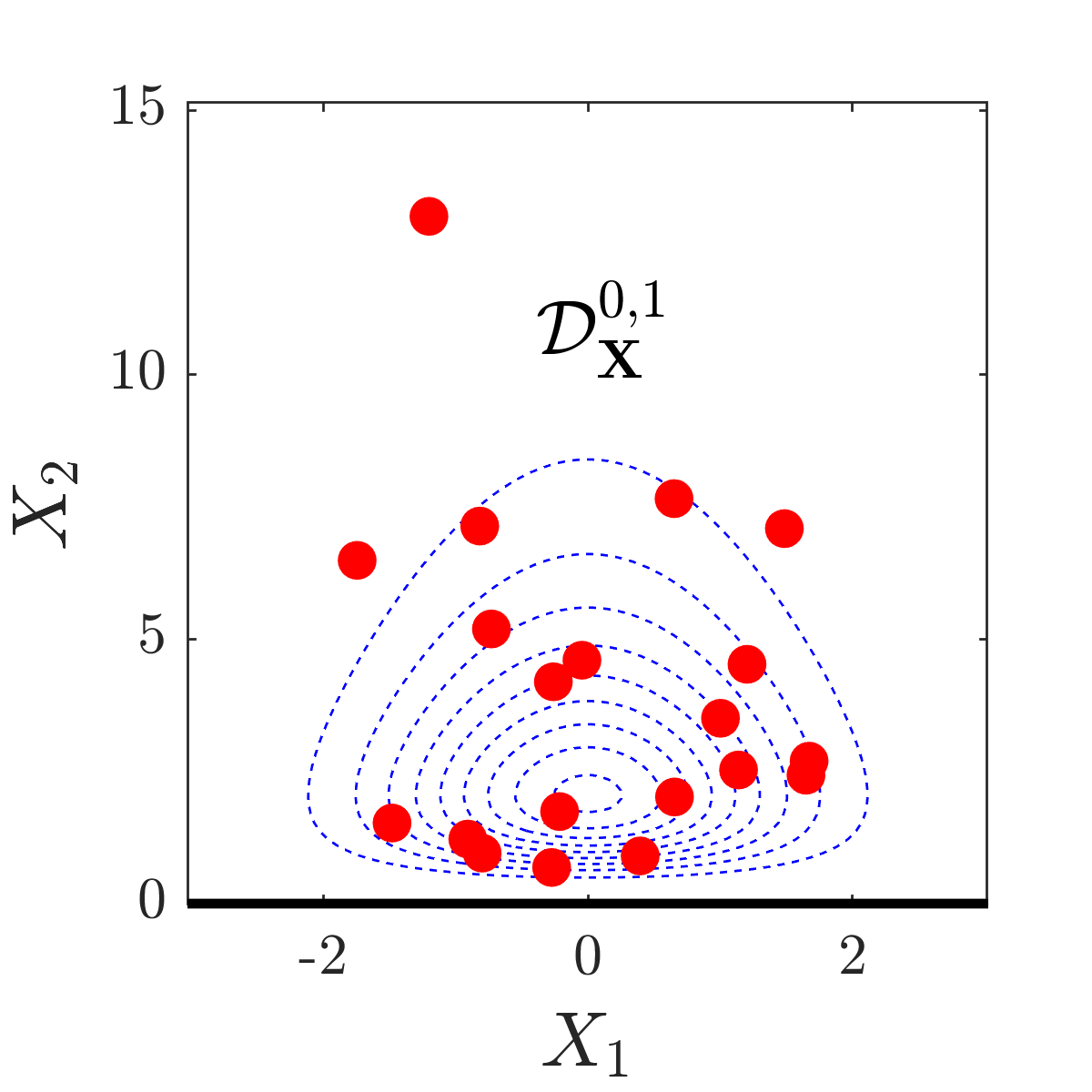}
		\end{minipage}
	}%
	\subfloat[$\ell=1$]{
	    \begin{minipage}{0.33\linewidth}
		\includegraphics[width=\linewidth,clip=true,trim=0 0 0 0]{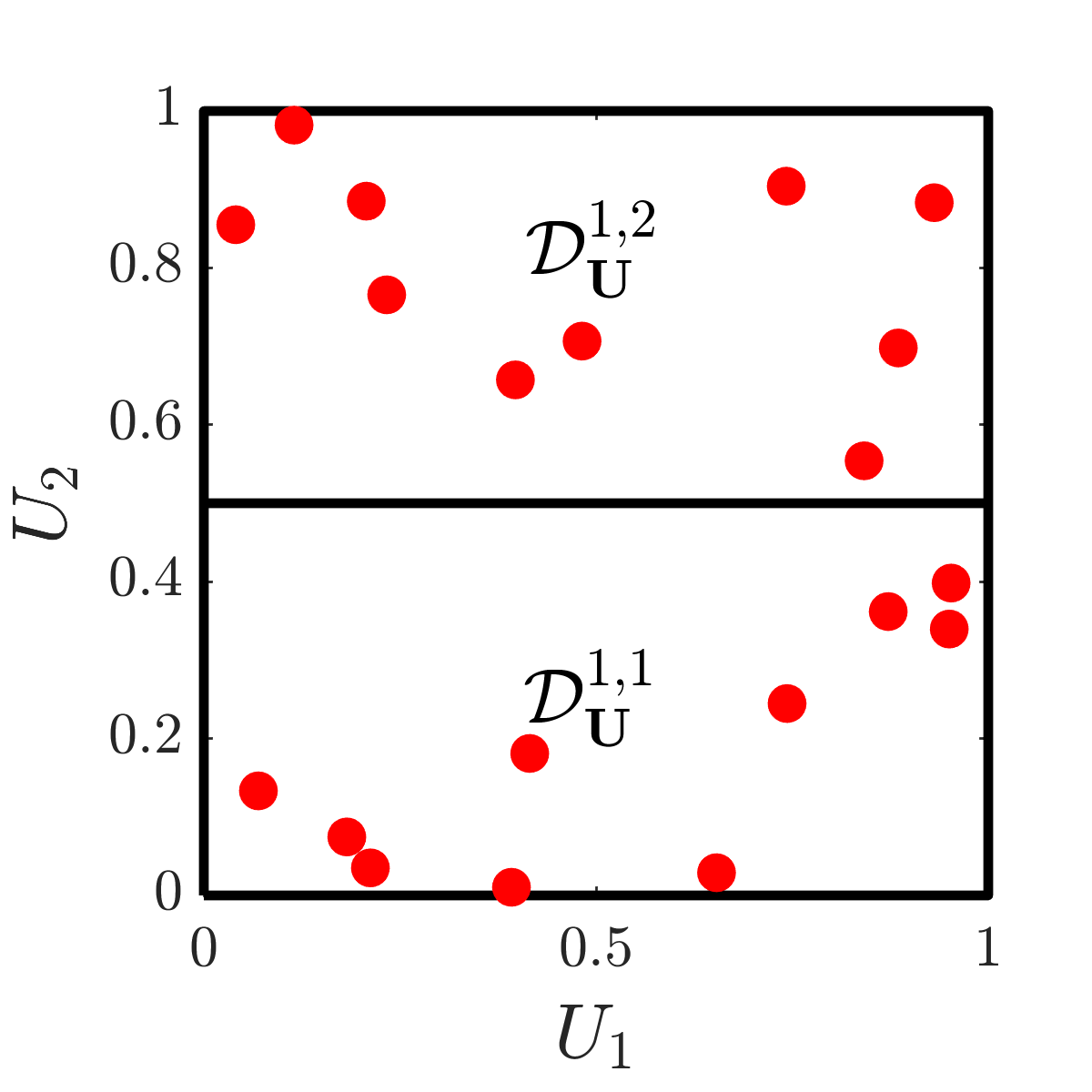}
		
		\includegraphics[width=\linewidth,clip=true,trim=0 0 0 0]{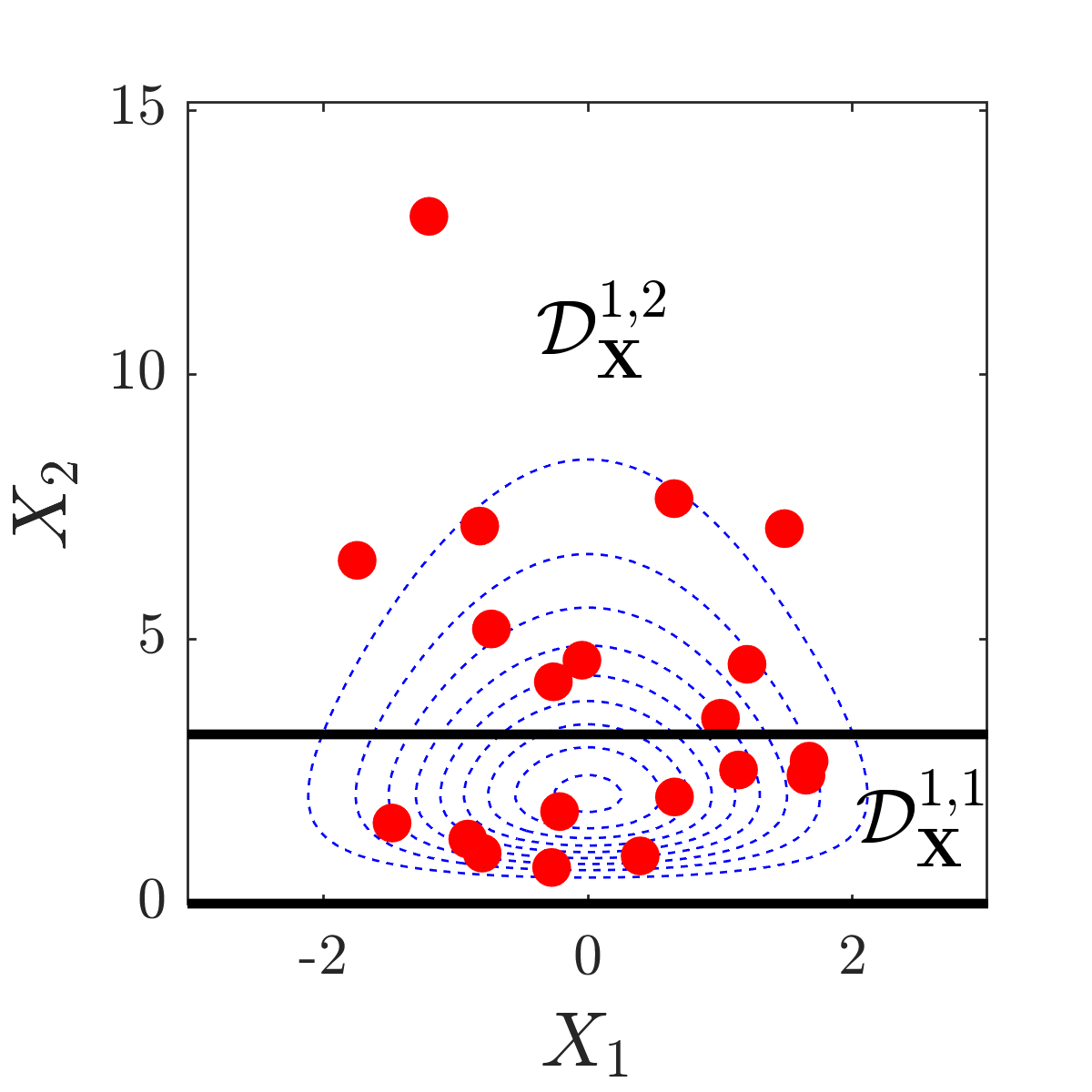}
		\end{minipage}
	}%
	\subfloat[$\ell=2$]{
	    \begin{minipage}{0.33\linewidth}
		\includegraphics[width=\linewidth,clip=true,trim=0 0 0 0]{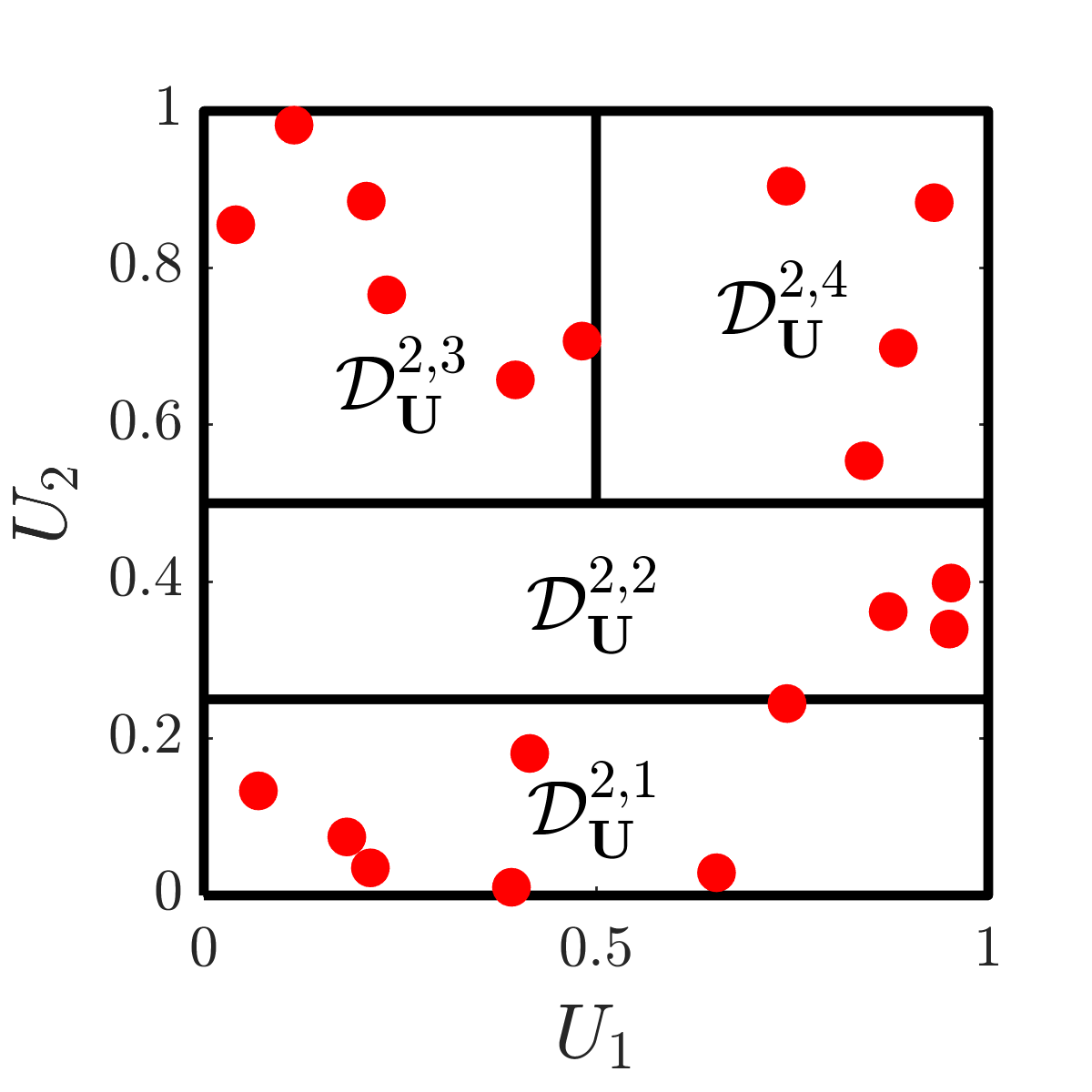}
		
		\includegraphics[width=\linewidth,clip=true,trim=0 0 0 0]{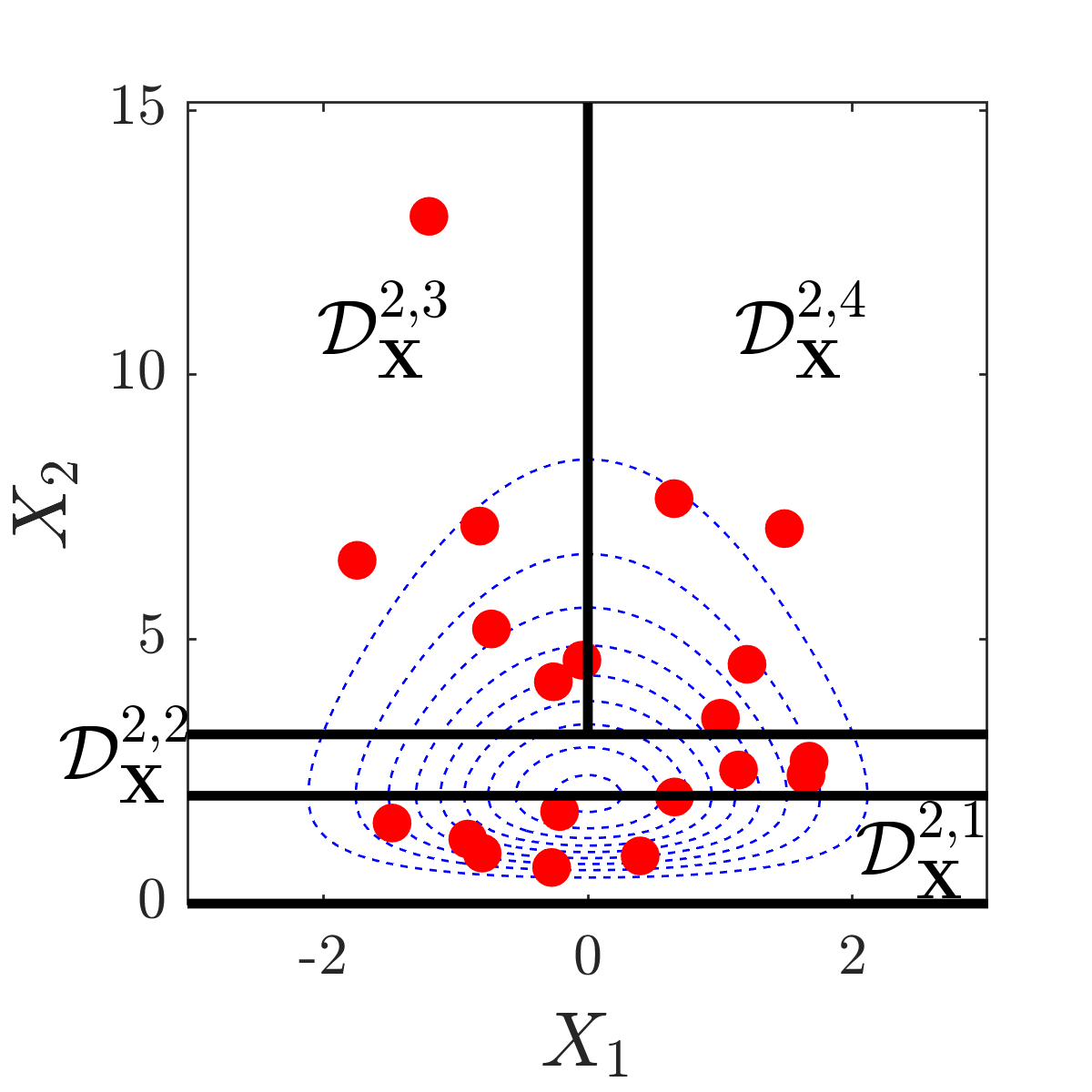}
		\end{minipage}
	}%
	\caption{Graphical representation of the partitioning strategy described in section \ref{sec:SSE partitioning strategy} for a two-dimensional problem with independent random variables. Upper row: partitioning in the quantile space; Lower row: partitioning in the original space. Red dots show a random sampling from the original distributions in both spaces, and serve as a visual aid to recognize the mapping between the two probability spaces from Eq.~\eqref{eqn:Rosenblatt}. The splitting direction in each subdomain is determined randomly in this example.}
\label{fig:SSE partitioning}
\end{figure}

Under very general conditions, it is possible to bijectively map any random vector $\vX$ with joint distribution $F_{\vX}$, to the uniform independent random vector $\vU \sim \cU(0,1)^M$ through an appropriate isoprobabilistic transform (\egr the Rosenblatt transform \cite{Rosenblatt1952,Torre2019PEM}):
\begin{equation}
\label{eqn:Rosenblatt}
    \begin{split}
    	\vU &= g(\vX)\\
    	\vX &= g^{-1}(\vU),
    \end{split}
\end{equation}
where $g(\cdot)$ denotes the isoprobabilistic transform.
This mapping simplifies the proposed partitioning strategy: splitting is performed in the uniformly distributed \textit{quantile space} $\vU$, and the resulting split domains $\cD_{\vU }^{\ell,p}$ are mapped back to the input space $\vX$ via the inverse transform (see Eq.~\eqref{eqn:Rosenblatt}). This has several computational benefits, including proper treatment of unbounded variables. Figure~\ref{fig:SSE partitioning} shows graphically a two-dimensional example of partitioning in the quantile (uniform) space $\vU$, and its corresponding mapping to unbounded random variables in the physical space $\vX$.

In the general case, a strategy is needed to choose a specific splitting direction $d^{\ell,p} \in \left\{1,\,\cdots,\, M \right\}$ for each existing subdomain $\cD_{\vX}^{\ell,p}$. Different heuristic reasoning can be used to make this choice, including purely random splitting (as in Figure~\ref{fig:SSE partitioning}), using the direction of maximum residual difference, or estimates of the variability of the $\widehat{\cR}_S^{\ell,p}$ in each direction (following the same rationale as in \cite{Shields2018VPaper}). The optimal criterion can be application-specific, because it may in principle depend on the chosen spectral representation.

\subsection{Sparse tree representation and expansion truncation}
\label{sec:SSE max L}
In a regression context, it is difficult to choose \textit{a priori} a truncation on the maximum number of levels $L$ in the expansion in Eq.~\eqref{eqn:SSE}. Because the samples may be unequally distributed, some subdomains at each level may be empty, or more formally $\cX^{\ell,p} = \emptyset $ for some combinations of $\ell$ and $p$. 
We therefore take a straightforward approach to obviate this issue, by initializing the residual at every level and subdomain to the null function, hence $\ve{a}^{\ell,p} = \ve{0}$. The coefficients are then updated only in the subdomains that satisfy $\left|\cX^{\ell,p}\right|\geq N_\text{min}$, where $N_\text{min}$ is a parameter of the SSE algorithm that represents the minimum number of points in a subdomain required to justify an expansion.
The SSE expansion is truncated when no new updates are possible given the current experimental design $\cX$, or more formally:
\begin{equation}
\label{eqn:SSE stopping L}
	L = \min\left\{\ell\, ~:~ \left|\cX^{\ell,p}\right| < N_\text{min}, \, \forall\, p \in \left\{1, \cdots, P_\ell\right\}  \right\} -1 .
\end{equation}

In addition to providing a suitable stopping criterion for the algorithm in Section~\ref{sec:SSE algo}, an added benefit of this strategy is that only the residual expansions that were effectively updated need to be stored in memory. This provides a degree of sparsity in the representation and potentially significantly reduces the memory fingerprint of the method, especially in the case of a large number of points in the experimental design. 

Note that, for an experimental design of size $N$ and a minimum number of points per expansion $N_{\rm min}$, the following holds:
\begin{equation}
2^{\bar{L}} N_{\rm min} \leq N,
\end{equation}
where $\bar{L}$ is the expected value of the maximum $L$ in $\eqref{eqn:SSE stopping L}$ and therefore
\begin{equation}
\label{eqn:SSE exp L}
\bar{L} \leq \left\lfloor\log_2\frac{N}{N_{\rm min}}\right\rfloor.
\end{equation}

%%%%%%%%%%%%%%%%%%%%%%%%%%%%%%%%%%%%%%%%%%%%%%%%%%%%%%%%%%%%%%%%%%%%%%%%%%%%
\subsection{Error estimation}
\label{sec:SSE error estimation}
% Why error measure

In the context of surrogate modeling, assessing the accuracy of the approximation is 
an important task.
Arguably the best known accuracy estimator in function approximation is the so-called
generalization error $E_{\mathrm{GEN}}$, which for SSE is given by
\begin{align}
E_{\mathrm{GEN}} &\eqdef 
\Esp{(\cm(\BParams)-\cm_{\text{SSE}}(\BParams))^2}.
\label{eq:SSE:basics:error:generror}
\end{align}
%
% Approximation
A direct estimation of this quantity is in general impossible, as it would require the availability of an extensive validation set.  
Instead, because we adopt a regression approach to calibrate the spectral decompositions in each subdomain, we estimate the generalization error through leave-one-out cross-validation \citep{Chapelle2002,Blatman2010Adaptive} that is available for each of the 
local expansions.

For notational convenience, we introduce here the set of \emph{terminal domains} $\cD_\cT =
\left\{\cD_{\vX}^{L,1},\cdots,\cD_{\vX}^{L,P_L}\right\}$, \textit{i.e.} those domains that belong to the last expansion level $L$ in Eq.~\eqref{eqn:SSE}.
By definition $\cD_\cT$ is a complete partition of the input domain $\cD_{\vX}$.
Because of the sequential nature of SSE, which locally refines the previous approximation level with the expansion of the residual in the current subdomain, an accurate estimate of the local generalization error in each of the terminal domains would then suffice to provide an estimate of the overall $E_{\mathrm{GEN}}$ of the full SSE.
If we denote the local residual error: 
\begin{equation}
\label{eqn:SSE local error}
	E_{\mathrm{GEN}}^{\ell,p} = \Esp{\left(\cR^{\ell}(\vX^{\ell,p}) - \widehat{\cR}^{\ell,p}_S(\vX^{\ell,p})\right)^2},
\end{equation}
then the global generalization error is simply given by the average error in each terminal domain:
\begin{equation}
\label{eq:SSE generror terminal domains}
E_{\mathrm{GEN}} = \Esp{E_{\mathrm{GEN}}^{L,p}} = \sum\limits_{p=1}^{P_L} E_{\mathrm{GEN}}^{L,p}\cdot \cV^{L,p},
\end{equation}
which is equal to the sum of the terminal domain errors weighted by the corresponding probability mass in Eq.~\eqref{eqn:pMass partition}.

To provide an estimator based on the available experimental design $\cX$, we only need an estimator of $E_{\mathrm{GEN}}^{L,p}$. 
Arguably the most common tool for the estimation of generalization error in regression problems is $k$-fold cross-validation (see, \textit{e.g.}, \cite{Vapnik2013}). 
The special case of $k = N$ is also known as \textit{leave-one-out} error and marked $E_\text{LOO}$.  For ordinary least square  regression it can be calculated analytically from the expansion coefficients and the basis functions \citep{Chapelle2002,Blatman2010Adaptive}. 

Given the sparse tree representation described in Section~\ref{sec:SSE max L}, it cannot be guaranteed that the residual $\cR^{\ell}$ is expanded in every terminal domain. Therefore, during the splitting phase of the SSE algorithm (Step~\ref{algo:2c} of the algorithm in Section~\ref{sec:SSE algo}) we initialize the error of all the subdomains $\cD^{\ell+1,\left\{s_1,\cdots,s_{N_S}\right\}}$ of the current subdomain $\cD^{\ell,p}$ to the leave-one-out error of the latter $E_{\rm LOO}^{\ell,p}$: 
\begin{equation}
E_{\rm LOO}^{\ell+1,\left\{s_1,\cdots,s_{N_S}\right\}} = E_{\rm LOO}^{\ell,P}. 
\end{equation}

Then, we update the error estimate in each subdomain during Step~\ref{algo:2a} only if the conditions for its expansion hold (see Section~\ref{sec:SSE max L}).
As a result, every terminal domain is either assigned its own leave-one-out error if it contains a residual expansion, or inherits the leave-one-out error from the last ancestor domain that was expanded.

By using the leave-one-out error in each terminal domain as an estimator of its generalization error ${\widehat{E}_{\rm GEN}^{L,p} =  \widehat{E}_{\rm LOO}^{L,p}}$, the empirical estimator of Eq.~\eqref{eq:SSE generror terminal domains} reads:
\begin{equation}
\label{eqn:SSE generror estimator}
\widehat{E}_{\mathrm{GEN}} =  \sum\limits_{p=1}^{P_L} \widehat{E}_{\rm LOO}^{L,p}\cdot \cV^{L,p}.	
\end{equation}
In most metamodeling applications, it is customary to normalize the estimated error by the variance of the experimental design, to obtain a dimensionless error measure. The relative error is thus defined as:
\begin{equation}
\label{eqn:SSE epsilonLOO}
\widehat{\epsilon}_{\mathrm{GEN}} = \frac{1}{\Var{\cY}} \sum\limits_{p=1}^{P_L} \widehat{E}_{\rm LOO}^{L,p}\cdot \cV^{L,p}.	
\end{equation}

%% file: 03_Applications.tex
\section{Applications}
\label{sec:Applications}
In this section we aim at showing the performance of SSE on a set of applications that can prove challenging for standard metamodeling techniques.
Because of its widespread use in the uncertainty quantification of engineering models, we choose as a spectral decomposition technique polynomial chaos expansions \cite{Xiu2002,BlatmanJCP2011} (hereinafter PCE).
This choice is also quite convenient due to several specific properties of PCE, that combine well with SSE.

\subsection{Synergies with polynomial chaos expansions}
\label{sec:synergies with pce}
By using the same notation as in Section~\ref{sec:SSE main}, and assuming that $\vX$ has independent components, the truncated polynomial chaos expansion of a finite variance model can be written as \cite{LeGratiet2016}:
\begin{equation}
\label{eqn:PCE}
	\cM_{\mathrm{PCE}}(\vX) = \sum\limits_{\ve{\alpha}\in \cA} a_{\ve{\alpha}} \Psi_{\ve{\alpha}}(\vX),
\end{equation}
where $\alpha$ is a multi-index that identifies the polynomial degree in each variable, $\cA$ is a suitable truncation set (\textit{e.g.} $\cA=\cA^{M,d}$ containing all multivariate polynomials with degree $\le d$), and the $\Psi_{\ve{\alpha}}(\vX)$ form an orthogonal basis of multivariate polynomials. 
The latter can be obtained via tensor product of univariate polynomials as follows:
\begin{equation}
\label{eqn:Psi}
	\Psi_{\ve{\alpha}}(\vX) = \prod\limits_{i = 1}^{M} \Phi_{\alpha_i}^{(i)}(\vX_i),
\end{equation}
where $\Phi^{(i)}_{\alpha_i}$ is a polynomial of degree $\alpha_i$ that belongs to the family of univariate polynomials orthogonal with respect to the input PDF of $X_i \sim f_{X_i}(x_i)$ and the inner product in Eq.~\eqref{eqn:innerProd}.

An interesting property of the univariate polynomials that synergizes well with our proposed SSE, is that it is possible to construct polynomials orthogonal to almost any input PDF through Gram-Schmidt orthogonalization (for an extensive review, see \cite{Gautschi2004,Ernst2012}). 
In the context of SSE, this property has a powerful implication: in each subdomain $\cD^{\ell,p}$ the basis elements $\Psi^{\ell,p}_{\ve{\alpha}}(\vX^{\ell,p})$ in Eq.~\eqref{eqn:SSE residual expansion} are still polynomial functions of the original input variables $\vX$.

This property, together with the analytical integrability of polynomials, allows us to derive several statistics of interest of PCE-based SSE analytically. 
Let us first introduce the notion of \textit{flattened representation}: because SSE is a polynomial in the original variables in every level and subdomain, this also holds for the terminal domains introduced in Section~\ref{sec:SSE error estimation}. 
Therefore, one can project the full SSE in Eq.~\eqref{eqn:SSE} as a local PCE onto each terminal domain:
\begin{equation}
\label{eqn:SSE flattened}
	\cm^F_{\rm SSE}(\vX^{L,p}) = \sum\limits_{\ve{\alpha}\in\cA^{\cT}} c^{p}_{\ve{\alpha}} \Psi_{\ve{\alpha}}^{L,p}(\vX^{L,p}),
\end{equation}
where $\cA^{\cT}$ is a suitable truncation set for the projection to be exact, and the $c^p_{\ve{\alpha}}$ are the corresponding coefficients. 
The latter can easily be computed either analytically or exactly through quadrature.
Note that, while the basis elements in the PCE in Eq.~\eqref{eqn:SSE flattened} correspond to the $\Psi^{k,P}_j$ in Eq.~\eqref{eqn:SSE residual expansion} (they only depend on the input PDF in Eq.~\eqref{eqn:partition pdf}), in general the coefficients will not be the same, \textit{i.e.} $c^p_{\alpha} \neq a^{L,p}_ {\alpha}$ in Eq.~\eqref{eqn:SSE residual expansion}.

Because PCE contains as a basis element the constant term, it is straightforward to demonstrate that the expected value of Eq.~\eqref{eqn:SSE} reads:
\begin{equation}
\label{eqn:SSE expectation}
\Esp{\cM_{\rm SSE}(\vX)} = \sum\limits_{p=1}^{P_L} c_{\ve{0}}^{p}\,\cV^{L,p}, 
\end{equation}
which is the weighted mean of all the mean values of the flattened representation in Eq.~\eqref{eqn:SSE flattened}.

Similarly, the variance can be calculated as:
\begin{equation}
\label{eqn:SSE variance}
\Var{\cM_{\rm SSE}(\vX)}
= 
\left(\sum_{p=1}^{P_L}\cV^{L,p}\sum_{\ve{\alpha}\in\cA^{\cT}}	
\left(c_{\ve{\alpha}}^{p}\right)^2\right) - 
\Esp{\cM_{\rm SSE}(\vX)}^2.
\end{equation}

A number of other quantities of engineering interest (\textit{e.g.} conditional variances, Sobol' sensitivity indices, etc.) can be derived similarly from the flattened representation. A selection of those is reported in \ref{app:postProcessingSSE}.

From a technical perspective, the flattened representation in Eq.~\eqref{eqn:SSE flattened} contains all the information needed to evaluate Eq.~\eqref{eqn:SSE} on a new point, but at a much lower storage cost, as only the final sets of coefficients $\ve{c}_{\ve{\alpha}}^p$ and basis indices $\cA^{\cT}$ need to be stored. 
This has additional advantages during the prediction of the response on new points, because it only requires the prediction of a single local expansion in the appropriate terminal domains, rather than that of all of its ancestors as in the original formulation in Eq.~\eqref{eqn:SSE}. More formally, for a point $\vx_0\in\cD_{\vX}$ it is sufficient to find $p_0\in 1,\cdots,P_L$ for which $\vx_0\in\cD_{\vX}^{L,p_0}$ and evaluate the flattened SSE from Eq.~\eqref{eqn:SSE flattened} for $p=p_0$.

\subsection{Example applications and testing strategy}
% overview
To compare the performance of SSE over sparse PCE, we choose four reference problems of increasing complexity: (i) a one-dimensional analytical function with localized non-polynomial behavior, (ii) a $100$-dimensional analytical function with decreasing parametric importance in higher dimensions, (iii) an $8$-dimensional engineering model describing the performance function of a damped oscillator and (iv) a three-dimensional discontinuous engineering model describing the snap-trough behavior of a truss structure.

% SSE settings
%For implementation reasons, the SSE construction sequence we use in the applications does not follow strictly the order presented in Section~\ref{sec:SSE algo}: the levels $\ell$ and subdomains $p$ are not investigated blindly until the convergence criterion is satisfied. 
%The partitioning order prioritizes the subdomains with maximum local error measure $\widehat{E}_{\rm LOO}^{L,p}\cdot \cV^{L,p}$, which are therefore expanded first. 
%This does, however, not affect the final SSE, as the convergence criterion (minimum number of experimental design points in each subdomain) is independent of the order they are considered.

% Splitting strategy
Among the ingredients identified in Section~\ref{sec:SSE algo} is a partitioning strategy, to choose the splitting direction in every subdomain. 
After extensive testing, we found that splitting according to the direction of highest variability of $\widehat{\cR}_S^{\ell,p}$ proved to be the most effective, especially for smaller experimental designs. 
We therefore split each subdomain $\cD_{\vX}^{\ell,p}$ into two subdomains with equal probability mass, {\em i.e.} $N_S = 2$, along the direction that has the maximum first order Sobol' index\cite{Sobol1993}, as analytically derived from the coefficients of $\widehat{\cR}_S^{\ell,p}$ \citep{SudretRESS2008b}.

% Other SSE parameters
In all applications we compare the convergence behavior of SSE vs. its spectral counterpart PCE as a function of the experimental design (ED) sizes $\Ntot$. 
To assess the robustness of the results, we consider $10$ independent replications of each ED, and provide the results in Tukey box-plots. 
For each experimental design size, SSE construction is terminated for $N_{\rm min} = \min\{5M,50\}$ (see Section~\ref{sec:SSE max L}). 

% Spectral expansion settings
As a spectral technique, we adopt the adaptive sparse-PCE based on LARS approach developed in \cite{BlatmanJCP2011} in its numerical implementation in UQLab \citep{MarelliUQLab2014,UQdoc_13_104}.
Each $\widehat{\cR}_S^{\ell,p}$ is therefore a degree- and $q$-norm-adaptive polynomial chaos expansion. 
We further introduce a rank truncation of $r=2$ to cope with high dimensional problems, \textit{i.e.} we limit the maximum number of input interactions \citep{UQdoc_13_104} to 2 variables at a time. The truncation set for each spectral expansion (Eq.~\eqref{eqn:PCE}) thus reads:
\begin{equation}
    \cA^{M,p,q,r} = \{\ve{\alpha} \in \mathbb{N}^{M} : \vert\vert\ve{\alpha}\vert\vert_q \le p, \vert\vert\ve{\alpha}\vert\vert_0 \le r\},
\end{equation}
where
\begin{equation}
    \vert\vert\ve{\alpha}\vert\vert_q = \left(\sum_{i=1}^M \alpha_i^q\right)^{\frac{1}{q}}, q\in (0,1]; \quad \vert\vert\ve{\alpha}\vert\vert_0 = \sum_{i=1}^M 1_{\{\alpha_i>0\}}.
\end{equation}

The $q$-norm is adaptively increased between $q = \{0.5,\cdots,0.8\}$ while the maximum polynomial degree is adaptively increased in the interval $p = \{0,1,\cdots,\maxDegSSE\}$, where the maximum degree $\maxDegSSE$ is a parameter for each case study.

% PCE settings
In all examples the SSE performance is compared to standard polynomial chaos expansions on the same ED.  These PCEs are constructed with the same adaptive approach used for the SSE expansions. Their maximum degree, however, is denoted by $\maxDegPCE$. 
It is set to the highest value our computational budget admitted for a given dimensionality but at least to $\maxDegPCE > 2\cdot \maxDegSSE$. 

% relative SSE error measure
We compare the performance of SSE and PCE in terms of the relative mean squared error (MSE) $\eta$, a well known error estimator defined as
\begin{equation}
\label{eq:Ex:MSE}
\eta \eqdef \frac{\Esp{\left(\cM (X) - \tilde{\cM}(X)\right)^2}}{\Var{\cM (X)}}
\end{equation}
where $\tilde{\cM}$ is either the PCE or SSE surrogate model. This error measure was estimated with standard Monte Carlo simulations using a  large sample of size $N=10^6$.

\subsection{Application 1: one-dimensional analytical function}
\label{sec:Applications:ex1}
\begin{figure}
	\centering
	\subfloat[$\ell=0$]{
		\begin{minipage}{0.48\linewidth}
			\label{fig:Ex1:detailedSteps:1}
			\includegraphics[width=\linewidth,clip=true,trim=0 10 0 0]{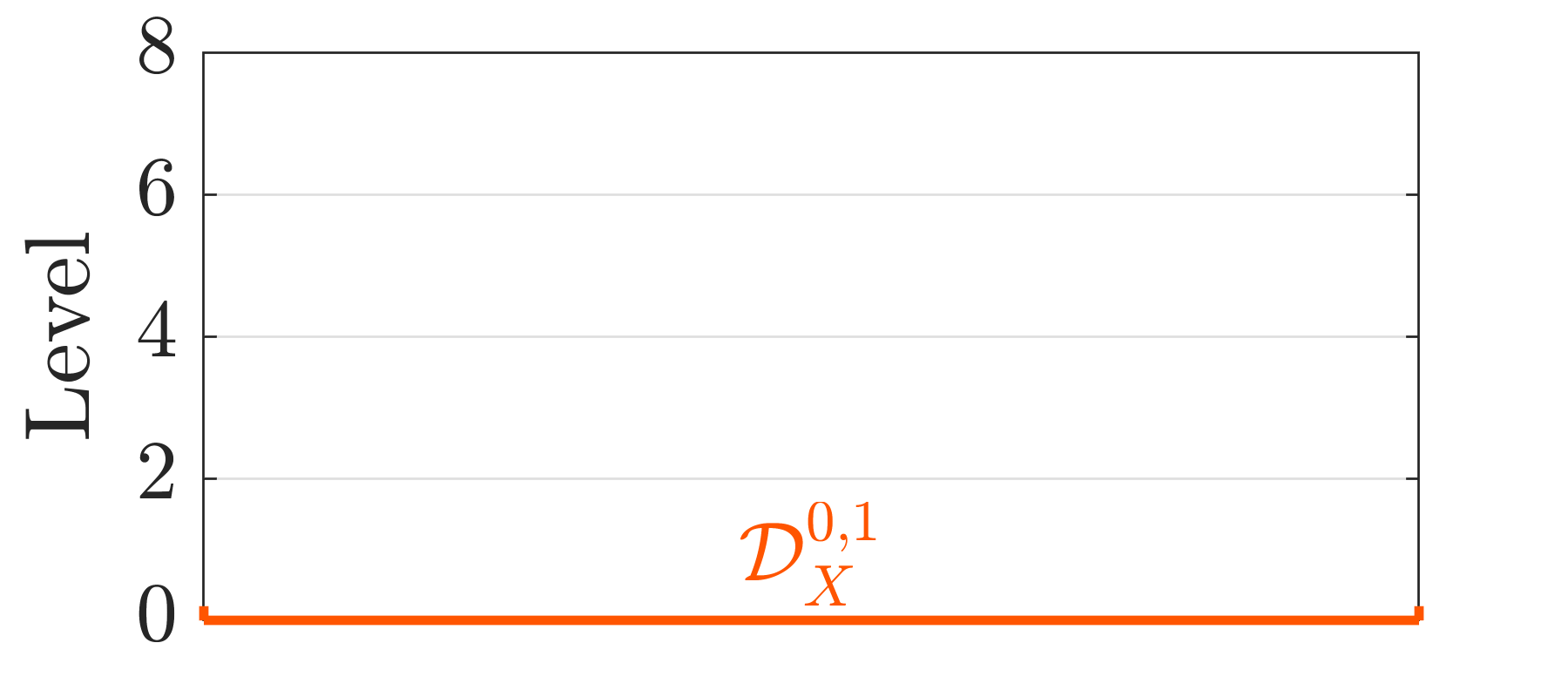}
			\includegraphics[width=\linewidth,clip=true,trim=0 12 0 10]{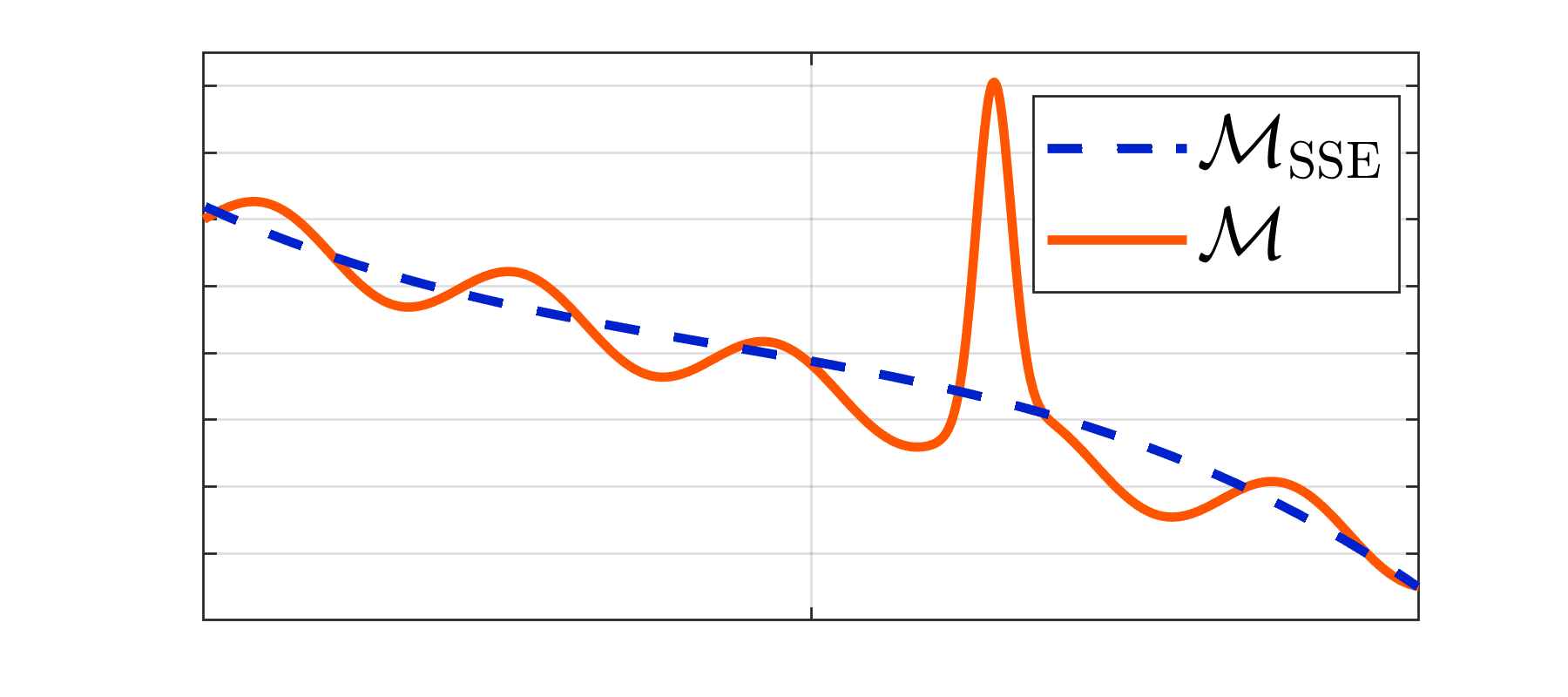}
			\includegraphics[width=\linewidth,clip=true,trim=0 0 0 10]{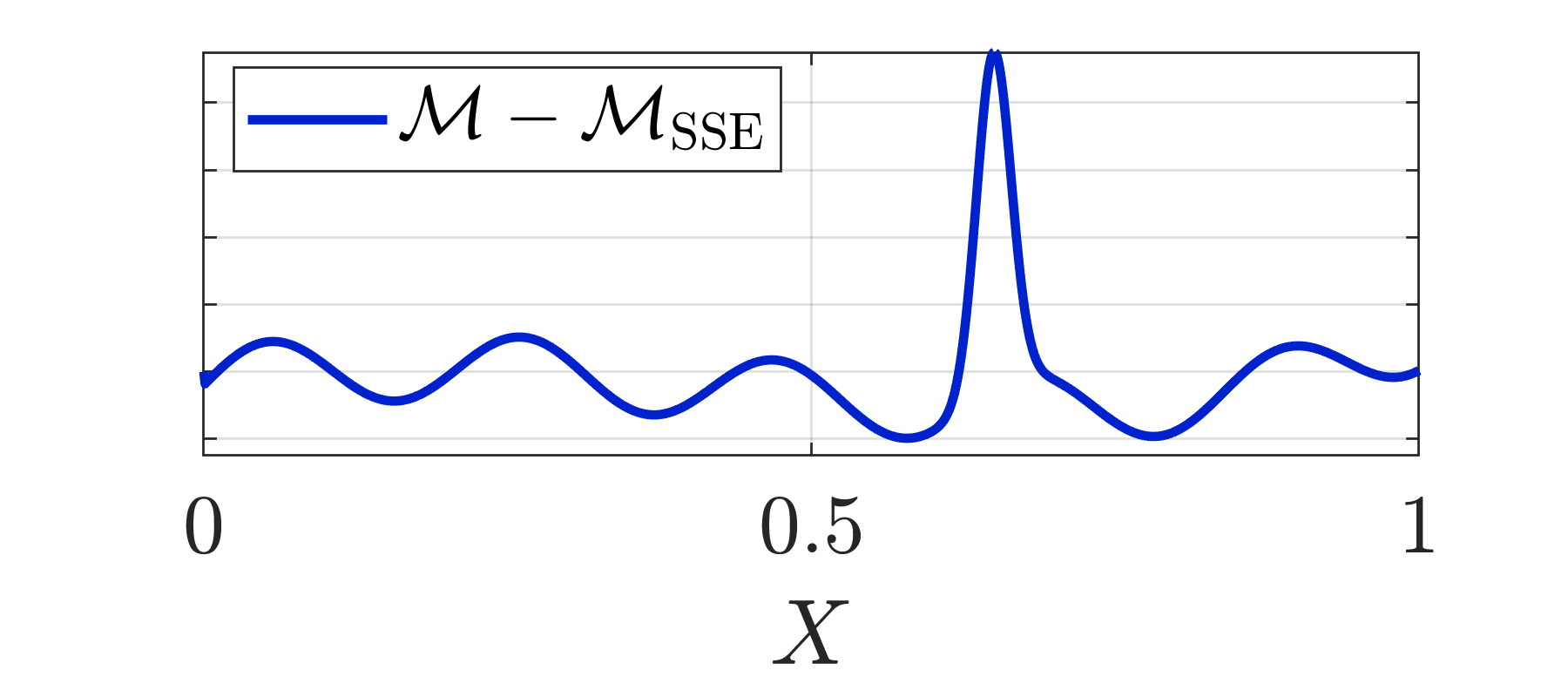}
		\end{minipage}
	}%
	\subfloat[$\ell=1$]{
		\begin{minipage}{0.48\linewidth}
			\label{fig:Ex1:detailedSteps:2}
			\includegraphics[width=\linewidth,clip=true,trim=0 10 0 0]{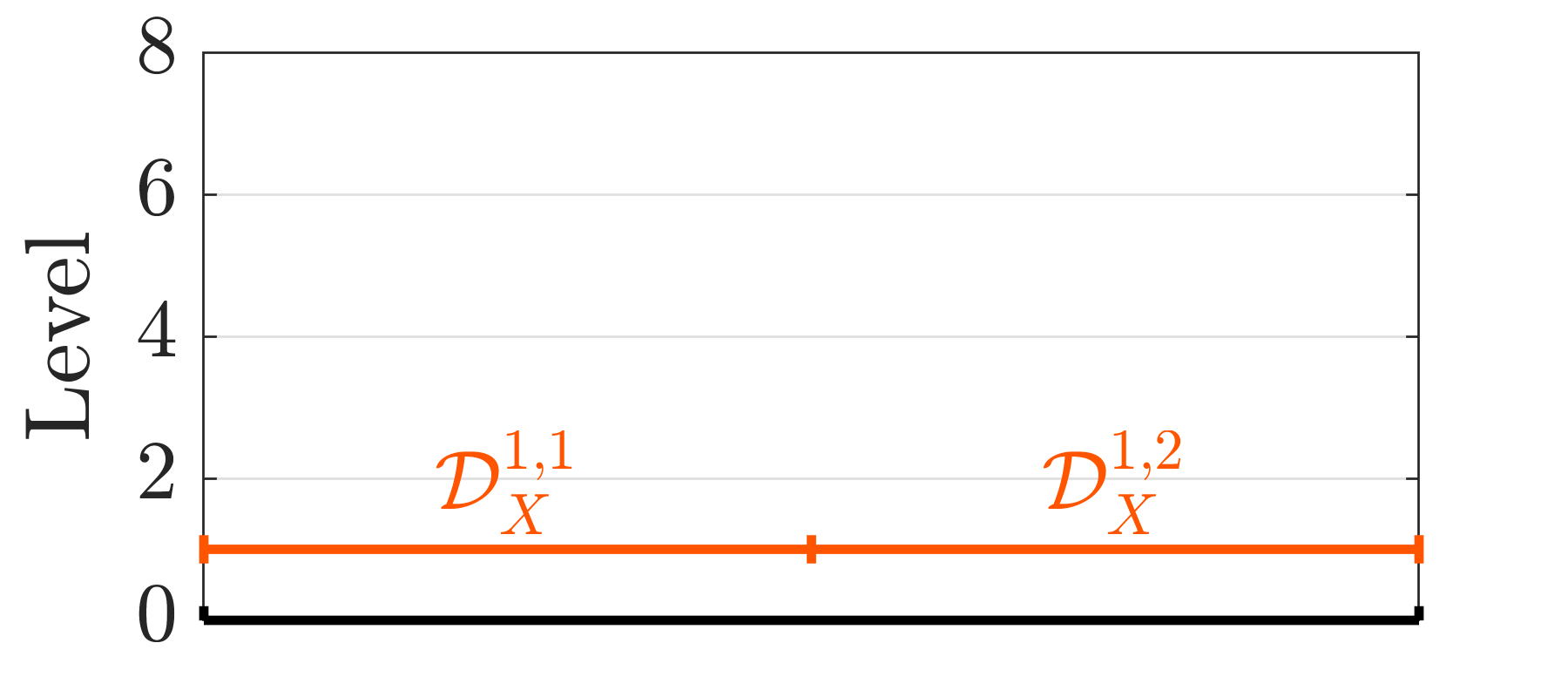}
			\includegraphics[width=\linewidth,clip=true,trim=0 12 0 10]{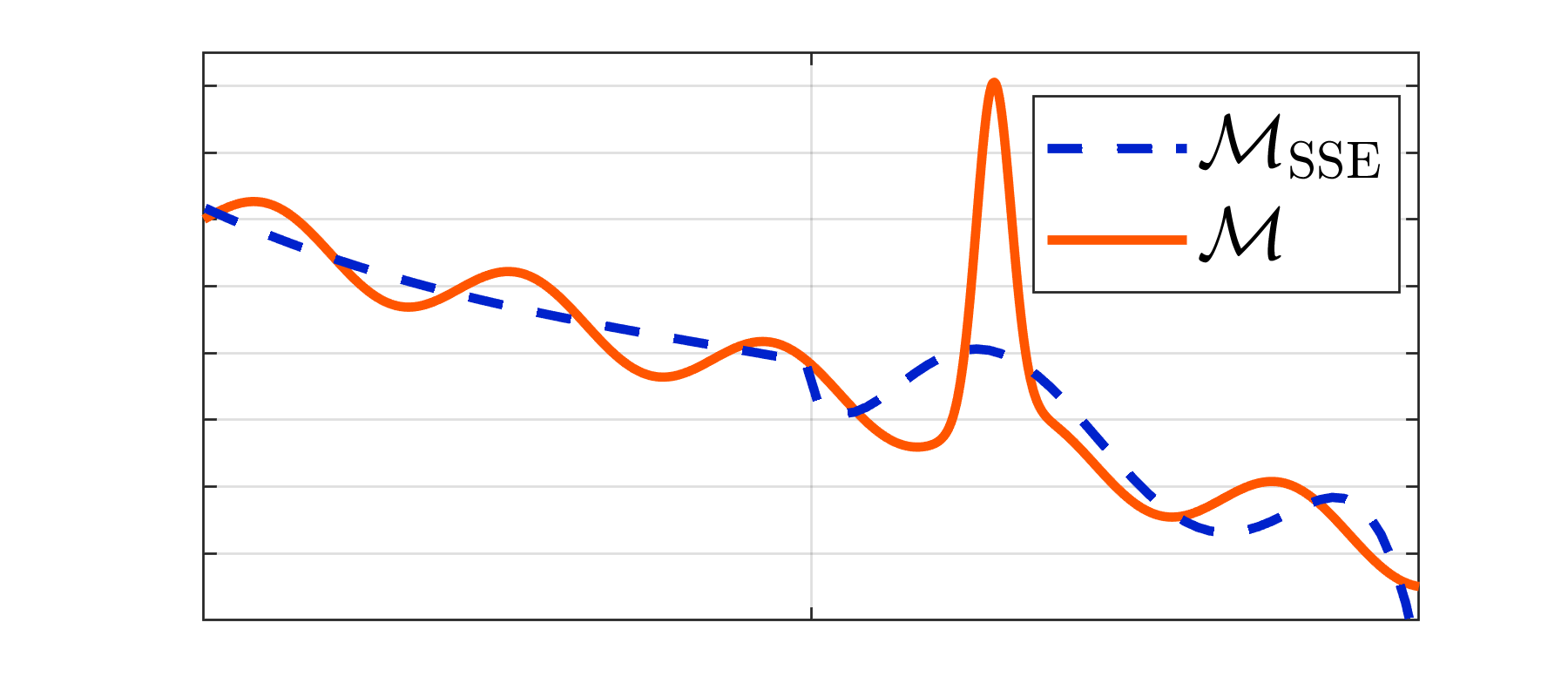}
			\includegraphics[width=\linewidth,clip=true,trim=0 0 0 10]{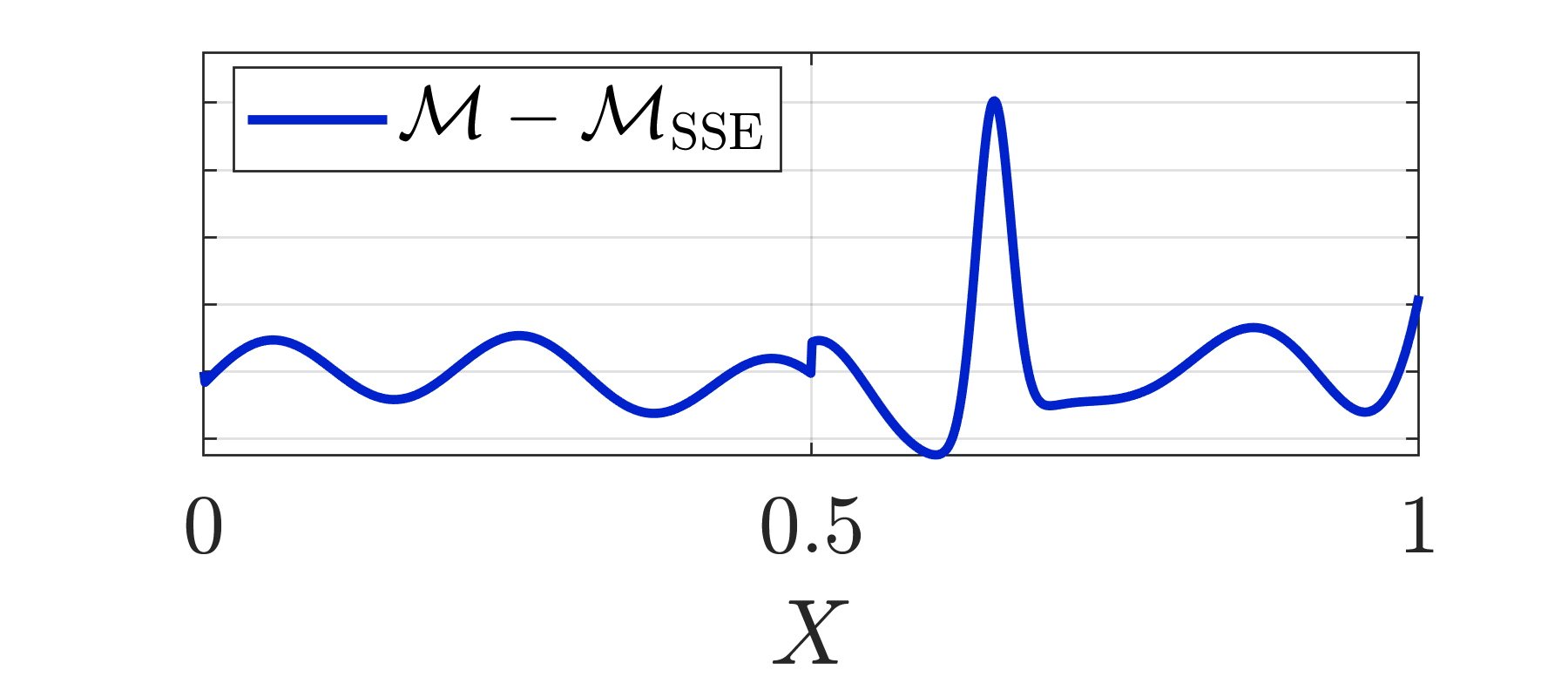}
		\end{minipage}
	}%

	\subfloat[$\ell=2$]{
		\begin{minipage}{0.48\linewidth}
			\label{fig:Ex1:detailedSteps:3}
			\includegraphics[width=\linewidth,clip=true,trim=0 10 0 0]{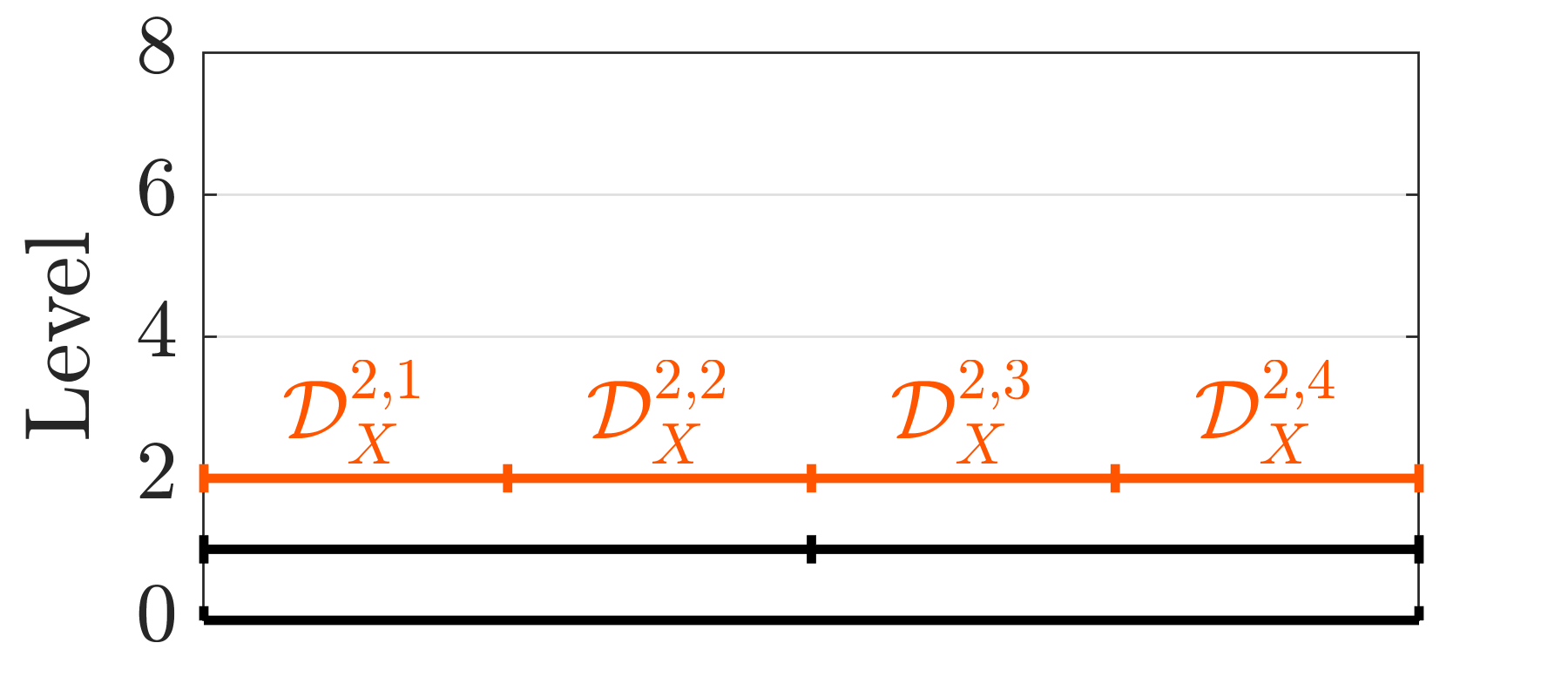}
			\includegraphics[width=\linewidth,clip=true,trim=0 12 0 10]{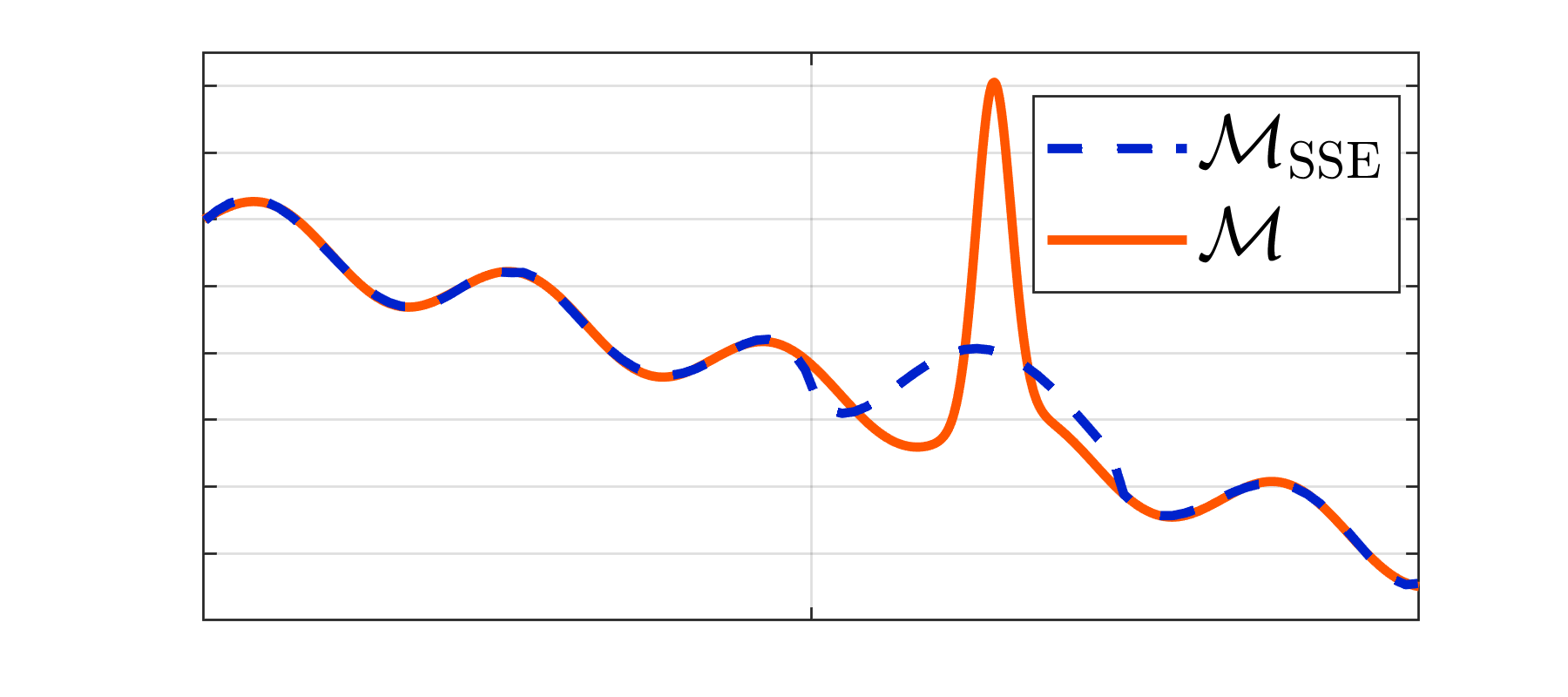}
			\includegraphics[width=\linewidth,clip=true,trim=0 0 0 10]{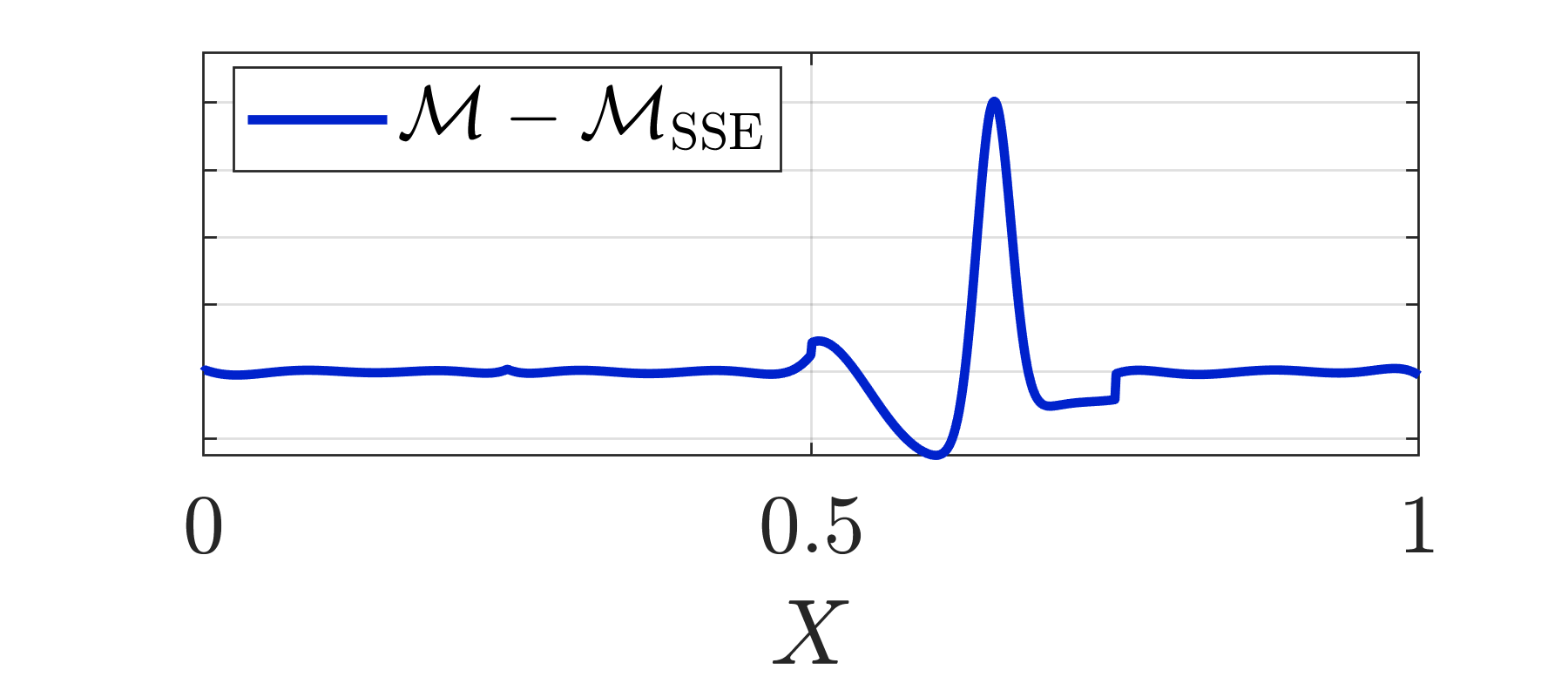}
		\end{minipage}
	}%
	\subfloat[$\ell=5$]{
		\begin{minipage}{0.48\linewidth}
			\label{fig:Ex1:detailedSteps:4}
			\includegraphics[width=\linewidth,clip=true,trim=0 10 0 0]{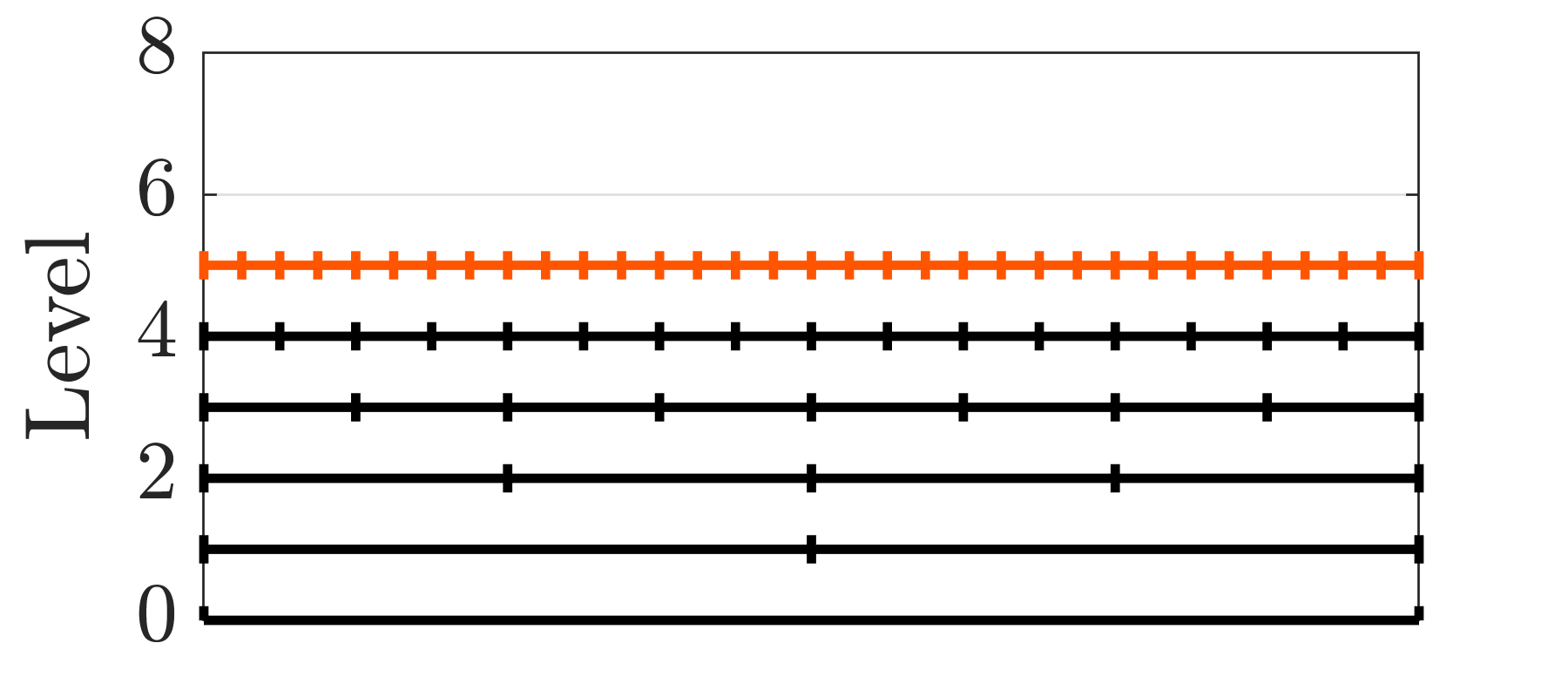}
			\includegraphics[width=\linewidth,clip=true,trim=0 12 0 10]{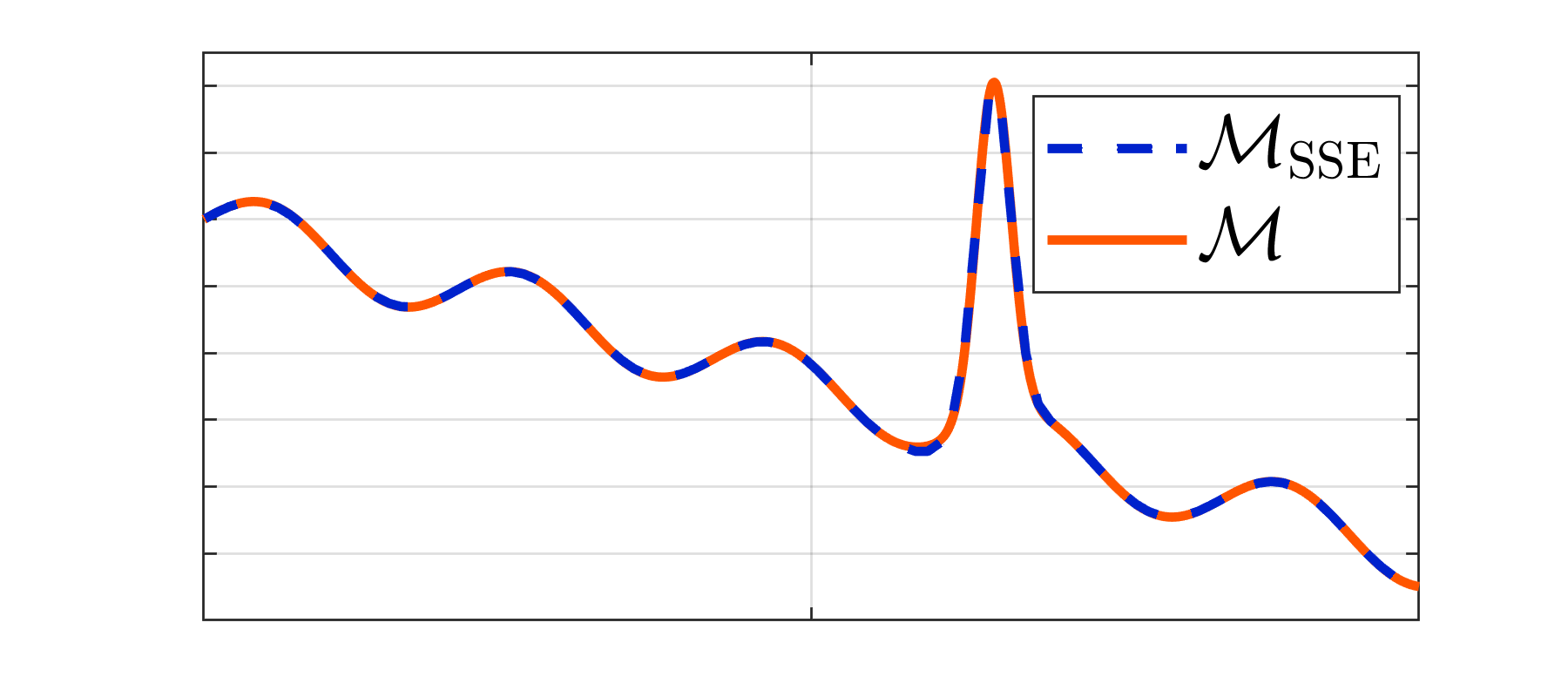}
			\includegraphics[width=\linewidth,clip=true,trim=0 0 0 10]{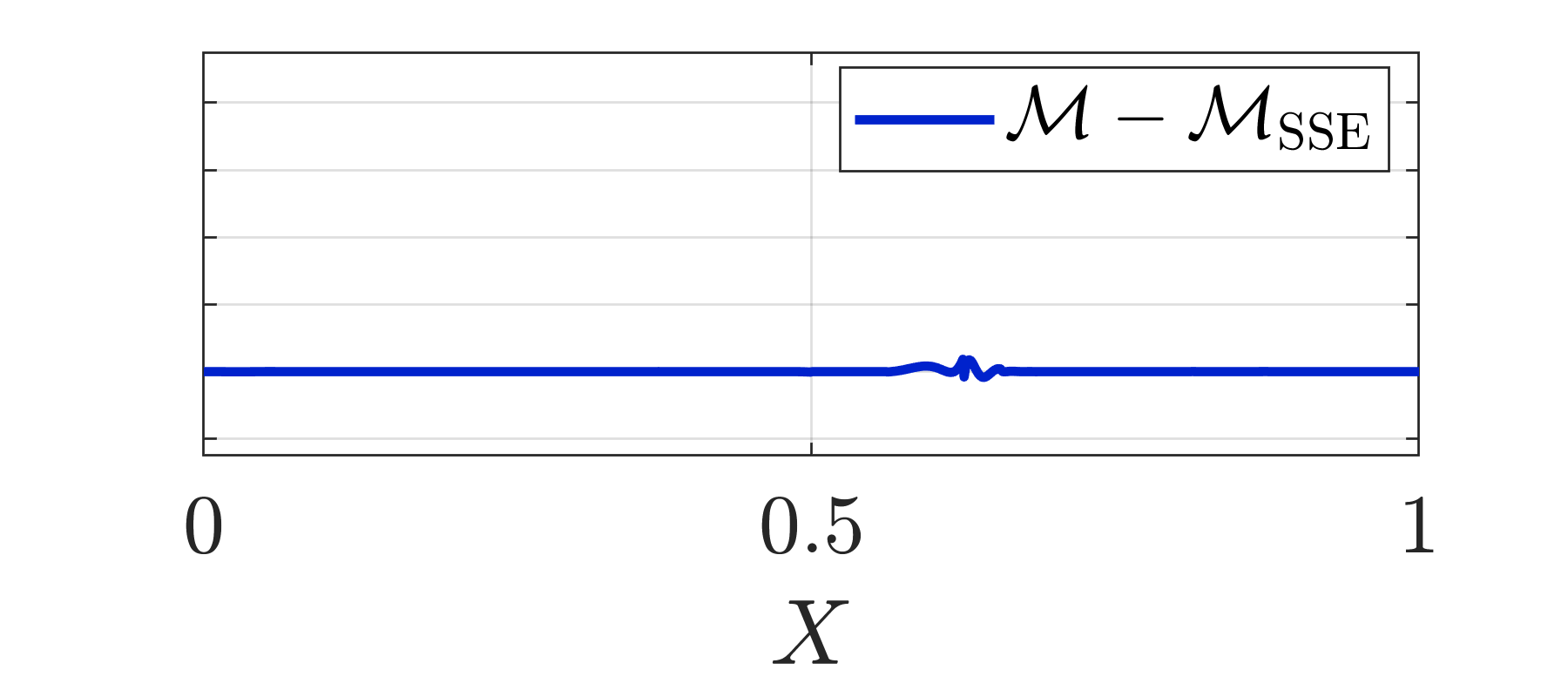}
		\end{minipage}
	}%
	\caption{\emph{One-dimensional analytical function}: selected steps of the SSE construction and the resulting domains, residuals and total approximation. The terminal domains (Eq.~\eqref{eqn:SSE flattened}) are highlighted in orange.\label{fig:Ex1:detailedSteps}}
\end{figure}

We first present a simple one-dimensional example that is meant to illustrate how SSE behaves with a model on which PCE is expected to fail. The model is given by:
\begin{equation}
\label{eq:application:1D:Def}
\cM(X) = -X + 0.1\sin{(30X)} + \exp{(-(50(X-0.65))^2)},
\end{equation}
where  $X\sim\cU(0,1)$ is a uniformly distributed random variable. The first two terms in Eq.~\eqref{eq:application:1D:Def} (polynomial and sinusoidal) can be accurately approximated by a low degree PCE, while the third term (squared exponential) causes PCE to require extremely high degree due to the localized peak it introduces at $x = 0.65$ (see Figure~\ref{fig:Ex1:detailedSteps}).
%
% explain figure setup
In the same Figure we detail four SSE refinement stages, with $\Ntot=200$ and $\maxDegSSE = 5$. 
For every step we show on the top panel a graphical representation of the various subdomains identified by the algorithm, with the subdomains of level $\ell$ highlighted in orange.
In the middle panel we plot the true model (orange solid line) and the current SSE approximation $\cM_{\mathrm{SSE}}$ as a dashed blue line.
In the bottom panel, we plot the corresponding residual $\cM(X)-\cM_{\mathrm{SSE}}(X)$ as a solid blue line in the same vertical scale as in the middle panel, for comparison.
	
% detailed iteration explanation
In the first step in Figure~\subref*{fig:Ex1:detailedSteps:1} the main trend of the function is identified, leaving a residual that mainly consists of the sine oscillation and the exponential peak. 
In the following step (Figure~\subref*{fig:Ex1:detailedSteps:2}) the approximation is not greatly improved in the subdomain $\cD_X^{1,1}:[0,0.5]$, because the available maximum degree $\maxDegSSE$ is not sufficiently high, resulting in a mostly constant polynomial correction. 
In subdomain $\cD_X^{1,2}:[0.5,1]$, the same problem is observed and the insufficient maximum degree results only in a small global improvement. 
In the next step (Figure~\subref*{fig:Ex1:detailedSteps:3}), the residual in $\cD_X^{2,1}$, $\cD_X^{2,2}$ and $\cD_X^{2,4}$ is significantly reduced to a very small oscillation around $0$. After the final step (Figure~\subref*{fig:Ex1:detailedSteps:4}), the overall approximation is has a high accuracy. 

From the residual progression it can be seen that the algorithm needs more levels to accurately approximate the target function near regions of high complexity, {\em i.e.}, near the exponential peak.
While this property does not affect the convergence when an experimental design of fixed size is chosen, it can be exploited in adaptive experimental design settings \citep{Wagner2020JCP}.

\begin{figure}
	\centering
	\includegraphics[width=0.7\linewidth]{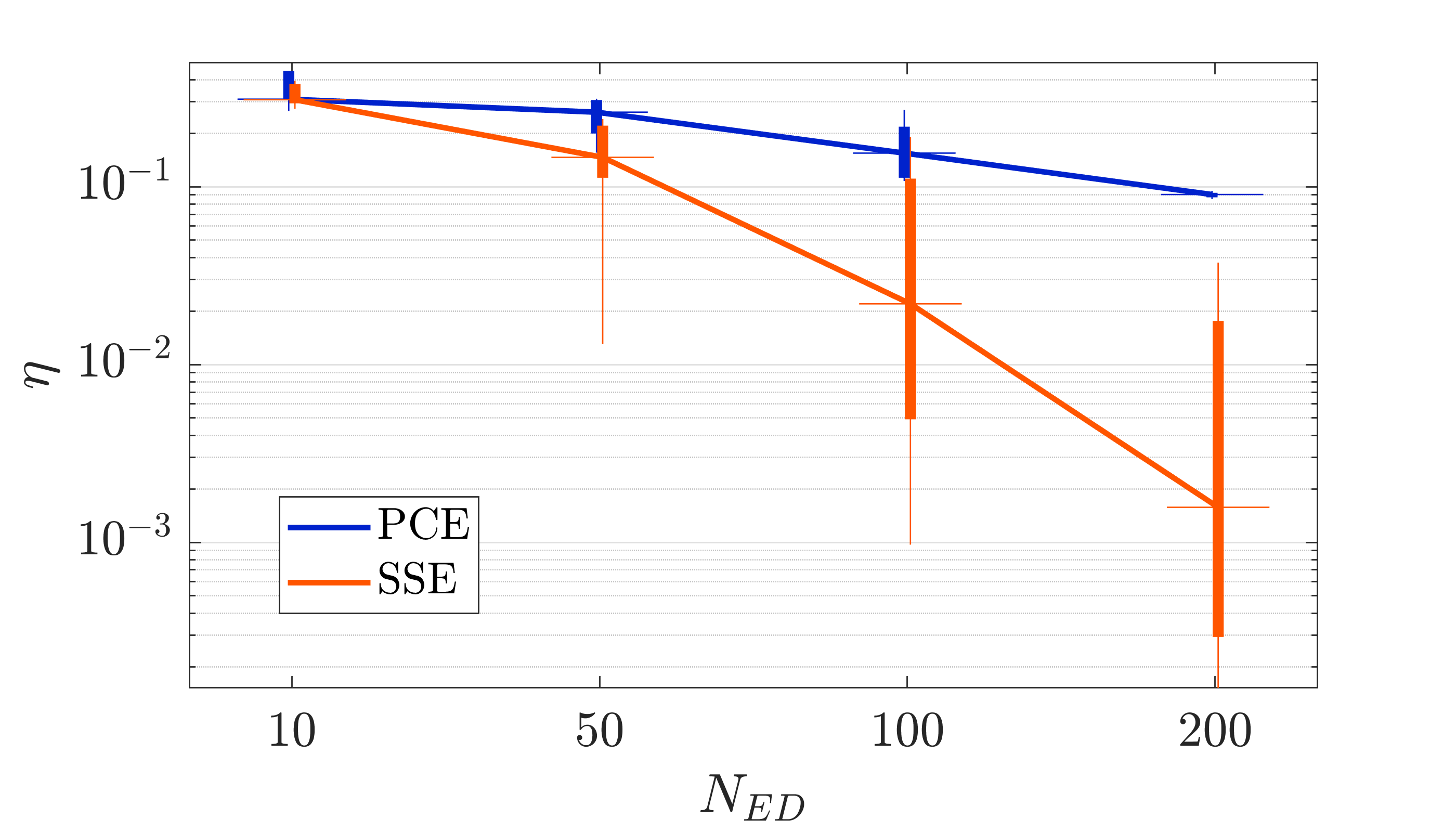}
	\caption{\emph{One-dimensional analytical function}: comparison of RMSE convergence between PCE and SSE as a function of the number of points in the experimental design. 
	A slight horizontal offset is added to improve readability. \label{fig:Ex1:convergence}}
\end{figure}

% different experimental design sizes comparison
As expected, the final SSE accuracy increases with the size of the experimental design. In Figure~\ref{fig:Ex1:convergence}, we compare SSE and PCE on a set of experimental design sizes of $\Ntot=\{10, 50, 100, 200\}$ in terms of  their relative mean squared error (MSE, Eq.~\eqref{eq:Ex:MSE}). PCE is constructed with a maximum adaptive degree of $\maxDegPCE=20$. 
%
% comment on convergence
At the extremely small experimental design of $\Ntot=10$, the SSE approach is comparable to PCE. 
As the available experimental design points increase, SSE exhibits faster convergence in RMSE than PCE, and from $\Ntot=50$ onwards SSE consistently outperforms PCE in this problem. 
At larger experimental designs, SSE can accurately reproduce the localized behavior of this test function, while not being constrained by the  global nature of PCE basis functions defined on the full domain. At the final ED size of $\Ntot=200$ the SSE relative MSE is at least one order of magnitude smaller than the PCE error. There is considerable variability in the relative MSE between individual realizations. 
This can be attributed to the squared exponential peak in Eq.~\eqref{eq:application:1D:Def}: depending on the input realizations in the experimental design, it is captured better or worse by the available data.

\subsection{Application 2: 100-dimensional analytical function}
\label{sec:Applications:ex2}
% explain function
With this example we want to explore the scalability of SSE in high dimensional problems. 
This example uses a variant of a test function introduced in \cite{ZhouPhD}. We modified the function to have a high nominal dimensionality ($M=100$), and relatively low effective dimensionality meaning that the majority of the variability can be attributed to a small number of input parameters. It takes the form
\begin{equation}
\label{eq:applications:ZhouFun:Def}
\begin{split}
\cM(\ve{X}) = \frac{10^M}{2} \left[\varphi(10\cdot(\ve{X}-1/3)) + 
\varphi(10\cdot(\ve{X}-2/3))\right],\\ \quad \text{where} ~~ 
\varphi(\ve{x})\eqdef (2\pi)^{-M/2}\exp{\left(-\frac{1}{2}\sum_{i=1}^M a_i^2x_i^2\right)}.
\end{split}
\end{equation} 

The factor $\ve{a}=(a_1,\cdots,a_M)$ modifies the original function and serves as a dimension-dependent weight that decays exponentially, as:
\begin{equation}
a_i = e^{-(i-1)}, \quad \text{with} \quad i \in \{1,\cdots,M\}.
\end{equation}

% input
The input random vector $\ve{X}$ is distributed according to a multivariate standard uniform distribution with independent marginals, $\ve{X}\sim f_{\vX}(\vX) = \prod_{i=1}^{100}\cU(0,1)$.

\begin{figure}[ht]
	\centering
	\subfloat[$X_{1}$ and $X_{2}$]{
		\begin{minipage}{0.5\linewidth}
			\includegraphics[width=0.9\linewidth]{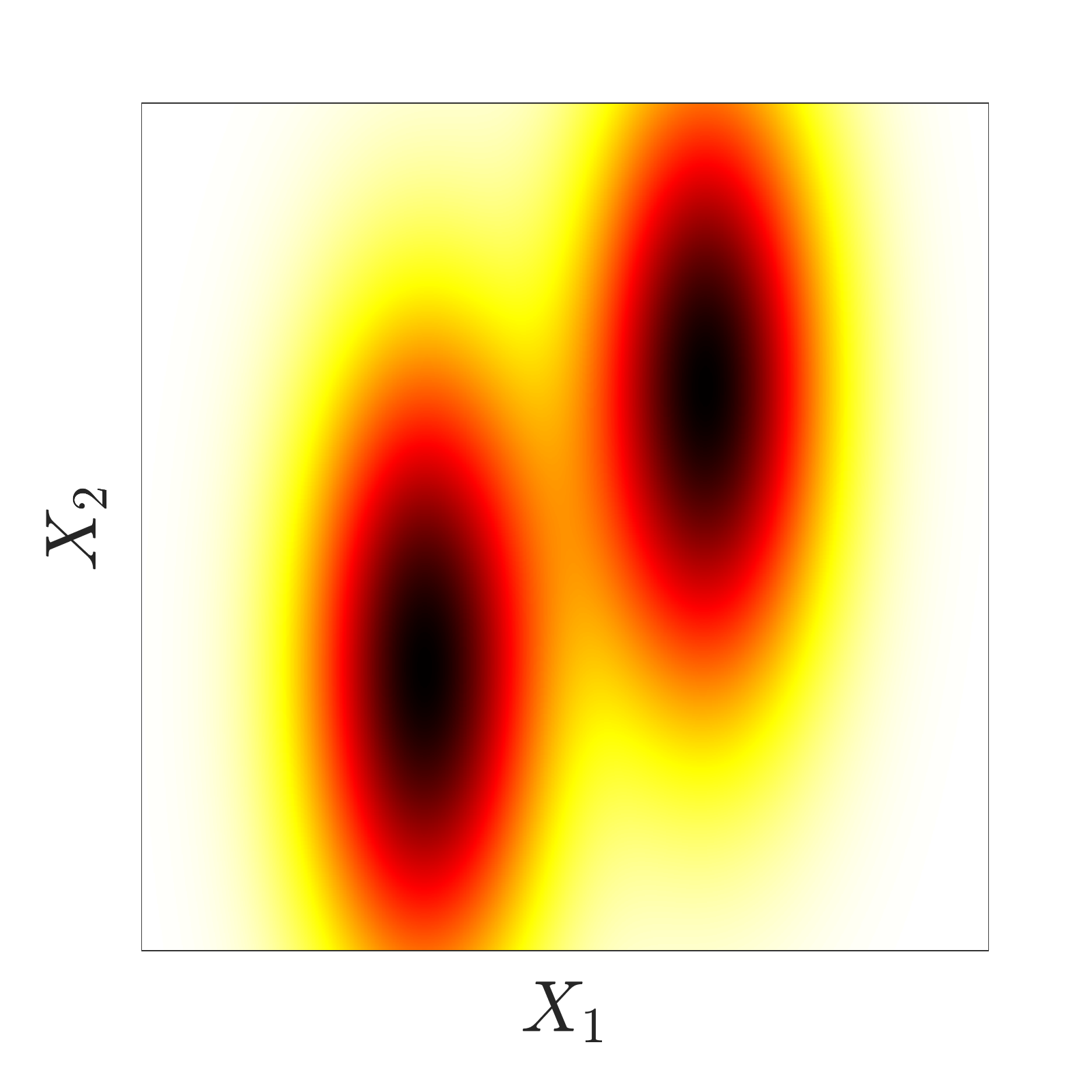}
			\label{fig:Ex2:contour:1}
		\end{minipage}
	}%
	\subfloat[$X_{1}$ and $X_{10}$]{
		\begin{minipage}{0.5\linewidth}
			\includegraphics[width=0.9\linewidth]{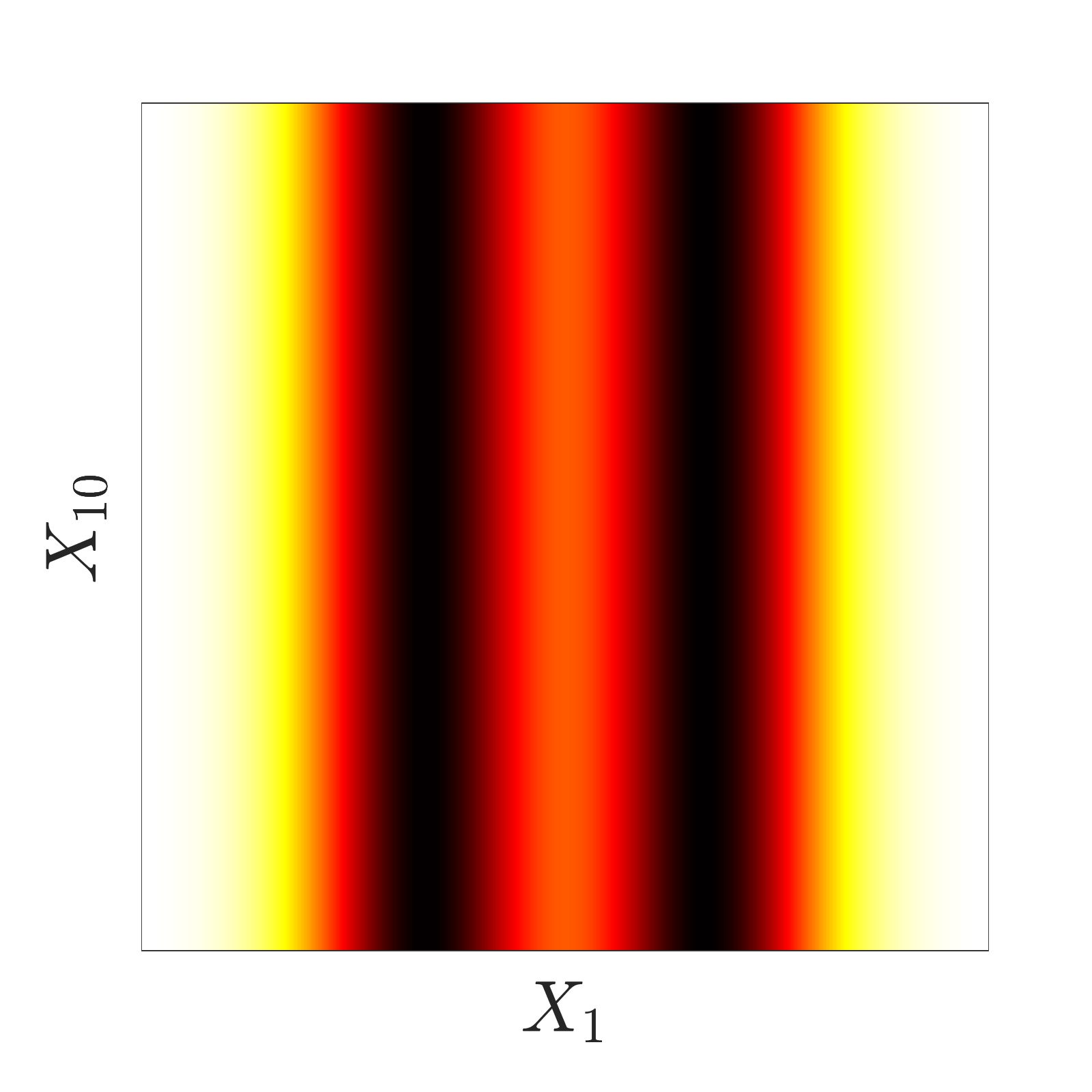}
			\label{fig:Ex2:contour:2}
		\end{minipage}
	}%
	\caption{\emph{100-dimensional analytical function}: bivariate contour cross-section plots. \label{fig:Ex2:contour}}
\end{figure}

% contour plots
Two contour cross-section plots of this function are shown between dimensions $\{X_1,X_2\}$ in Figure~\subref*{fig:Ex2:contour:1} and between $\{X_1,X_{10}\}$ in Figure~\subref*{fig:Ex2:contour:2}. The effect of the decay factor $\ve{a}$ is clearly visible and results in a \emph{close-to} constant behavior along the parametric dimension $X_{10}$.

\begin{figure}
    \centering
	\includegraphics[width=0.7\linewidth]{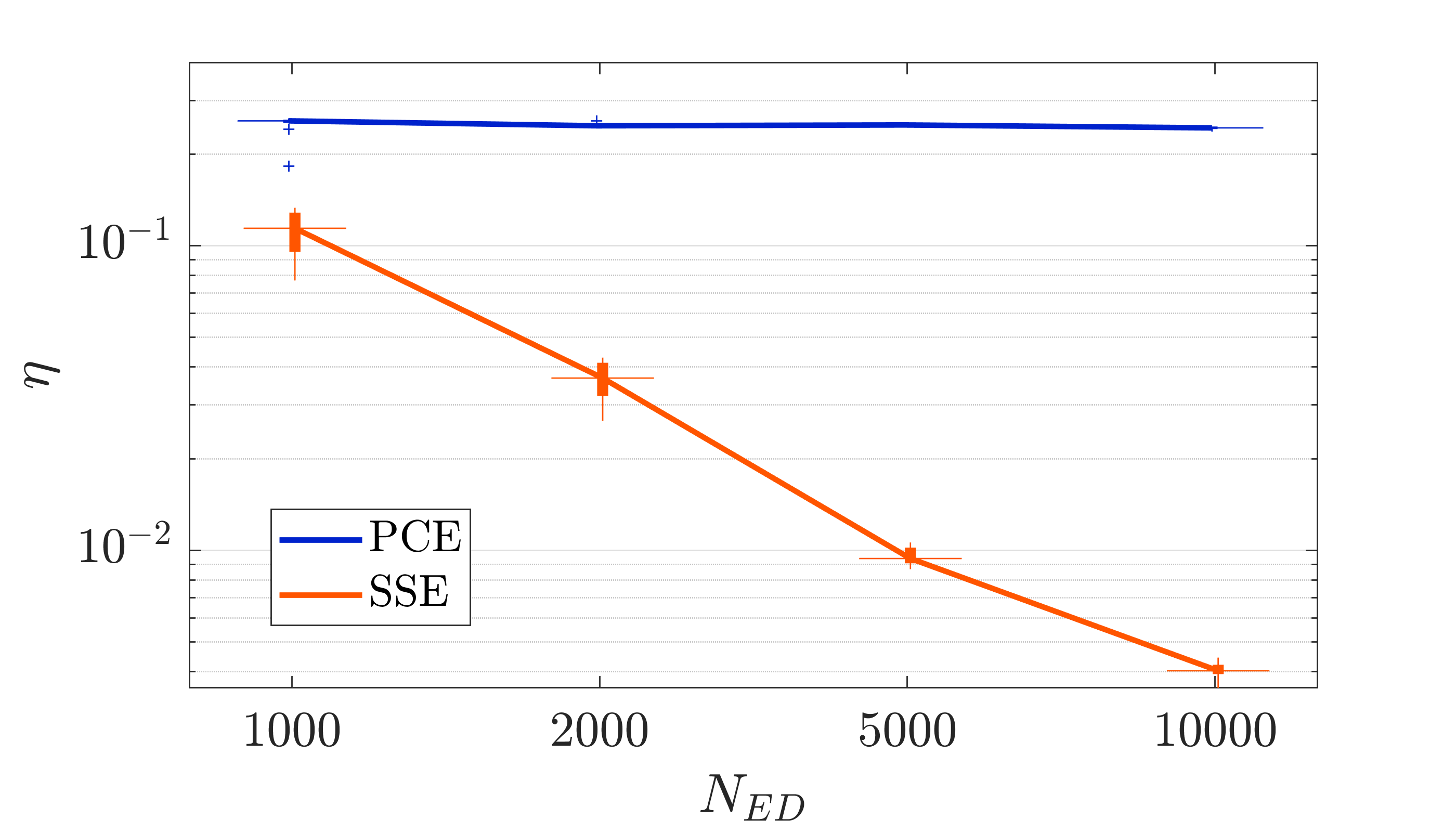}
	\caption{\emph{100-dimensional analytical function}: comparison of RMSE convergence between PCE and SSE as a function of the number of points in the experimental design. 
	A slight horizontal offset is added to improve readability. \label{fig:Ex2:convergence}}
\end{figure}

% convergence
To manage the computational complexity, in this example we limit the maximum polynomial degree in SSE to $\maxDegSSE=2$ and compute the SSE with total experimental design sizes of $\Ntot = \{1{,}000; 2{,}000; 5{,}000; 10{,}000\}$. 
For PCE, we choose a maximum degree of $\maxDegSSE=7$, which is the maximum degree we could run on a standard desktop computer with 16GB of RAM before incurring memory issues. 
The resulting comparison between PCE and SSE with respect to the relative MSE is plotted in Figure~\ref{fig:Ex2:convergence}.
%
% comment
In this scenario, the SSE algorithm outperforms PCE on all investigated experimental designs. 
In fact, PCE seems to benefit from increasing the experimental design only marginally, with a relative error of $\eta \approx 0.25$ for all considered values of $\Ntot$. 
SSE shows instead a convincing convergence behavior, by reducing its residual by almost two orders of magnitude across the various experimental designs. 
Given the high dimensionality of this problem, the approximation power of sparse PCE is limited by the curse of dimensionality, rather than the lack of data.
By reducing the complexity of the spectral representation at each level, SSE can better exploit informative datasets without incurring similar computational bottlenecks.

\subsection{Application 3: damped oscillator}
\label{sec:Applications:ex3}
\begin{figure}
	\centering
	\includegraphics[width=0.7\linewidth]{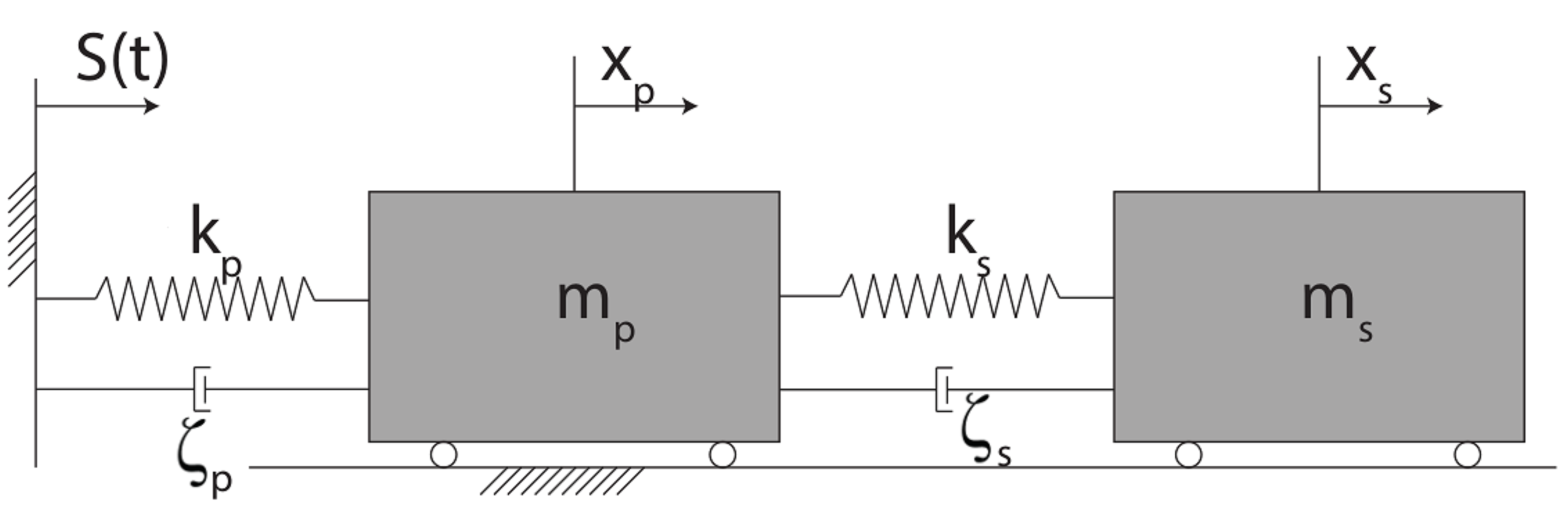}
	\caption{\emph{8-dimensional damped oscillator}: model setup. \label{fig:Ex3:setup}}
\end{figure}

Damped oscillators are a class of engineering models that is commonly used in structural reliability problems \citep{DubourgThesis}. 
This class of problems is known to be often difficult to surrogate, due its high non-linearity and often local behavior.
A sketch of the oscillator considered in this example is displayed in Figure~\ref{fig:Ex3:setup}. 
It consists of a primary and secondary system with masses $m_p, m_s$, stiffnesses $k_p, k_s$ and damping ratios $\zeta_p, \zeta_s$. The subscripts $p$ and $s$ denote the \emph{primary} and \emph{secondary} system properties, respectively. 
%
%The model of interest is its limit state function, which describes the state of the system through its sign. Negative values correspond to system failure, while positive values represent safe operation. 

In this example we consider the limit state function of the damped oscillator given by
\begin{equation}
\cM(\ve{X}) = F_s - p\cdot k_s \sqrt{\mathbb{E}_{S}\left[x_S^2\right]},
\end{equation}
where $F_s$ is the force capacity of the secondary spring, $p$ is the \emph{so-called} peak factor and $x_S$ is the relative displacement between the primary and secondary systems. The mean-square relative displacement of the secondary spring under a white noise base acceleration $S$ is analytically given by:
\begin{equation}
\mathbb{E}_{S}\left[x_S^2\right] = \pi \frac{S_0}{4\zeta_s \omega_s^3}\frac{\zeta_a \zeta_s}{\zeta_p \zeta_s(4\zeta_a^2+\theta^2) +\gamma\zeta_a^2}  \frac{(\zeta_p \omega_p^3+\zeta_s \omega_s^3) \omega_p}{4\zeta_a \omega_a^4},
\end{equation}
where $S_0$ is the white noise intensity, $\omega_p=\sqrt{k_p/m_p}$ and $\omega_s=\sqrt{k_s/m_s}$ are the natural frequencies of the two subsystems, and the further abbreviations are used: $\gamma=m_s/m_p$, $\omega_a=(\omega_p+\omega_s)/2$, $\zeta_a=(\zeta_p+\zeta_s)/2$ and $\theta=(\omega_p-\omega_s)/\omega_a$. 

All variables but the peak factor (set to $p = 3$) are modelled as independent random variables and are summarized in the random vector $\ve{X} = \{m_p,m_s,k_p,k_s,\zeta_p,\zeta_s,S_0,F_s\}$. Their marginal distributions are lognormal, with the parameters given in Table~\ref{tab:ex3:marginals}.
\begin{table}
	\centering
	\begin{tabular}{cllrc}
		\hline
		\textbf{Variable} & \textbf{Description} & \textbf{Distribution} & \textbf{Mean} & \textbf{C.O.V.}\\
		\hline
		$m_p$ & primary mass & Lognormal & $1.50$ &	$0.1$\\
		$m_s$ & secondary mass & Lognormal & $0.01$ &	$0.1$\\
		$k_p$ & primary spring stiffness & Lognormal & $1.00$ &	$0.2$\\
		$k_s$ & secondary spring stiffness& Lognormal & $0.01$ &	$0.2$\\
		$\zeta_p$ & primary damping ratio & Lognormal & $0.05$ &	$0.4$\\
		$\zeta_s$ & secondary damping ratio & Lognormal & $0.02$ &	$0.5$\\
		$S_0$ & white noise intensity & Lognormal & $100.00$ &	$0.1$\\
		$F_s$ & secondary spring force capacity & Lognormal & $15.00$ &	$0.1$\\
		\hline
	\end{tabular}
	\caption{\emph{8-dimensional damped oscillator}: marginal distributions. \label{tab:ex3:marginals}}
\end{table}
%

\begin{figure}
	\centering
	\includegraphics[width=0.7\linewidth]{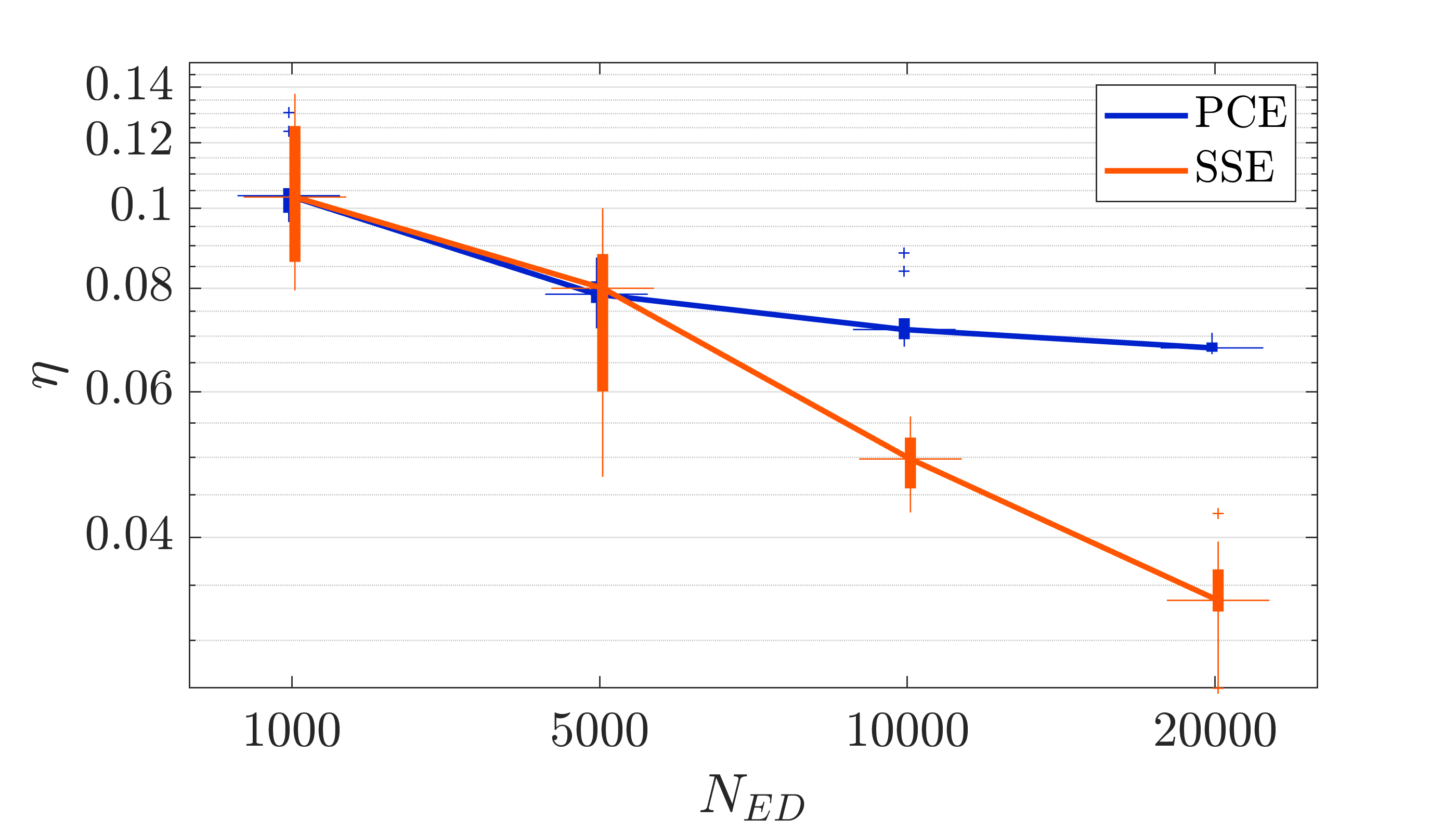}
	\caption{\emph{8-dimensional damped oscillator}: comparison of RMSE convergence between PCE and SSE as a function of the number of points in the experimental design. 
	A slight horizontal offset is added to improve readability.
	\label{fig:Ex3:convergence}}
\end{figure}

% Convergence
For the convergence study, we choose experimental design sizes of $\Ntot = \{1{,}000; 5{,}000; 10{,}000; 20{,}000\}$. The maximum degrees are set to $\maxDegSSE=4$ and $\maxDegPCE=10$ for SSE and PCE, respectively. Figure~\ref{fig:Ex3:convergence} summarizes the results.
This benchmark is known to be quite difficult to approximate with standard surrogate model techniques, as it is clear from the scale of the RMSE in Figure~\ref{fig:Ex3:convergence}.
For the two smaller experimental design sizes $\Ntot  = \{1{,}000; 5{,}000\}$, both PCE and SSE perform quite poorly, with SSE showing a similar median behavior, but much higher variability. 
For larger experimental designs, however, the SSE performance improves significantly over that of PCE, until at $\Ntot=20{,}000$ the relative MSE of SSE is half that of PCE.
This behavior is in line with the previous findings: provided enough information, SSE can provide higher expressive power than its static counterpart.

\subsection{Application 4: truss with discontinuous snap-through behavior}
\label{sec:Applications:ex4}
\begin{figure}
	\centering
	\subfloat[Before snap-through]{
		\begin{minipage}{0.5\linewidth}
				\includegraphics[width=0.9\linewidth]{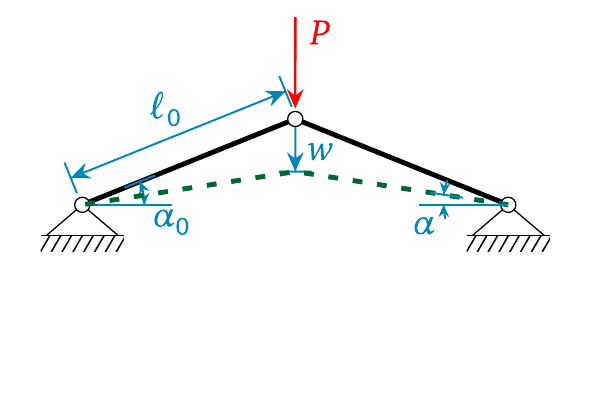}
			\label{fig:ex4:setup:1}
		\end{minipage}
	}%
	\subfloat[After snap-through]{
		\begin{minipage}{0.5\linewidth}
				\includegraphics[width=0.9\linewidth]{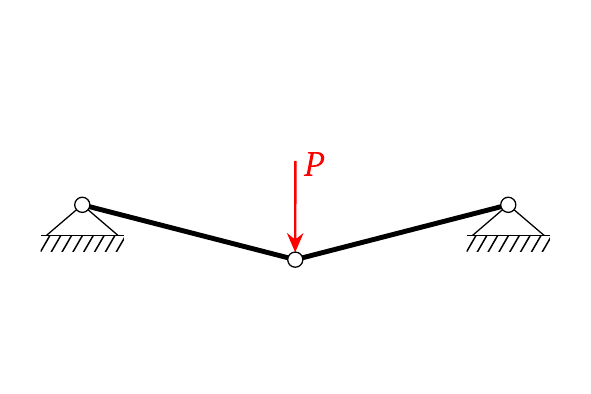}
			\label{fig:ex4:setup:2}
		\end{minipage}
	}%
	
	\caption{\emph{3-dimensional snap through truss}: illustration of the truss structure subject to snap-through}
	\label{fig:ex4:setup}
\end{figure}

% intro
As a last example, we address another problem of engineering interest: the geometrically non-linear two-bar truss structure shown in Figure~\ref{fig:ex4:setup}. 
The structure itself is defined by the initial inclination $\alpha_0$ and length $\ell_0$ of the two bars. 
A peculiarity of this structure is that it exhibits the so-called \emph{snap-through} behavior. 
At first, the vertical displacement $w$ of such a structure typically increases linearly with an increasing load $P$ (Figure~\subref*{fig:ex4:setup:1}).
Once a specific critical load is exceeded, the structure \emph{snaps through} to another equilibrium point, at which the load can be increased further (Figure~\subref*{fig:ex4:setup:2}). 
The main implication of this kind of behavior is that it is discontinuous in some critical regions of the input space, which are in general unknown \textit{a priori}.

% model definition
The vertical displacement $w$ of the truss tip is related to the angle $\alpha$ by
\begin{equation}\label{eq:14}
\cM(\ve{X}) = w = \ell_0 \cos\alpha_0 (\tan\alpha_0 - \tan\alpha(\ve{X})).
\end{equation}
At the same time, $\alpha$ needs to satisfy the following constitutive equation that depends on the random vector $\ve{X} = \{P, E, A\}$:
\begin{equation}\label{eq:13}
P = - 2 E A \tan\alpha (\cos\alpha_0 - \cos\alpha).
\end{equation}
For a given realization of $\ve{X}$, this equation can be solved numerically for $\alpha$, the value of which then is used in Eq.~\eqref{eq:14} to estimate the corresponding vertical displacement.

% SSE/PCE setup
In this study we set the constants $l_0 = 5~\mathrm{m}$ and $\alpha_0 = 10^\circ$, and treat the parameters $\ve{X}$ as independent random variables with marginals listed in Table~\ref{tab:ex4:marginals} \citep{Moustapha2019}. 
We investigate experimental designs of sizes $\Ntot=\{100; 500; 1{,}000; 2{,}000; 5{,}000\}$ and set the maximum polynomial degrees to $\maxDegSSE=4$ and $\maxDegPCE=10$ for SSE and PCE, respectively.
\begin{table}
	\centering
	\begin{tabular}{cllrc}
		\hline
		\textbf{Variable} & \textbf{Description} & \textbf{Distribution} & \textbf{Mean} & \textbf{C.O.V.}\\
		\hline
		$P$ & load & Gumbel & $430$ & $0.20$ \\
		$E$ & Young's modulus & Lognormal & $210$ & $0.10$ \\
		$A$ & cross sectional area & Gaussian & $10$ & $0.05$ \\
		\hline
	\end{tabular}
	\caption{\emph{3-dimensional snap-through truss}: marginal distributions.}
	\label{tab:ex4:marginals}
\end{table}

\begin{figure}
	\centering
	\includegraphics[width=0.7\linewidth]{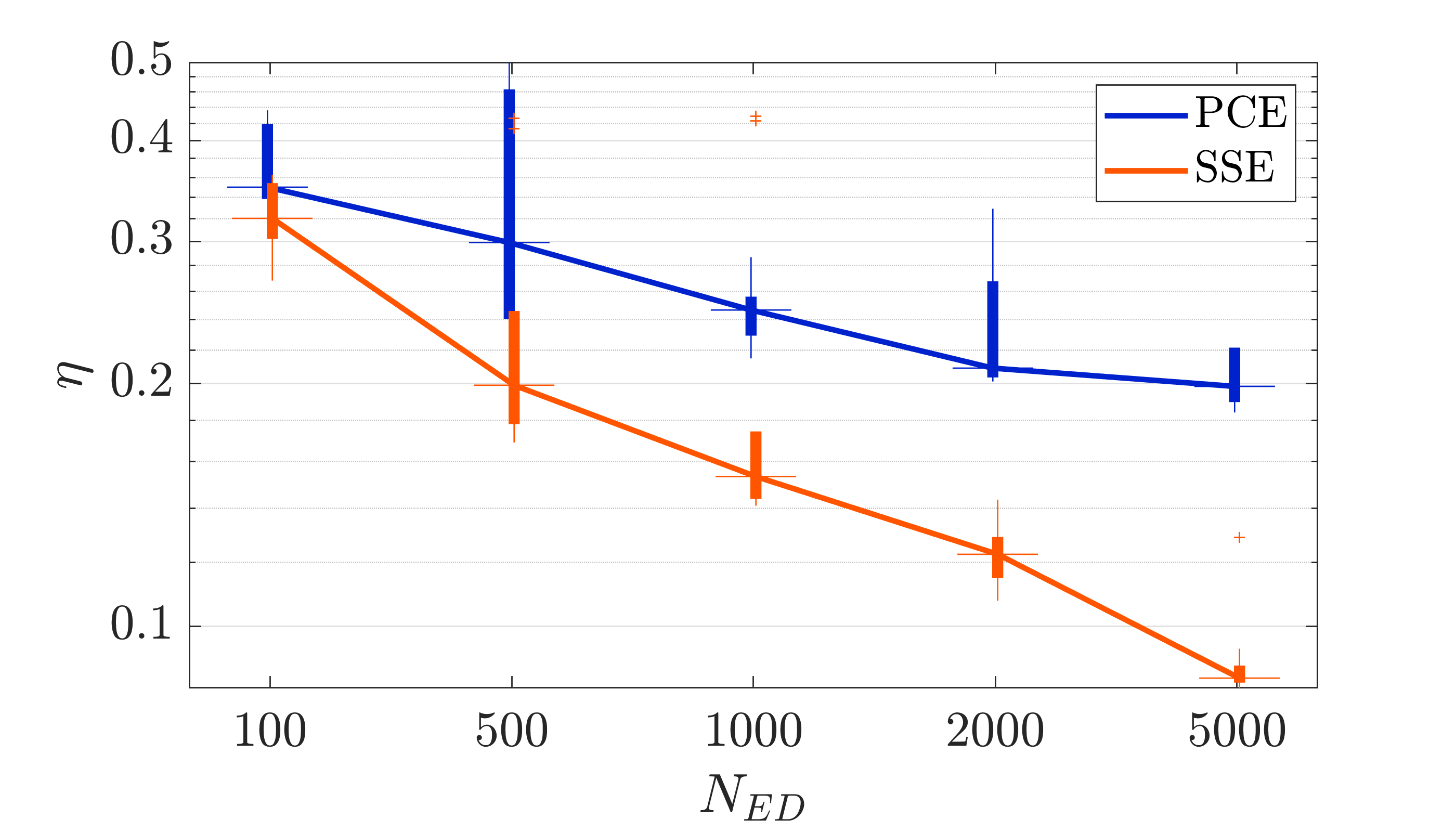}
	\caption{\emph{3-dimensional snap-through truss}: comparison of RMSE convergence between PCE and SSE as a function of the number of points in the experimental design. 
	A slight horizontal offset is added to improve readability.\label{fig:Ex4:convergence}}
\end{figure}

\begin{figure}
    \captionsetup{justification=centering}
	\centering
	\subfloat[PCE point-wise error:\newline ${(\cM(\vx)-\cM_{\mathrm{PCE}}(\vx))^2/\Var{\cM}}$]{
		\includegraphics[height=5cm,clip=true,trim=10 10 20 0]{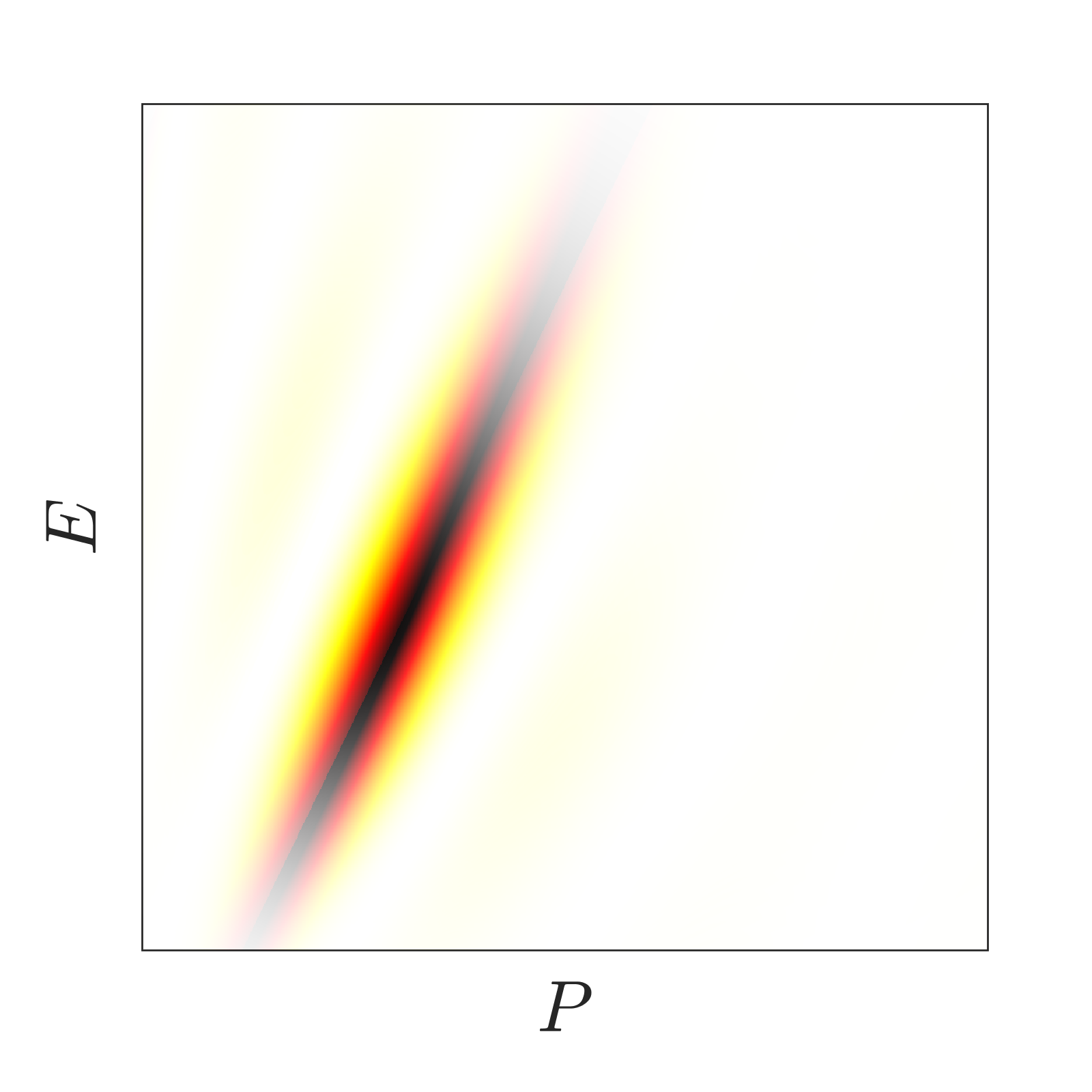}
		\label{fig:ex4:comparison:PCEDelta}
	}%
	\subfloat[SSE point-wise error:\newline ${(\cM(\vx)-\cM_{\mathrm{SSE}}(\vx))^2/\Var{\cM}}$]{
		\includegraphics[height=5cm,clip=true,trim=10 10 20 0]{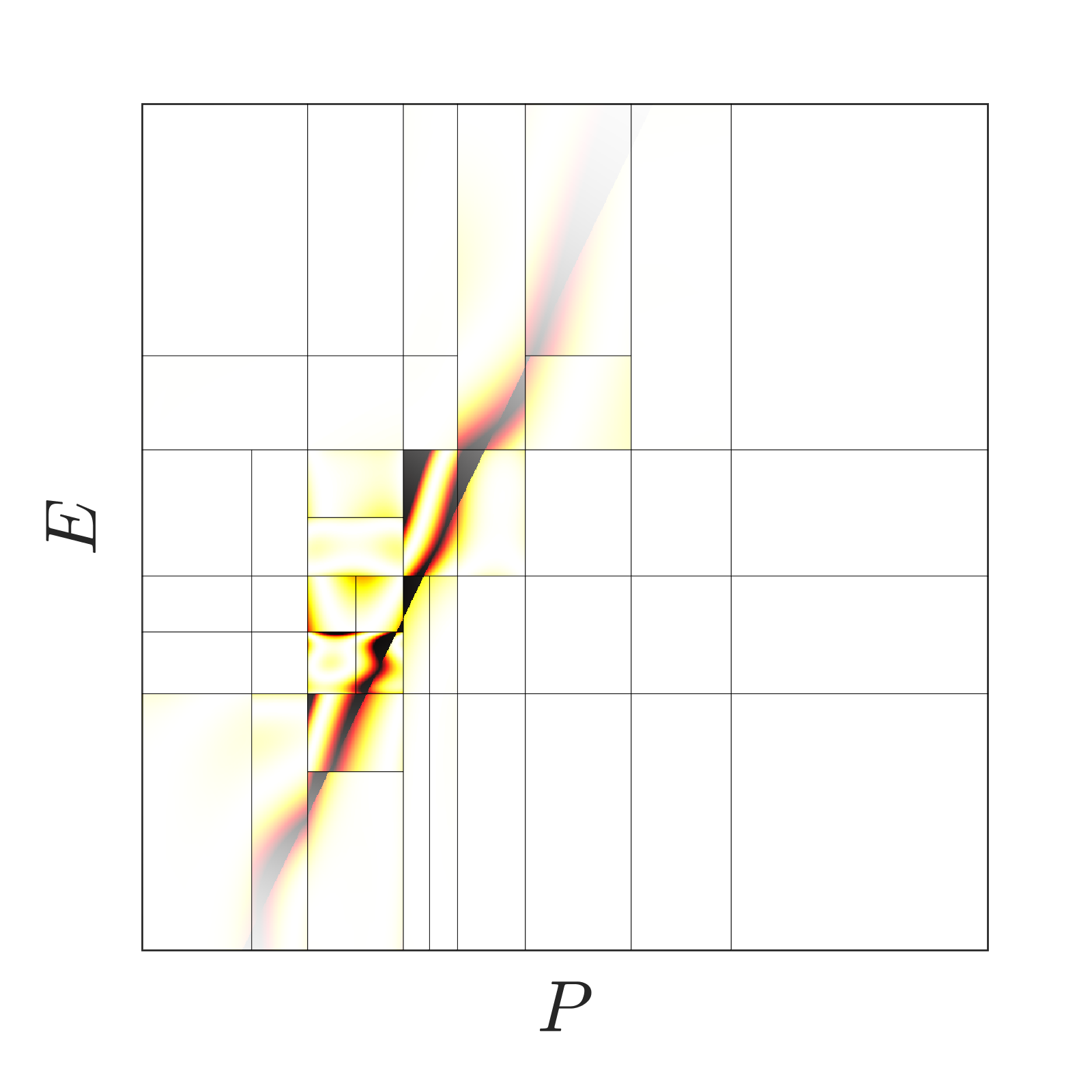}
		\label{fig:ex4:comparison:SSEDelta}
	}%
	\subfloat[SSE domain-wise error estimator:\newline ${\widehat{E}_{\mathrm{LOO}}^{\ell,p}/\Var{\cM}}$]{
		\includegraphics[height=5cm,clip=true,trim=10 10 20 0]{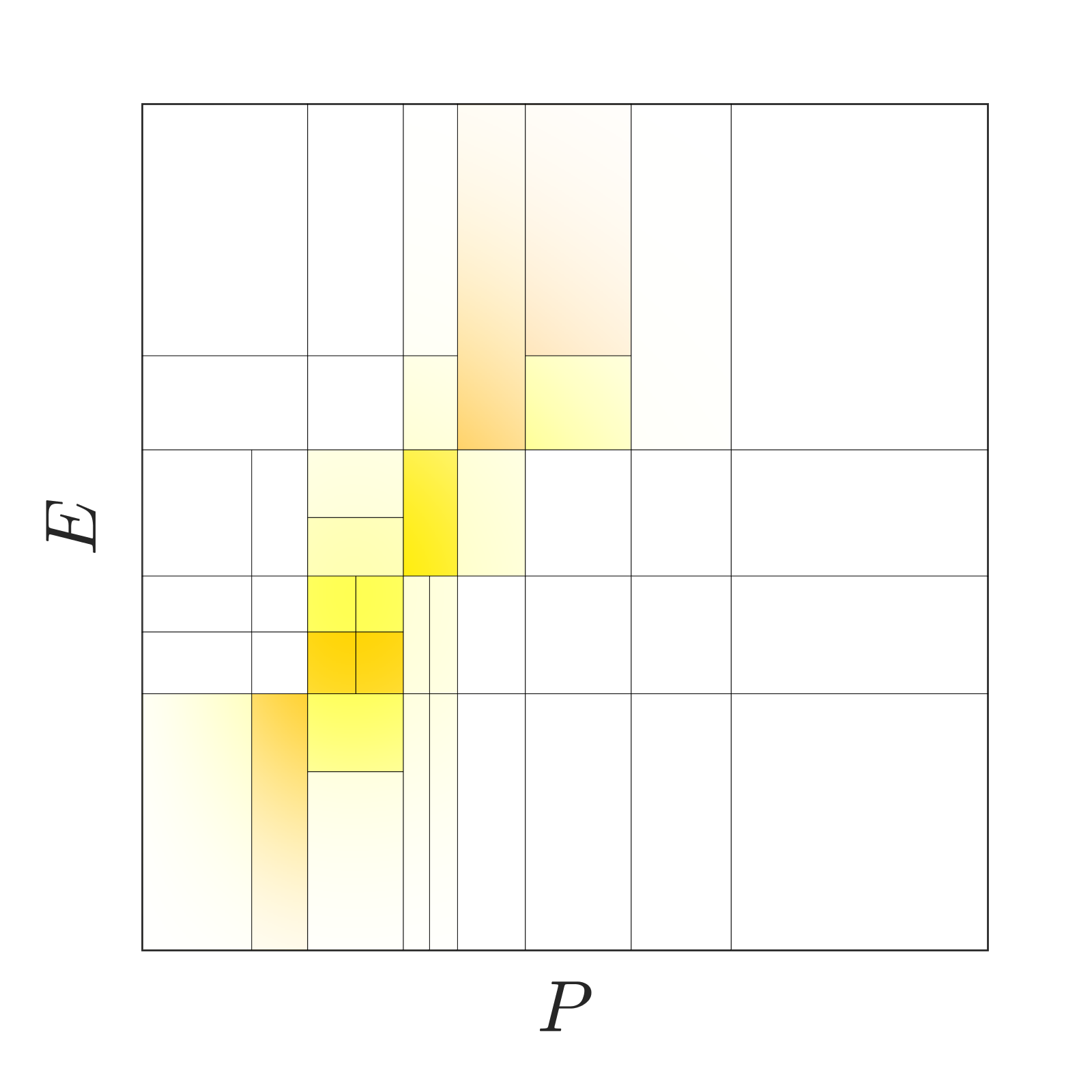}
		\label{fig:ex4:comparison:LOO}
	}%
	\subfloat{
		\includegraphics[height=5cm,clip=true,trim=200 10 0 0]{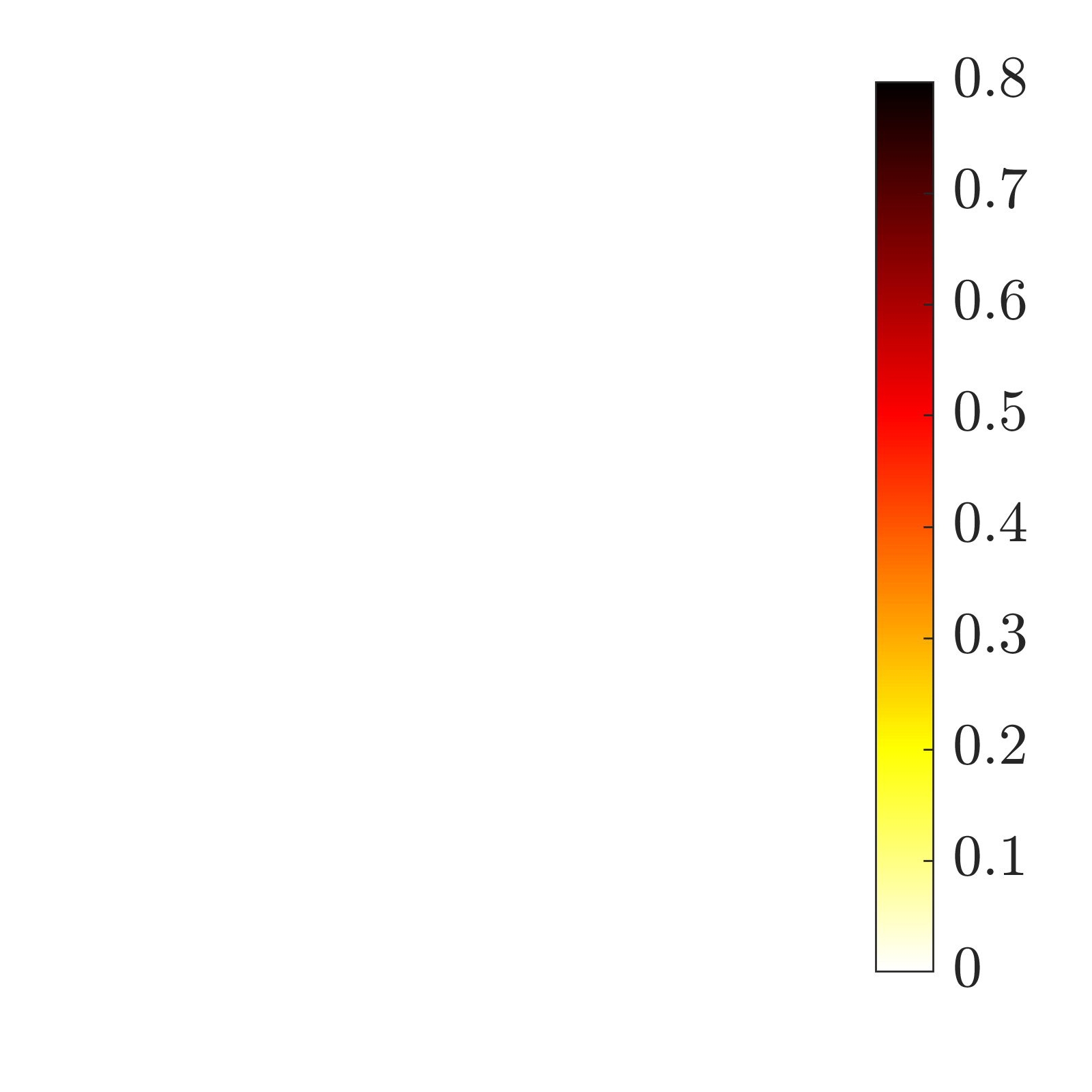}
	}%
	\caption{\emph{3-dimensional snap-through truss}: comparison between PCE/SSE pointwise error and SSE domain-wise error estimator from Eq.~\eqref{eqn:SSE generror estimator}) normalized by the model variance. The fine lines show the SSE subdomains. \label{fig:ex4:comparison}}
\end{figure}

Figure~\ref{fig:Ex4:convergence} summarizes the convergence behavior of PCE and SSE in this benchmark.
For all experimental designs, SSE outperforms sparse PCE.
The dispersion of the RMSE is also significantly improved. 
These observations can be explained with the well-known \emph{Gibbs phenomenon} in spectral representations,  that leads to large discrepancies close to discontinuities. 
The effect is far less severe (although still present) for SSE than for PCE because it is restricted to those subdomains that actually contain the discontinuity. 
This behavior is investigated more closely in Figure~\ref{fig:ex4:comparison}, where a cross section through $\cM$ is shown. It is created by setting $A$ to its mean value and drawing a map proportional to the point-wise discrepancy in the remaining directions.
We adjust transparency of the model response to reflect the underlying joint PDF: solid colors correspond to high probability, fading ones to low probability. Figures~\subref*{fig:ex4:comparison:PCEDelta} and \subref*{fig:ex4:comparison:SSEDelta} show the relative point-wise error at an experimental design size of $\Ntot = 5{,}000$ for PCE and SSE, respectively. Figure~\subref*{fig:ex4:comparison:LOO} shows instead the domain-wise error estimator $\widehat{E}^{\ell,p}_{\rm LOO}$ from Eq.~\eqref{eqn:SSE generror estimator}. 

% discuss
All plots clearly show the effect of the function discontinuity. For PCE the Gibbs phenomenon is clearly visible. 
It causes a large error near the discontinuity, with an oscillating error at large distances to the discontinuity. 
Naturally, SSE also suffers from the same problem, but its effect is only localized close to the discontinuity and it does not affect further regions. 
Furthermore, the available domain-wise local error estimator in Figure \subref*{fig:ex4:comparison:LOO} gives a clear indication of local loss of accuracy of SSE. 
In practical applications this information can be crucial, as it allows one to assess the confidence of the surrogate model predictions, and could be used to adaptively enrich the experimental design close to critical regions \citep{Wagner2020JCP}.

\subsection{Error estimation accuracy}
\label{sec:Applications:Error}

For practical applications, it is important that surrogate models offer insights into their prediction accuracy. Most surrogate models offer global confidence bounds on their predictions that are typically computed through cross-validation techniques ({\em e.g.}, leave-one-out error), while some techniques also offer point-wise confidence bounds ({\em e.g.}, Kriging \cite{Santner2003} and bootstrap PCE \citep{MarelliSS2018}). 

\begin{figure}
	\centering
	\subfloat[One-dimensional analytical function]{
		\begin{minipage}{0.4\linewidth}
			\includegraphics[width=\linewidth]{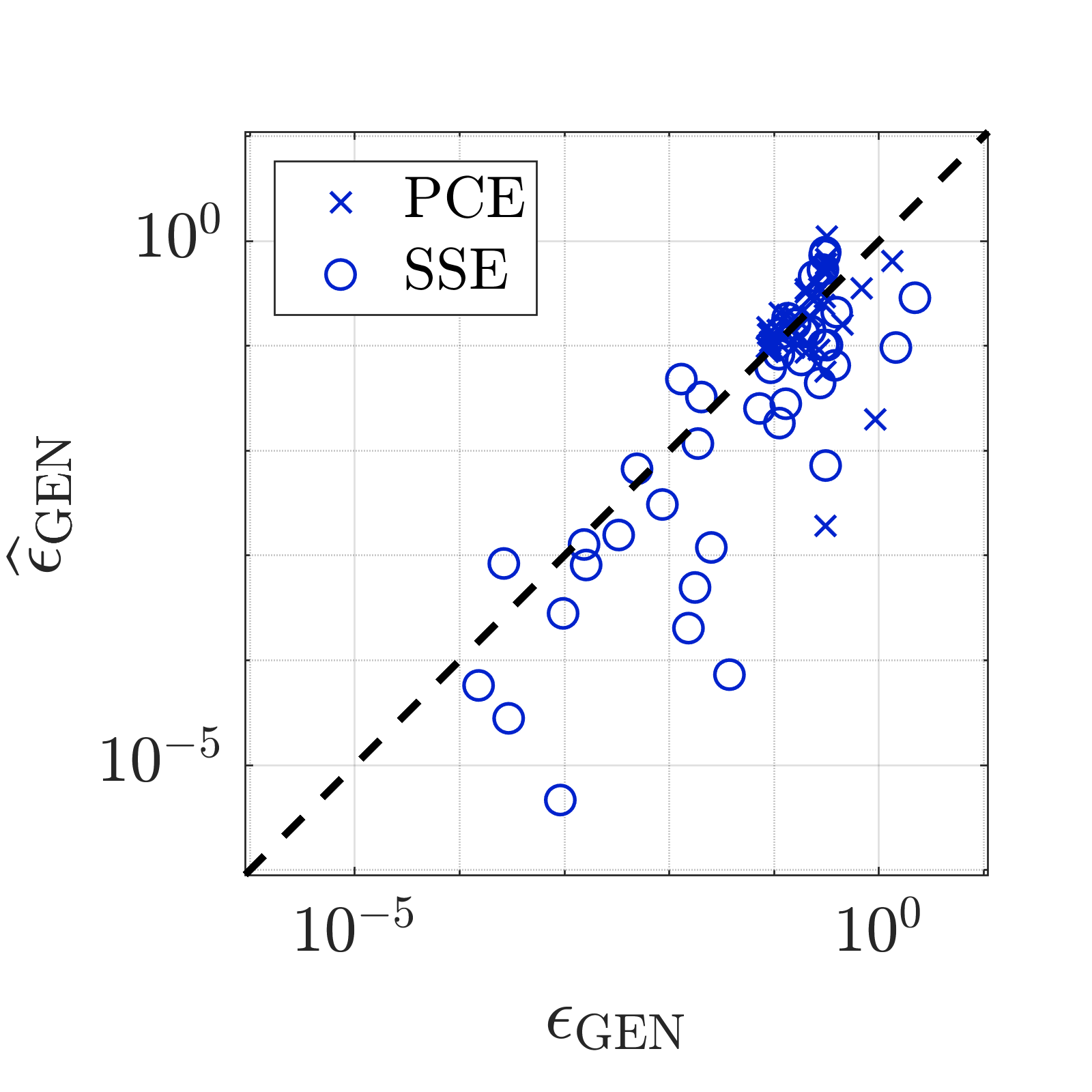}
			\label{fig:RMSvsLOO:ex1}
		\end{minipage}
	}%
	\subfloat[100-dimensional analytical function]{
		\begin{minipage}{0.4\linewidth}
			\includegraphics[width=\linewidth]{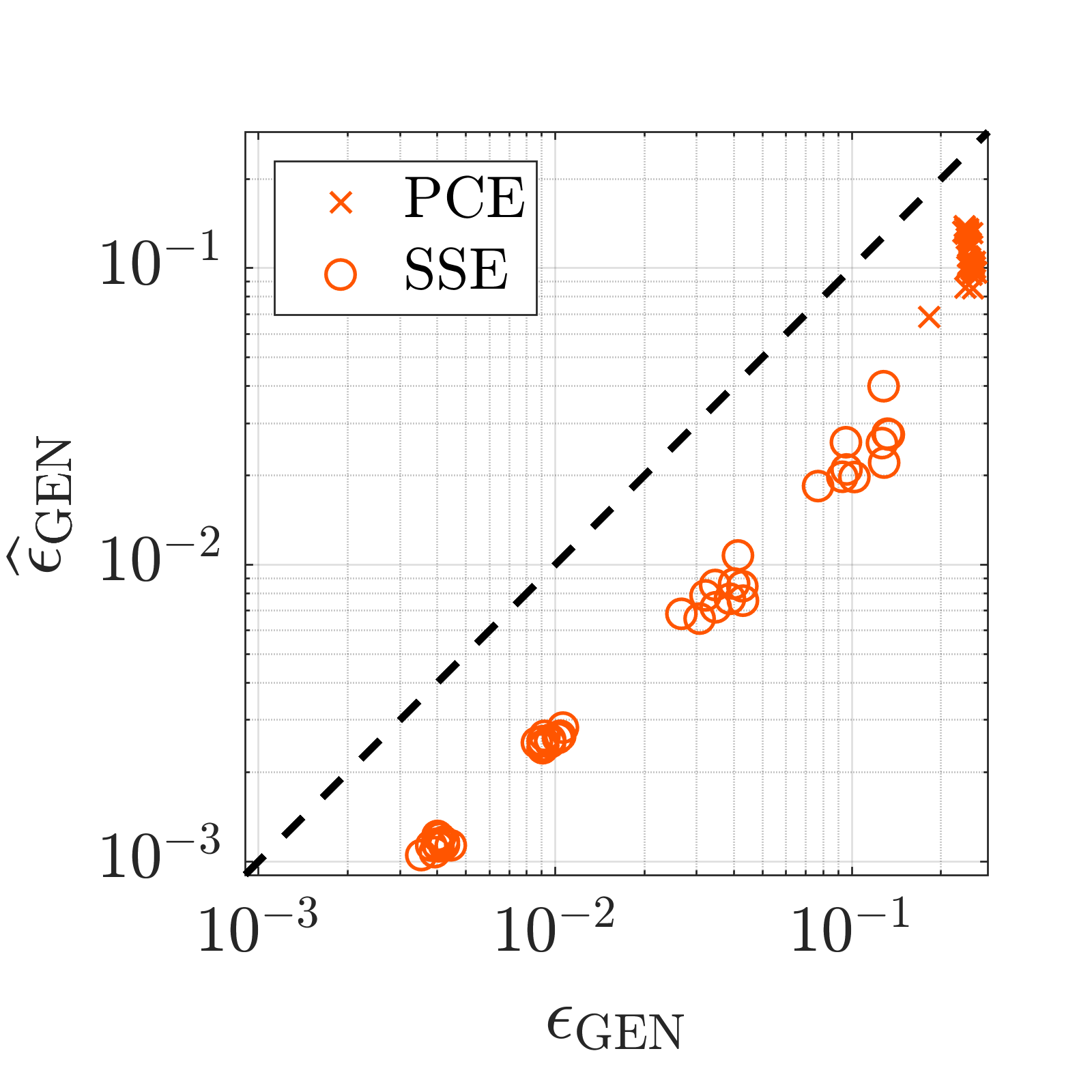}
			\label{fig:RMSvsLOO:ex2}
		\end{minipage}
	}%
	
	\subfloat[8-dimensional damped oscillator]{
		\begin{minipage}{0.4\linewidth}
			\includegraphics[width=\linewidth]{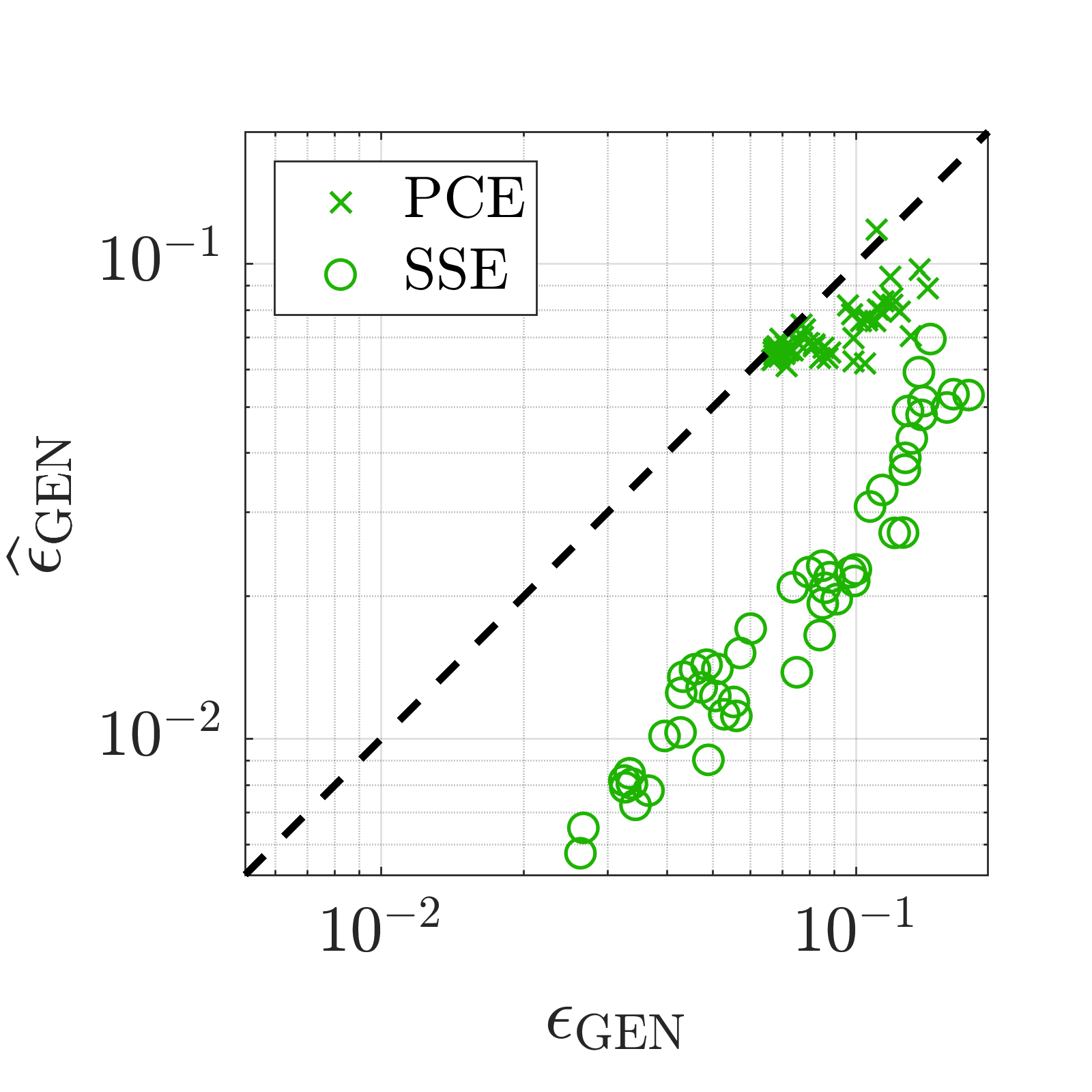}
			\label{fig:RMSvsLOO:ex3}
		\end{minipage}
	}%
	\subfloat[3-dimensional snap-through truss]{
		\begin{minipage}{0.4\linewidth}
			\includegraphics[width=\linewidth]{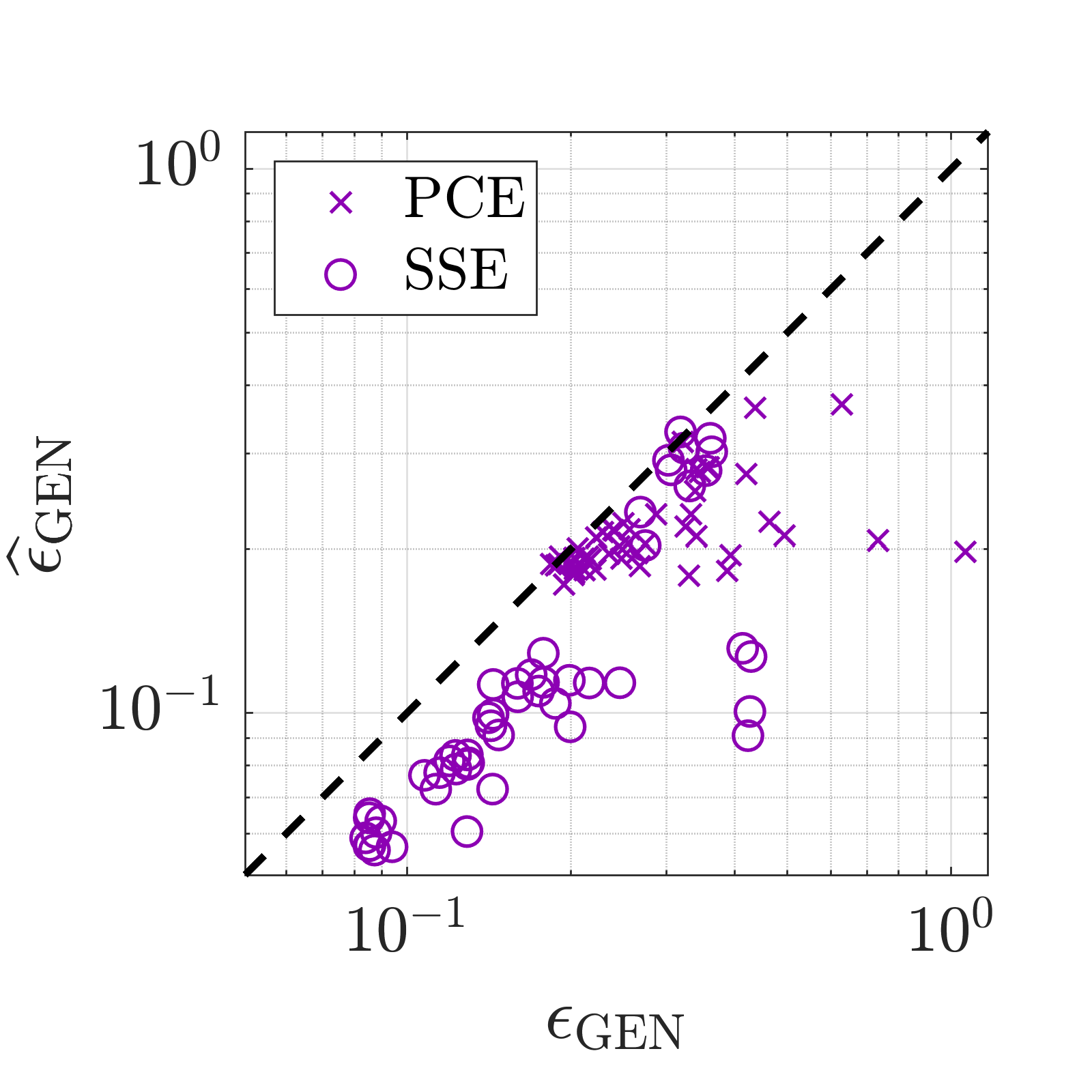}
			\label{fig:RMSvsLOO:ex4}
		\end{minipage}
	}%
	\caption{Comparison of the relative generalization error $\epsilon_{\mathrm{GEN}}$ with the relative generalization error estimator $\widehat{\epsilon}_{\mathrm{GEN}}$ which for SSE is given by Eq.~\eqref{eqn:SSE epsilonLOO} and for PCE is the relative leave-one-out error $E_\text{LOO}/\Var{Y}$.\label{fig:RMSvsLOO}}
\end{figure}

For SSE, the domain-wise error estimators of the local expansions give some insight into the local accuracy of SSE, as shown in the last case study (Section~\ref{sec:Applications:ex4}). The weighted sum of those domain-wise estimators can be used as a global estimator of the generalization error as proposed in Section~\ref{sec:SSE error estimation}, Eq.~\eqref{eqn:SSE epsilonLOO}. 
To assess the accuracy of this global estimator, in Figure~\ref{fig:RMSvsLOO} we plot it against the relative MSE on a validation set for all presented case studies. The diagonal dashed line corresponds to perfect error estimation: $\eta = \widehat{\varepsilon}_{\mathrm{GEN}}$. 
%
%
% comment on figure
In all the applications $\widehat{\varepsilon}_{\mathrm{GEN}}$ significantly underestimates the true error. For comparison, we show the corresponding LOO estimator from PCE, which also exhibits a bias towards lower errors.
On the other hand, it is clear that there is a strong correlation the SSE estimators and the true validation error across all applications, which still makes $\widehat{\varepsilon}_{\mathrm{GEN}}$ a strong potential candidate for model selection in future extensions of the method to adaptively select the SSE hyperparameters ({\em e.g.}, adaptively choosing $\maxDegSSE$).

\subsection{Considerations on computational costs and scalability}
\label{sec:scalability}
For a metamodeling technique to be relevant in an engineering context, the associated training and prediction costs need to be negligible with respect to the costs of the underlying physics-based computational model.
In this sense, SSE performs quite well, as its computational costs scale primarily with the size of the experimental design. 
Indeed, because the residual expansions (Eq.~\eqref{eqn:SSE residual expansion}) can be chosen with very low degree or otherwise strict truncation, the driving cost is the total number of subdomains that are expanded in the SSE sparse tree (see Section~\ref{sec:SSE max L}). 
Interestingly, the expected number of expansions $N_E$ depends only on the ratio between the total experimental design size $N$ and the minimum number of points in each subdomain needed to perform a residual expansion, $N_{\rm min}$, as in (see also Eq.~\eqref{eqn:SSE exp L}):
\begin{equation}
\label{eqn:expected number of PCE}
N_{\rm PCE} = \sum\limits_{\ell = 0}^{\bar{L}} 2^\ell = 2^{\bar{L}+1} -1 = 2^{\lfloor\log_2\left({N}/{N_{\rm min}}\right)\rfloor+1}-1
\leq 2N/N_{min} -1.
\end{equation}
In other words, the computational complexity increases at most linearly with the number of points in the experimental design.
Additionally, the storage costs can be further reduced to $2^{\bar{L}} = N/N_{\rm min}$ expansions when using the flattened representation in Eq.~\eqref{eqn:SSE flattened}. 
Therefore, as required, the training and evaluation costs of SSE are normally negligible with respect to those needed to produce a training set for any realistic engineering application.

%% file: 04_Conclusions.tex
\section{Conclusions}
\label{sec:Conclusions}

In an effort to extend the applicability of the powerful class of spectral decomposition-based metamodels, we propose a novel metamodel technique called \emph{stochastic spectral embedding} (SSE), that exploits both recent advances in UQ (sparse spectral expansions) and in machine learning (regression trees). 
While our presentation was general in nature, we showed how well this approach synergizes with sparse polynomial chaos expansions. 
We also provided analytical formulas to calculate several statistical properties of the resulting model by means of the so-called \textit{flattened representation}, which has additional benefits in terms of computational costs.

We tested the performance of SSE on both simple test functions and engineering-like examples of varying dimensionality and complexity, using varying experimental design sizes, and compared it to our best performing sparse PCE. 
Its generalization capabilities, especially for highly complex models and large experimental designs, outperform PCE in most cases.

We also demonstrated that the associated computational costs of the training of SSE scale linearly in expectation with the number of points in the experimental design. 
This compares favorably with most metamodeling techniques common in the UQ community (\textit{e.g.} PCE or Kriging).

This performance, however, comes at the cost of trading the continuity of PCE for the piecewise continuity of SSE. This also implies the loss of the effective generalization error estimate provided by $\widehat{\epsilon}_{\rm LOO}$ in linear regression. 
To mitigate this issue, we proposed the error estimate $\widehat{\epsilon}_{\rm GEN}$ (Eq.~\eqref{eqn:SSE epsilonLOO}). Despite its absolute scale being biased towards lower values, it still shows high correlation with the actual generalization error for all experimental design sizes and dimensions.
This is a promising property for further research into providing automatic selection of the hyperparameters of the algorithm (which at the moment are the maximum degree of the residual expansions $\maxDegSSE$, as well as the minimum number of points in each subdomain $N_{\rm min}$ required to expand the residual) and further enhance its performance.

Additional research is ongoing towards the use of the approximate local error measures provided by $\widehat{E}_{\rm GEN}^{\ell,p}$ (Eq.~\eqref{eqn:SSE local error}) for goal-oriented adaptive experimental design construction, a topic explored by the authors in \cite{Wagner2020JCP}.

%% file: 05_Appendix.tex
\section{Postprocessing PCE-based SSE}
\label{app:postProcessingSSE}

If polynomials chaos expansions are used to construct the residual expansions, several quantities of engineering interest can be computed analytically as a post-processing step of the final SSE.
In this section we derive expressions for (i) \emph{conditional expectations}, (ii) \emph{partial variances} and (iii) Sobol' indices.

\subsection{Conditional expectations}
\label{sec:SSE:Postprocessing:ConditionalExpectation}
Conditional expectations describe the expectation of a multivariate function of a random vector $\vX \sim f_{\vX}(\vx)$, conditioned on a subset of $\ve{X}$ assuming a fixed value. Let $\vX=\{X_i\}_{i = 1,\cdots,M}\in\cD_{\vX}$ be an independent random vector with PDF $f_{\vX}(\vx) = \prod_{i=1}^M f_{X_i}(x)$. Denote further by $\iu$ and $\iv$ two disjoint index sets such that $\iu\cup\iv = \{1,\dots,M\}$ and by $\vX_{\iu}\eqdef\{X_i\}_{i\in\iu}\in\cD_{\vX_{\iu}}$ a random sub-vector with PDF $f_{\vX_{\iu}}(\vx_{\iu}) = \prod_{i\in\iu} f_{X_i}(x)$. 
Additionally define the complementary random vector 
$\vX_{\iv}\eqdef\{X_i\}_{i\in\iv}\in\cD_{\vX_{\iv}}$.
% with PDF $f_{\vX_{\iv}}(\vx_{\iv}) = \prod_{i\ni\iu}^M f_{X_i}(x)$. 
The conditional expectation of 
$\cM_{\mathrm{SSE}}(\vX)$ {\em w.r.t.} $\vX_{\iu}$ can then be written as 
\begin{equation}
\label{eq:SSE:conditionalExpectation}
\mathbb{E}\left[\cM_{\mathrm{SSE}}(\vX)\right\vert \vX_{\iu}] \eqdef %\mathbb{E}_{\vX_{\iv}}\left[\cM_{\mathrm{SSE}}(\vX)\right] \eqdef 
\int_{\cD_{\vX_{\iv}}}\cM^F_{\mathrm{SSE}}(\vx)f_{\vX}(\vx)\,\di{\vx_{\iv}},
\end{equation}
where we used the \emph{flattened representation} from Eq.~\eqref{eqn:SSE flattened}. This corresponds to marginalizing over the parameters $\vX_{\iv}$.
Due to the local orthonormality of the SSE representation, an analytical 
expression for this integral can be found as
\begin{equation}
\label{eq:SSE:SSEconditionalExpectationExpress}
\mathbb{E}\left[\cM_{\mathrm{SSE}}(\vX)\right\vert \vX_{\iu}] =
\sum_{p=1}^{P_{L}}\cV^{L,p}_{\iv}\sum_{\ve{\alpha}\in\cA^{\cT}_{\iv=0}} 
c_{\ve{\alpha}}^{p}\Psi_{\ve{\alpha},\iu}^{L,p}(\vX_{\iu}^{L,p}),
\end{equation}
where $\cV_{\iv}^{L,p}\eqdef\int_{\cD^{L,p}_{\vX_{\iv}}} 
f_{\vX_{\iv}}(\vx_{\iv})\,\di{\vx_{\iv}}$ is the input mass in 
the marginalized dimensions and 
$\cA^{\cT}_{\iv=0}\eqdef\{\ve{\alpha}\in\cA^{\cT}:\alpha_i= 0 \Leftrightarrow 
i\in\iv\}$. Further, $\vX_{\iu}^{L,p}\in\cD_{\vX_{\iu}^{L,p}}$ is an auxiliary random variable that is only defined in the $(L,p)$-subdomain and $\Psi_{\ve{\alpha},\iu}^{L,p}(\vX_{\iu}^{L,p})\eqdef\prod_{i\in\iu}\Phi_{\alpha_i}^{(i),L,p}(X_{i}^{L,p})$
is a polynomial basis function of the non-marginalized variables. 

The marginalization process in Eq.~\eqref{eq:SSE:conditionalExpectation} creates an additional problem: this expression generally contains overlapping subdomains $\{\cD_{\vX_{\iu}}^{L,p}\}_{p=1,\cdots,P_L}$ due to the fact that terminal subdomains in the full input space are not necessarily terminal subdomains in the lower dimensional, conditional expectation input space defined by $\vu$. 
However, because the basis functions are polynomials, it is once again possible to perform the \textit{flattening} process (see Section~\ref{sec:SSE error estimation}) and obtain disjoint subdomains. 
By denoting as $\cP\subseteq\{1,\cdots,P_L\}$ the set of terminal subdomains in the conditional input variables $\vX_{\vu}$, we can rewrite Eq.~\eqref{eq:SSE:SSEconditionalExpectationExpress} as
\begin{equation}
\label{eq:SSE:SSEconditionalFlat}
\mathbb{E}\left[\cM_{\mathrm{SSE}}(\vX)\right\vert \vX_{\iu}] =
\sum_{p\in\cP}\sum_{\ve{\alpha}\in\cA^{\cP}_{\iv=0}} 
d_{\ve{\alpha}}^{p}\Psi_{\ve{\alpha},\iu}^{L,p}(\vX_{\iu}^{L,p}),
\end{equation}
where $\cA^{\cP}_{\iv=0}$ is a suitable multi-index set allowing an exact representation of the polynomials and $d_{\ve{\alpha}}^{p}$ are the corresponding coefficients.

\subsection{Partial variance and Sobol' indices}
\label{sec:SSE:Postprocessing:PartialVariance}

The Sobol' Hoeffding decomposition \citep{Sobol1993,LeGratiet2016} of the SSE representation $\cM_{\mathrm{SSE}}$ reads
\begin{equation}
\label{eq:SSE:HoeffdingSobolDecomp}
\cM_{\mathrm{SSE}}(\vX) = 
\cM_{\mathrm{SSE}}^0 + 
\sum_{\substack{\iu\subset\{1,\cdots,M\} \\ 
		\iu\neq\emptyset}}\cM^{\iu}_{\mathrm{SSE}}(\vX_{\iu}),
\end{equation}
where $\cM^0_{\mathrm{SSE}}\eqdef\Esp{\cM_{\mathrm{SSE}}(\vX)}$ and the 
remaining terms can be computed recursively by
\begin{align}
\label{eq:SSE:RecursiveDecomposition:first}
\cM^i_{\mathrm{SSE}}(X_{i}) &= 
\mathbb{E}_{\vX_{\sim i}}\left[\cM_{\mathrm{SSE}}(\vX)\right] - 
\cM^0_{\mathrm{SSE}},\\
\cM^{ij}_{\mathrm{SSE}}(\vX_{ij}) &= 
\mathbb{E}_{\vX_{\sim ij}}\left[\cM_{\mathrm{SSE}}(\vX)\right] - 
\cM^i_{\mathrm{SSE}}(X_{i}) - \cM^j_{\mathrm{SSE}}(X_{j}) -
\cM^0_{\mathrm{SSE}},\\
\cdots &= \cdots\\
\cM^{\iu}_{\mathrm{SSE}}(\vX_{\iu}) &= 
\mathbb{E}_{\vX_{\iv}}\left[\cM_{\mathrm{SSE}}(\vX)\right] - 
\sum_{\substack{\iw\subset\iu \\ \iw\neq\emptyset}}
\cM^{\iw}_{\mathrm{SSE}}(\vX_{\iw}) -
\cM^0_{\mathrm{SSE}}.
\end{align}

The decomposition of Eq.~\eqref{eq:SSE:HoeffdingSobolDecomp}, allows the definition of the so-called \emph{partial variance}, \ies the fraction of the variance $\Var{\cM_{\mathrm{SSE}}(\vX)}$ that can be attributed to $\vX_{\iu}$, defined by
\begin{equation}
\label{eq:SSE:partialVariance}
V_{\iu} \eqdef
\int_{\cD_{\vX_{\iu}}} 
\left(\cM_{\mathrm{SSE}}^{\iu}(\vx_{\iu})\right)^2
f_{\vX_\iu}(\vx_{\iu})\,\di{\vx_{\iu}}.
\end{equation}

Using Eq.~\eqref{eq:SSE:RecursiveDecomposition:first}, the so-called \emph{first order partial variance} can therefore be written as
\begin{equation}
\label{eq:SSE:firstOrder}
V_{i} = \int_{\cD_{X_{i}}}  
\left(
\mathbb{E}_{\vX_{\sim i}}\left[\cM_{\mathrm{SSE}}(\vX)\right] - 
\cM^0_{\mathrm{SSE}}
\right)^2f_{X_i}(x_i)
\,\di{x_{i}}.
\end{equation}

With the expression for the conditional expectation from Eq.~\eqref{eq:SSE:SSEconditionalFlat}, this integral can be solved analytically as
\begin{align}
V_{i} &= \int_{\cD_{X_{i}}}  
\left(
\sum_{p\in\cP}\sum_{\ve{\alpha}\in\cA^{\cP}_{\sim i=0}} 
d_{\ve{\alpha}}^{p}\Phi_{\alpha_i}^{(i),L,p}(X_{i}^{L,p}) - \cM^0_{\mathrm{SSE}}
\right)^2f_{X_i}(x_i)
\,\di{x_{i}}\\
&= \sum_{p\in\cP}\cV_i^{L,p}\sum_{\ve{\alpha}\in\cA^{\cP}_{\sim 
		i=0}}\left(d_{\ve{\alpha}}^{p}\right)^2 - 
\left(\cM^0_{\mathrm{SSE}}\right)^2.
\end{align}

With the availability of the partial variances in Eq.~\eqref{eq:SSE:firstOrder} and the total variance from Eq.~\eqref{eqn:SSE variance}, one can analytically derive the \emph{first order Sobol' indices}:
\begin{equation}
S_{i} \eqdef \frac{V_{i}}{\Var{\cM_{\rm SSE}(\vX)}}.
\end{equation}

Higher order indices can be computed in a similar way.